%% file: main.tex
\documentclass[twocolumn,tighten,astrosymb,trackchanges]{aastex631}
\input{affiliations}

\usepackage{CJK}
\usepackage{rotating}
\usepackage{multirow}
\newcommand\kms{\,km~s$^{-1}$}
\newcommand\mch{$M_{\text{Ch}}$}
\newcommand\um{\,$\mu$m}
\let\tablenum\relax
\usepackage{siunitx}
\usepackage{amsmath,amsthm,amsfonts,amssymb,amscd}
\usepackage{booktabs}
\usepackage{lipsum}
\usepackage{graphicx}
\usepackage{makecell}
\usepackage{siunitx}
\definecolor{maroon}{rgb}{0.760,0.118,0.337}

\definecolor{darkaqua}{rgb}{0.0,0.45,0.65}

\def\msun{\hbox{\,M$_{\odot}$}}

\def\cm{\mbox{\,cm}}
\def\cm3{\mbox{\,cm$^{-3}$}}

\shorttitle{\textit{JWST} Stable Nickel}
\shortauthors{Kwok et al.}

\begin{document}
\begin{CJK*}{UTF8}{gbsn}
\title{\textit{JWST} Spectroscopy of SN\,Ia 2022aaiq and 2024gy: Evidence for Enhanced Central Stable Ni Abundance and a Deflagration-to-Detonation Transition}

\correspondingauthor{Lindsey A. Kwok}
\email{lindsey.kwok@northwestern.edu}

\input{authors}

\begin{abstract}
We present optical$+$near-infrared (NIR)$+$mid-infrared (MIR) observations of the normal Type~Ia supernovae (SN\,Ia) 2022aaiq and 2024gy in the nebular phase, continuously spanning 0.35--28\um. Medium-resolution \textit{JWST} spectroscopy reveals novel narrow ($v_{\mathrm{FWHM}}<1500$\kms) [\ion{Ni}{2}]~1.94 and 6.64\um\ cores in both events. The MIR  [\ion{Ni}{2}]~6.64~\micron\ line exhibits a distinct narrow core atop a broader base, indicating a central enhancement of stable Ni. This structure points to high central densities consistent with a near-Chandrasekhar-mass (\mch) progenitor or a high-metallicity sub-\mch\ progenitor. From detailed line-profile inversions of SN\,2024gy, we derive emissivity profiles for stable iron-group elements (IGEs), radioactive material, and intermediate-mass elements (IMEs), revealing spatially distinct ejecta zones. The [\ion{Ni}{3}]~7.35~\micron\ line shows a shallow-to-steep slope transition---a ``broken-slope'' morphology---that matches predictions for delayed detonation explosions with separated deflagration and detonation ashes. We also reanalyze and compare to archival \textit{JWST} spectra of SN\,2021aefx and the subluminous SN\,2022xkq. From the stable Ni luminosities, we infer that SN\,2024gy produced $\sim$5--10 times more stable Ni mass than SN\,2022xkq, favoring a near-\mch\ scenario for SN\,2024gy and sub-\mch\ scenario for SN\,2022xkq. These results demonstrate that resolved line profiles, now accessible with \textit{JWST}, provide powerful diagnostics of explosion geometry, central density, and progenitor mass in SN\,Ia.

\end{abstract}

\keywords{Supernovae (1668), Type Ia supernovae (1728), White dwarf stars (1799)}

\section{Introduction \label{sec:intro}}

Type Ia supernovae (SN\,Ia) are thermonuclear explosions of white dwarfs (WDs; \citealt{Hoyle1960}). Their nucleosynthetic yields depend sensitively on the density of different layers of the WD during the explosion. At the highest densities ($\gtrsim10^{8}$\,g\,cm$^{-3}$, and assuming $T \gtrsim 5\times10^9$\,K, $Y_e = 0.5$), nuclear burning produces stable iron-group elements (IGEs). Slightly lower densities yield radioactive IGEs such as Fe, Co, and Ni, while intermediate-mass elements (IMEs; e.g., Si, Ar, Ca) form at still lower densities. The lowest-density regions contribute low-mass elements (LMEs; Ne, Mg, O) as well as possible unburned C/O material \citep[for a review, see][]{Seitenzahl2017}.

The central density of the exploding WD is set primarily by its mass, owing to electron degeneracy pressure. Near Chandrasekhar-mass (\mch) WDs ignite at higher central densities than sub-\mch\ WDs. As a result, at fixed progenitor metallicity, sub-\mch\ explosions undergo fewer electron captures and produce smaller yields of stable IGEs \citep{Hoflich2004, Blondin2018, Wilk2018, Shingles2020}. The presence and abundance of stable Ni are particularly diagnostic of the explosion conditions: it requires the highest densities and temperatures, and some explosion mechanisms therefore yield little of it at a given metallicity or $^{56}$Ni mass \citep{Lach2020, Blondin2022, Pakmor2024}. The most common isotope is $^{58}$Ni, though $^{60}$Ni can be significant at the highest densities and in certain double-detonation models \citep{Gronow2020, Gronow2021, townsley_double_2019}. Observations of stable Ni in SN\,Ia ejecta thus directly probe both progenitor mass and explosion mechanism.

At late times ($>$100 days after peak brightness; nebular phase), the ejecta expand and dilute enough to become optically thin \citep{Bowers1997, Branch2008, Silverman2013, Friesen2014, Black2016}. Nebular spectra reveal emission lines that trace the geometry of the emitting regions and allow measurements of elemental distributions and kinematics \citep[see][for a review]{jerkstrand_spectra_2017}. By this stage, all $^{56}$Ni has decayed (half-life 6.1 days), so Ni emission originates only from stable isotopes.

Stable Ni has been studied in both optical and ground-based near-infrared (NIR) spectra, particularly through the [\ion{Ni}{2}]~7378~\AA\ and [\ion{Ni}{2}]~1.94\um\ lines \citep{Maeda2010a, Blondin2018, Maguire2018, Flors2018, Diamond2018, Dhawan2018, Flors2020, Blondin2022, Kumar2025}. The 1.94\um\ feature was clearly detected in SN\,2014J \citep{Diamond2018, Dhawan2018}, and \citet{Kumar2025} extended such analyses to a larger sample using improved telluric corrections. These analyses reveal diversity: for example, \citet{Flors2020} favored sub-\mch\ progenitors, while \citet{Kumar2025} argued that narrow NIR Ni lines in subluminous events suggest near-\mch\ explosions.

A major difficulty in the analysis is line blending. The optical [\ion{Ni}{2}]~7378~\AA\ line overlaps with neighboring features, while the NIR [\ion{Ni}{2}]~1.94\um\ line, though less blended \citep{Wilk2018}, falls near a strong atmospheric telluric band (for low-redshift SN). High signal-to-noise ratio (S/N) and excellent telluric correction are required for robust detection \citep{Kumar2025}. \textit{JWST} removes these limitations, providing access to the 1.94\um\ feature without atmospheric contamination and, critically, new stable Ni diagnostics in the mid-infrared (MIR). These include [\ion{Ni}{2}]~6.64\um, [\ion{Ni}{3}]~7.35\um, [\ion{Ni}{4}]~8.40\um, and [\ion{Ni}{3}]~11.00\um. The MIR also isolates IME lines such as [\ion{Ca}{4}]~3.21\um, [\ion{Ar}{2}]~6.98\um, [\ion{Ar}{3}]~8.99\um, and [\ion{S}{4}]~10.51\um, and LME tracers such as [\ion{Ne}{2}]~12.88\um\ and [\ion{Mg}{2}]~9.76\um\ \citep{Gerardy2007, Kwok2023, DerKacy2023, Blondin2023, DerKacy2024, Ashall2024, Siebert2024, Kwok2024, Kwok2025}. These lines are all significantly more isolated than optical and NIR lines. Together, these provide the clearest view yet of the chemical structure of SN\,Ia ejecta.

In delayed detonation (DDT) models, the explosion begins with a subsonic deflagration that releases energy and causes the star to expand before a detonation front ignites. The high densities during the deflagration phase enhance the production of stable IGEs, while the subsequent detonation synthesizes material at lower densities. As a result, deflagration and detonation ashes occupy distinct regions of the ejecta, with different proportions of stable and radioactive Ni \citep{Seitenzahl2013, Pakmor2024}.

In contrast, sub-\mch\ double-detonation models involve an initial He-shell detonation that triggers a secondary detonation in the C/O core. Stable Ni is synthesized primarily in the central detonation ashes, colocated with radioactive IGEs \citep{Gronow2021, Pakmor2024}. These differing burning pathways lead to distinct distributions of nucleosynthetic products, which in turn imprint characteristic signatures on nebular emission-line profiles that are now accessible with \textit{JWST}.

In this paper we present late-time, medium-resolution NIR$+$MIR spectroscopy of the normal SN\,Ia 2022aaiq and 2024gy from \textit{JWST} General Observer (GO) programs 2072 \citep{Jha2021} and 4516 \citep{Jha2023}. We focus on the morphologies of the [\ion{Ni}{2}] and [\ion{Ni}{3}] lines and their implications for explosion models. In both events, [\ion{Ni}{2}] shows a narrow core ($v \leq 1500$\kms) indicative of an enhanced central abundance of stable IGEs, while [\ion{Ni}{3}] exhibits a ``broken-slope'' profile, consistent with predictions of DDT models.

\begin{figure*}
    \centering
    \includegraphics[width=0.7\linewidth]{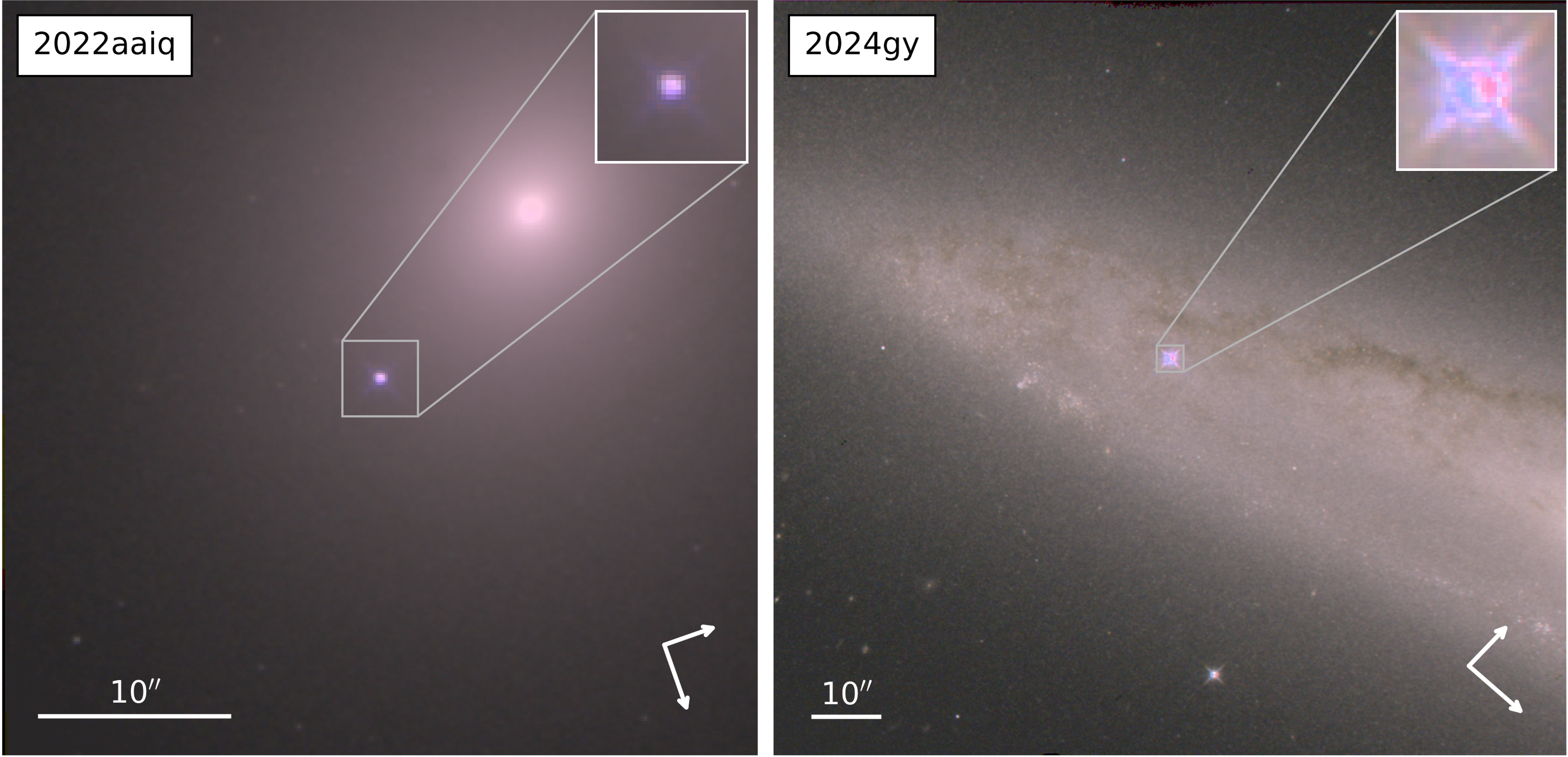}
    \caption{\textit{HST} WFC3/IR NIR images of SN\,2022aaiq (left) in its elliptical host galaxy, NGC~5631, and SN\,2024gy (right) in its spiral host galaxy, NGC~4216.  The RGB channels are mapped from F160W, F140W, and F105W images, respectively.  The images are 40\arcsec$\times$40\arcsec\ and  120\arcsec$\times$120\arcsec, respectively, with a scale bar for reference.  The orientation is marked by the compass rose with the longer and shorter arms representing north and east, respectively.  A 5\arcsec$\times$5\arcsec\ box is centered on each SN with a zoomed-in image of that region shown in the upper-right corner.}
    \label{fig:sn_images}
\end{figure*}

\section{Observations \label{sec:obs}}

SN\,2022aaiq (\autoref{fig:sn_images}, \textit{left}; $\alpha = 14^{\mathrm{h}} 26^{\mathrm{m}} 32^{\mathrm{s}}_{^{\centerdot}}011$, $\delta = +56^\circ35' 03\farcs17$, J2000) was discovered and reported to the Transient Name Server\footnote{\url{https://www.wis-tns.org}} (TNS) by P.\,Wiggins on 15 November 2022 (\citealt{Wiggins2022}; all dates in this work are given in UTC) and was classified as a Type Ia SN on 17~November~2022 by the UCSC team \citep{Siebert2022}. SN\,2022aaiq exploded in NGC~5631, an elliptical galaxy at redshift $z = 0.006485$ \citep{Cappellari2011}.

Discovery details for SN\,2024gy (\autoref{fig:sn_images}, \textit{right;} $\alpha = 12^{\mathrm{h}} 15^{\mathrm{m}} 51^{\mathrm{s}}_{^{\centerdot}}289$, $\delta = +13^\circ06' 56\farcs13$, J2000), hosted by the spiral galaxy NGC~4216, are given by \citet{Li2026}. We adopt a slightly different host-galaxy heliocentric redshift than \citet{Li2026}, preferring a recent measurement with optical lines at the host nucleus, $z = 0.000385\pm0.000045$ or $c\,z = 116\pm14$
\kms\ \citep{vandenBosch2015}. Our derived velocities for SN line features in this work are relative to the systemic velocity of the host galaxy, but owing to the unknown velocity of SN\,2024gy within its host, we adopt an additional uncertainty of $\sim\pm200$\kms\ when interpreting line shifts. Additional optical analysis of SN\,2024gy is given by \citep{Li2026} and J. Terwel et al. (under revision).

\autoref{fig:sn_images} shows SN\,2022aaiq and SN\,2024gy in their host galaxies with NIR imaging from the \textit{Hubble Space Telescope} (\textit{HST}). These observations were taken as part of program 17128 (PI: R.~Foley) with WFC3/IR in the F105W, F140W, and F160W passbands.

\subsection{\textit{JWST} Data}

SN\,2022aaiq and SN\,2024gy were each observed at two epochs with the \textit{JWST} Near-Infrared Spectrograph (NIRSpec) Fixed Slit (FS) G235M $+$ G395M gratings \citep{Jakobsen2022,Birkmann2022,Rigby2022} and the Mid-Infrared Instrument (MIRI) Medium Resolution Spectrograph (MRS) \citep{Kendrew2015,Kendrew2016,Rigby2022}, with the same settings and exposure times (MIRI/MRS, 2~hr; NIRSpec gratings, 0.5~hr). SN\,2022aaiq was observed by \textit{JWST} on 04~April~2023 at a phase relative to \textit{B}-band maximum (1~December~2022; \autoref{sec:bayesn}) of $+$125~days and 24~June~2023 at $+$207~days. SN\,2024gy was observed by \textit{JWST} on 09~June~2024 at $+$144~days and 22~December~2024 at $+$337~days post-peak (20~January~2024; \autoref{sec:bayesn}; \citealt{Li2026}). 

For comparison, we also include in our analysis MIRI/MRS data of the normal SN\,Ia\,2021aefx and the subluminous (photometrically transitional, spectroscopically SN 1991bg-like; \citealt{Pearson2024}) SN\,Ia\,2022xkq from \textit{JWST} program GO-2114 \citep{Ashall2021b} published by \cite{Ashall2024} and \cite{DerKacy2024}, respectively. In this work, we use the reductions of these data presented by \cite{Kwok2025}, where the MRS data were reprocessed to improve background subtraction. Differences in the rereduced spectra are discussed in \autoref{sec:obs_22xkq} and \autoref{sec:obs_21aefx}.

We processed the MIRI/MRS observations of SN\,2021aefx, SN\,2022xkq, SN\,2022aaiq, and SN\,2024gy using a consistent reduction workflow. Starting from the stage2 \texttt{*rate.fits} products, we reran the \textit{JWST} pipeline to generate a single, spatially aligned data cube combining Channels~1--4. This step follows Section~3 of the public M.~Shahbandeh reduction notebook.\footnote{\url{https://github.com/shahbandeh/MIRI_MRS/blob/main/MRS_reductions.ipynb}} We subtract the background and extract the one-dimensional (1D) SN spectra using \texttt{AstroBkgInterp} \citep{Nickson_AstroBkgInterp_2025}\footnote{\url{https://github.com/brynickson/AstroBkgInterp}} which performs interpolation-based background estimation optimized for point-source extraction. We used a fixed aperture radius of 5 pixels and a background annulus width of 6 pixels. This aperture captures the full SN flux well up to 14\um, but may underestimate the flux at longer wavelengths as the instrument point-spread function (PSF) broadens. While a smaller aperture at longer wavelengths loses a fraction of the SN flux, it significantly improves the isolation of SN emission lines from the high background in Channel~4. We therefore caution that the flux calibration is uncertain beyond $\sim$14\um\ and do not attempt detailed flux-dependent analysis at these wavelengths. For background subtraction, we adopted the recommended parameters: a polynomial fitting mode of degree 3, a bin size of 5, and a cube resolution of high. None of the four SN exhibits a complex background, and this procedure isolates the SN flux well. Finally, we trimmed each spectrum at wavelengths longer than 26\um, where the noise in Channel~4 becomes dominant and no reliable emission features are detected.

All \textit{JWST} data presented here were obtained from the Mikulski Archive for Space Telescopes\footnote{\url{https://mast.stsci.edu/portal/Mashup/Clients/Mast/Portal.html}} (MAST) at the Space Telescope Science Institute (STScI). The specific observations can be accessed via \dataset[DOI: 10.17909/tzsy-x483]{http://dx.doi.org/10.17909/tzsy-x483}. The data of SN\,2022aaiq, SN\,2024gy, SN\,2021aefx, and SN\, 2022xkq are shown in \autoref{fig:spec_opt+NIR+MIR}, \autoref{fig:MRS_IDs}, and \autoref{fig:NIR_IDs}, respectively. The \textit{JWST} spectra of SN\,2022aaiq and SN\,2024gy continuously span 1.7--28\um. In stitching together the spectra, we switch from NIRSpec to MIRI at 5\um.

\subsection{Ground-based Data}

\subsubsection{Optical Photometry}

Optical photometry for SN\,2022aaiq and SN\,2024gy was obtained through the Global Supernova Project (GSP) collaboration using the Las Cumbres Observatory (LCO; \citealt{brown_cumbres_2013}) 0.4~m and 1~m telescopes. Preprocessing, including bias correction and flat fielding, was handled by the \texttt{BANZAI} pipeline \citep{McCully2018}. Further data reduction was carried out using \texttt{lcogtsnpipe} \citep{Valenti2016}, a photometric reduction pipeline that uses point-spread-function (PSF) photometry \citep{Stetson1987} to calculate zero points, color terms, and extracted magnitudes.

In this work we do not analyze the light curves in detail. We use them to estimate distance and extinction for dereddening and calibrating the spectra used to measure stable Ni luminosities in \autoref{sec:ni_masses}. We estimate the distance to SN\,2024gy using \texttt{BayeSN}, a hierarchical Bayesian SN Ia light-curve model \citep{mandel_hierarchical_2022, Grayling2024}. We favor using \texttt{BayeSN} because SN\,2024gy is located near the center of NGC~4216 and it explicitly models host-galaxy dust extinction separately from intrinsic SN spectral energy distribution (SED) variations, unlike alternatives like SALT2 \citep{guy_salt2_2007}.

Using the BayeSN model trained by \cite{ward_bayesn_2023}, fitting the light curve of SN\,2024gy yields a distance modulus to NGC~4216 of $\mu = 31.18 \pm 0.11$ mag, which corresponds to a distance of $17.2 \pm 0.9$ Mpc that we adopt for our analysis of SN\,2024gy. This distance agrees within the uncertainties with measurements from \cite{Li2026}. The \texttt{BayeSN} fit to SN\,2024gy results in a host-galaxy extinction of $A_V = 0.72\pm 0.06$~mag (with $R_V = 2.55 \pm 0.10$) and a time of $B$-band maximum light of MJD $= 60329.19\pm 0.05$ corresponding to 20 January 2024 (\autoref{sec:bayesn}).

We similarly fit SN~2022aaiq with BayeSN, using LCO and Zwicky Transient Facility (ZTF) photometry, the latter obtained via the ZTF forced-photometry service \citep{masci_ztf_2019,masci_ztf_2023}. Given sparser data than for SN~2024gy, we fix $R_V$ to the BayeSN training sample mean value of 2.7 and find $A_V$ to be consistent with 0. We recover a distance modulus of $\mu = 32.76 \pm 0.10$ mag, corresponding to a distance of $35.6 \pm 1.7$ Mpc. This distance is longer than other redshift-independent distances reported in the NASA Extragalactic Database but is within $2\sigma$ agreement to the recent surface-brightness-fluctuation-based distance from \citet[][$\mu = 32.27 \pm0.14$ mag]{Tully2013}.\footnote{\url{https://ned.ipac.caltech.edu/byname?objname=NGC+5631&hconst=67.8&omegam=0.308&omegav=0.692&wmap=4&corr_z=1}} The time of maximum $B$-band light is MJD $= 59914\pm 0.09$, corresponding to 1 December 2022. We present the BayeSN fit alongside that of SN~2024gy in \autoref{sec:bayesn}.

\begin{figure*}
    \centering
    \includegraphics[width=\textwidth]{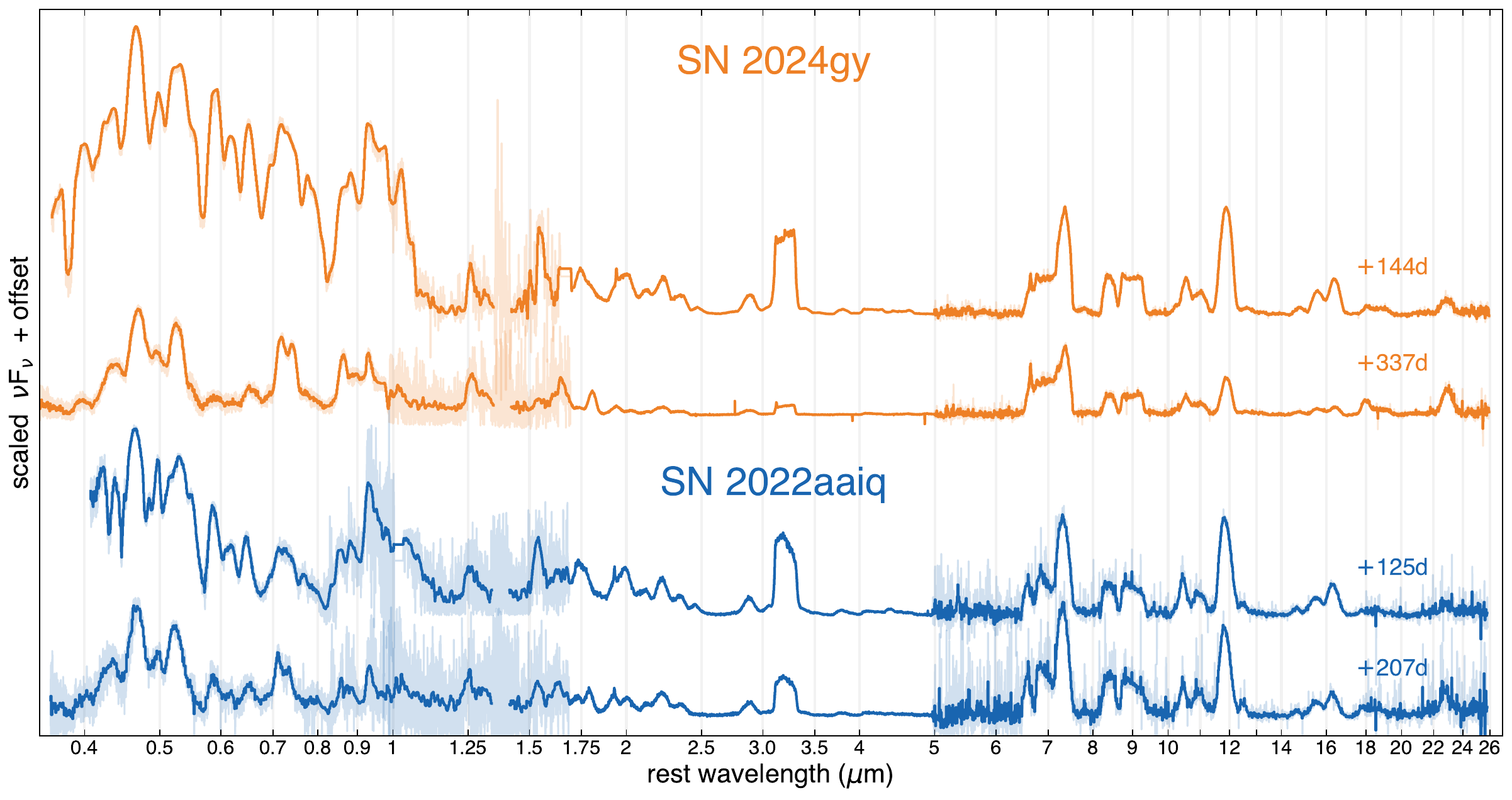}
    \caption{Panchromatic optical $+$ NIR $+$ MIR spectra of the normal SN Ia 2024gy (orange), at $+$144~d and $+$337~d post-maximum, and SN 2022aaiq (blue) at $+$125~d and $+$207~d post-maximum. The unbinned data are shown with low opacity. The spectra are scaled and offset, and the ordinate  is given in $\nu$F$\nu$ using an arcsinh scaling, for display purposes.}
    \label{fig:spec_opt+NIR+MIR}
\end{figure*}

\subsubsection{Optical Spectroscopy \label{sec:opt_spec}}

We obtained contemporaneous optical spectra at similar phases to the \textit{JWST} observations of SN\,2022aaiq and SN\,2024gy. Spectra of SN\,2024gy were taken with the Low Resolution Imaging Spectrometer \citep[LRIS, program PIs: A.~V.~Filippenko, A.~A.~Miller;][]{Oke95} on the Keck~I 10\,m telescope on 06~June~2024, 13~June~2024, 28~June~2024, and 01~January~2025. We combine the first two LRIS observations to complement the first \textit{JWST} epoch, and pair the last LRIS observation with the second \textit{JWST} epoch (\autoref{fig:spec_opt+NIR+MIR}). We analyze the 28~June~2024 LRIS spectrum, which has the highest S/N, in more detail in \autoref{sec:opt_ni}. This spectrum is reduced with the \texttt{PypeIt} package \citep{pypeit:joss_pub}, and we use \texttt{hostsub\_gp}\footnote{\url{https://github.com/slowdivePTG/HostSub_GP}} to remove the host-galaxy contamination \citep{Liu_hostsub_2025}.

Contemporaneous optical spectra for both \textit{JWST} observations of SN\,2022aaiq were taken with the Low Resolution Spectrograph 2 \citep[LRS2, program PI: J. C. Wheeler][]{chonis_lrs2:_2014} mounted on the Hobby-Eberly Telescope (HET) \citep{1998SPIE.3352...34R} located at McDonald Observatory on 31~March~2023 and 25~June~2023 (\autoref{fig:spec_opt+NIR+MIR}). The LRS2 Integral Field Unit (IFU) data were reduced by the \texttt{Panacea}\footnote{\url{https://github.com/grzeimann/Panacea}} pipeline. For each wavelength slice in the three-dimensional (3D) data cube, we mask out the SN position with a 3\arcsec\ diameter circular aperture and estimate the underlying galaxy flux by linearly interpolating from the surrounding pixels. This process is repeated independently for each slice. The resulting ``host-only'' cube is subtracted from the original data cube before extracting the 1D SN spectrum.

A contemporaneous optical spectrum of SN\,2022aaiq was also taken with the Echellette Spectrograph and Imager \citep[ESI; program ID: U177, program PI: R.~J.~Foley][]{Sheinis+2002} on the Keck~II 10\,m telescope on 27~March~2023. Observations were taken with the 0.75\arcsec slit, aligned to the parallactic angle \citep{Filippenko1982}. Spectra were reduced in a standard manner using the tools in our custom spectroscopy pipeline\footnote{\url{https://github.com/msiebert1/UCSC_spectral_pipeline/tree/esi}} \citep{Siebert2019}. All 10 orders were extracted, wavelength calibrated, flux calibrated, and telluric corrected, and then combined into a single continuous spectrum. This spectrum was combined with the optical HET spectrum and the first JWST epoch for SN\,2022aaiq.

We rescale the optical spectra to match photometry using a low-order spline function (``spectral mangling''). We correct for Galactic extinction using the F19 model from \citet{Fitzpatrick2019} with $R_V = 3.1$ and $E(B-V)=0.028$\,mag, and for host extinction using the parameters inferred from \texttt{BayeSN}. In \autoref{sec:opt_ni} we also utilize a Magellan/Baade optical spectrum of SN\,2022xkq from \cite{Pearson2024} taken on 5~March~2023 at a similar phase to the \textit{JWST} observation. This spectrum is also calibrated to photometry and dereddened for the extinction given by \cite{Pearson2024}.

\subsubsection{NIR spectroscopy}

Our contemporaneous ground-based NIR spectra for both \textit{JWST} epochs of SN\,2022aaiq were obtained from the Near-InfraRed Echellette Spectrometer (NIRES) on the Keck~II 10\,m telescope through the Keck Infrared Transient Survey (KITS) program \citep{Tinyanont2024}. SN\,2022aaiq was observed with NIRES on 30~March~2023 and 07~June~2023. SN\,2024gy was observed with NIRES through the KITS program on 15~December~2024, similar in phase to the second \textit{JWST} epoch. All NIRES spectra were reduced using procedures outlined by \cite{Tinyanont2024}. 

A NIR spectrum contemporaneous with the first \textit{JWST} epoch for SN\,2024gy was obtained on 05~June~2024 with the Folded-port InfraRed Echellette Spectrograph \citep[FIRE, program PI: A. Polin;][]{Simcoe2013} mounted on the 6.5\,m Magellan-Baade Telescope at Las Campanas Observatory in Chile. The FIRE spectrum was reduced using the IDL pipeline {\tt firehose} \citep{Simcoe2013}.

\begin{figure*}
    \centering
    \includegraphics[width=\textwidth]{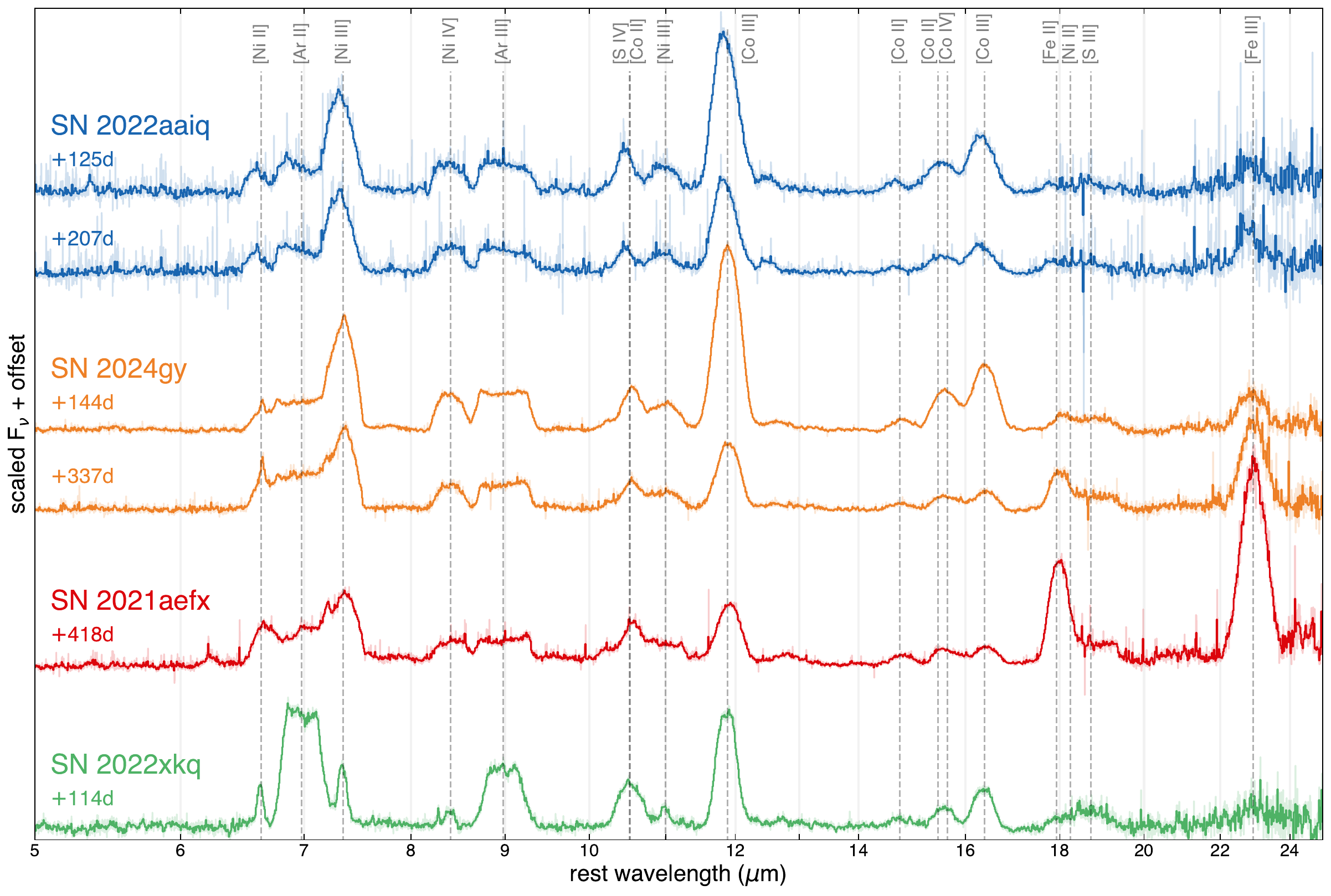}
    \caption{Comparison and identifications of prominent lines for \textit{JWST}/MIRI MRS spectra of normal SNe\,Ia 2022aaiq (blue), 2024gy (orange), and 2021aefx (red), and SN\,1991bg-like SN\,2022xkq (green). A narrow component of [\ion{Ni}{2}]~6.64\um\ is detected in SN\,2024gy. Low opacities show the unbinned data. Owing to differences in phase and distance, we scale the spectra and offset for display purposes. A linear (not arcsinh) flux scaling is used.}
    \label{fig:MRS_IDs}
\end{figure*}

\begin{figure*}
    \centering
    \includegraphics[width=\textwidth]{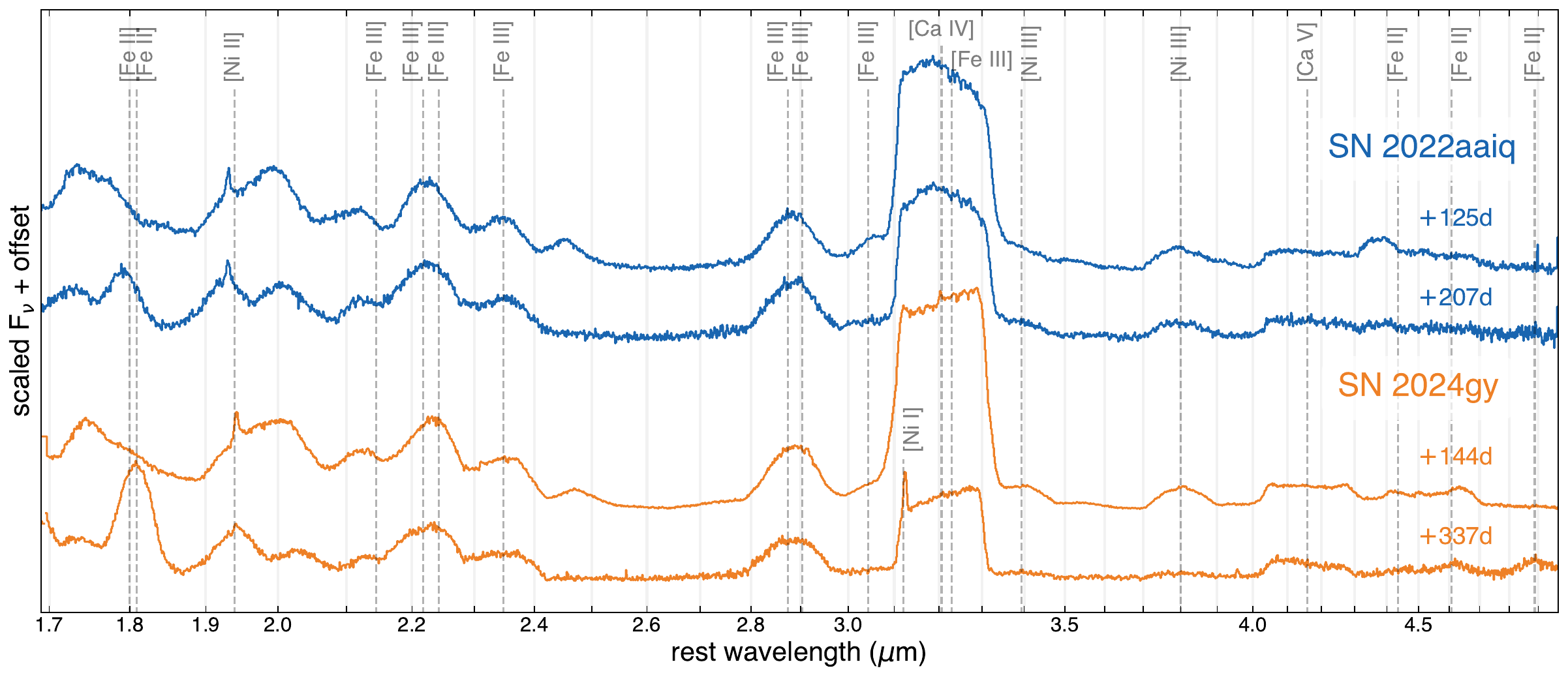}
    \caption{Comparison and identifications of prominent lines for the \textit{JWST}/NIRSpec G235M$+$G395M spectra of normal SN\,Ia 2022aaiq (blue), and 2024gy (orange). Narrow features from [\ion{Ni}{2}]~1.94\um\ are detected in both epochs for SN\,2022aaiq and SN\,2024gy. A narrow [\ion{Ni}{1}]~3.12\um\ spike is also detected in SN\,2024gy at $+$337~days. Owing to differences in phase and distance, we scale the spectra and offset for display purposes. An arcsinh flux scaling is also applied to more clearly show weaker lines.}
    \label{fig:NIR_IDs}
\end{figure*}

\section{Spectral Analysis \label{sec:spec_analysis}}

\subsection{Line Identifications}

We identify lines using the predictions of \citet{Blondin2023}. Most prominent MIR forbidden lines arise from ground-state or low-lying transitions. The four SN studied here (SN\,2022aaiq, 2024gy, 2021aefx, 2022xkq) all show the same dominant features, though relative strengths vary with SN and phase (see \autoref{fig:MRS_IDs}). Many lines have been previously reported \citep{Kwok2023, DerKacy2023, Blondin2023, DerKacy2024, Ashall2024}; here we highlight those clarified or newly detected in our data. Notably, the \textit{JWST} spectra of SN\,2022aaiq and SN\,2024gy reveal narrow components in several Ni lines.

\subsubsection{Narrow Stable Nickel}
\label{sec:narrow_Ni_ids}

Our medium-resolution \textit{JWST} spectra of SN\,2022aaiq and SN\,2024gy show a narrow [\ion{Ni}{2}]~1.94\um\ line rising above the surrounding broad emission (\autoref{fig:NIR_IDs} and \autoref{fig:Ni2_fits}). This feature, difficult to isolate from the ground owing to telluric absorption and line overlap, is cleanly detected and resolved here with high S/N, confirming the results of \citet{Kumar2025}. Its velocity offset is consistent with all other Ni lines across the NIR and MIR (\autoref{sec:appendix_all_Ni}), blueshifted by $\sim$1000\kms\ in SN\,2022aaiq and redshifted by $\sim$500\kms\ in SN\,2024gy, likely reflecting viewing-angle effects.

The [\ion{Ni}{2}]~6.64\um\ line also shows a narrow central component (consistent in velocity offset and width) superimposed on a broad base (\autoref{fig:MRS_IDs}, \autoref{fig:Ar_fits}, and \autoref{fig:Ni2_fits}). This narrow component is robustly detected in both epochs for SN\,2024gy and more tentatively in SN\,2022aaiq. Following \citet{Taubenberger2009}, we refer to a narrow feature atop a broader base as a ``narrow core'' profile. The more isolated MIR line clarifies that the [\ion{Ni}{2}]~1.94\um\ feature must also comprise both broad and narrow components. As \citet{Kumar2025} note, their detection criteria, which favor narrow features rising above broader emission, likely underestimate the incidence of [\ion{Ni}{2}]~1.94\um\ in their sample; future analyses should account for both components.

We also detect [\ion{Ni}{1}]~3.12\um\ for the first time in a normal SN\,Ia\footnote{\citet{Kwok2023} identified this line in SN\,2021aefx, but it was reclassified as [\ion{Ca}{4}]~3.21\um\ by \citet{Blondin2023}.}. It appears in the $+$337~day spectrum of SN\,2024gy as a narrow feature at the blue edge of [\ion{Ca}{4}]~3.21\um\ (\autoref{fig:NIR_IDs} and \autoref{fig:Ni1}). The wings of the [\ion{Ca}{4}] profile maintain the same shape, narrowing only marginally, from the previous epoch, indicating no accompanying broad [\ion{Ni}{1}] component at this phase. The [\ion{Ni}{1}] feature matches the offsets and widths of the narrow [\ion{Ni}{2}] components. By contrast, the [\ion{Ni}{3}]~3.80\um\ line of SN\,2024gy at $+$144~days shows a narrow-core profile with a less pronounced narrow spike atop an even broader base; this is less clear in the [\ion{Ni}{3}]~7.35\um\ line. We quantify kinematic properties in \autoref{sec:line_profiles} and interpret the central Ni emission in \autoref{sec:discussion}.

\subsubsection{SN\,2022aaiq and SN\,2024gy}

Beyond the narrow Ni features, both SN show several new or clarified lines. We detect [\ion{S}{4}]~10.51\um\ and [\ion{S}{3}]~18.71\um\ with flat-topped profiles similar to Ar (\autoref{fig:MRS_IDs}). The [\ion{S}{4}] feature produces the subtle shoulder near 10.2\um\ and contributes to blending near 11\um. In the $+$337~day spectrum of SN\,2024gy, [\ion{Fe}{2}]~17.93\um\ emerges atop the blue side of [\ion{S}{3}]~18.71\um, while [\ion{Fe}{3}]~22.92\um\ grows in strength in both SN. The predicted 14--17\um\ Co complex (comprised of [\ion{Co}{2}]~14.74\um, [\ion{Co}{2}]~15.46\um, [\ion{Co}{4}]~15.64\um, and [\ion{Co}{3}]~16.39\um; \citealt{Blondin2023}) is present in both objects.

Our NIRSpec observations also reveal weak lines in the range 3--5\um\ (\autoref{fig:NIR_IDs}). [\ion{Ni}{3}]~3.39 and 3.80\um\ are detected in both SN; at early phases, when [\ion{Ni}{3}]~3.80\um\ is measured with good S/N, it provides a relatively clean tracer of [\ion{Ni}{3}]. Several [\ion{Fe}{2}] lines appear near 4.5\um. We further identify a flat-topped feature near 4.15\um\ as [\ion{Ca}{5}]~4.16\um\ using the Atomic Line List\footnote{Atomic Line List: \url{https://linelist.pa.uky.edu/newpage/}} \citep{vanHoof2018}. Although highly ionized, this ground-state transition shows morphology consistent with that of other IME lines. [\ion{Ca}{5}] was not included in the CMFGEN models of \citet{Blondin2023}.

\subsubsection{SN\,2022xkq \label{sec:obs_22xkq}}

As background-subtraction techniques for MRS data have improved, we reassess several identifications for SN\,2022xkq from \citet{DerKacy2023} using our rereduction. We now detect weak [\ion{Ni}{4}]~8.40\um\ and [\ion{Ni}{3}]~11.00\um\ lines. The [\ion{Ni}{4}]~8.40\um\ profile shows an extremely narrow feature on its blue side (see \autoref{sec:appendix_all_Ni}). Lacking a convincing separate identification, we suggest it may be part of the [\ion{Ni}{4}] emission, potentially from a higher-velocity clump or nugget. We also detect the full Co complex between 14--17\um.

\begin{figure*}
    \centering
    \includegraphics[width=\textwidth]{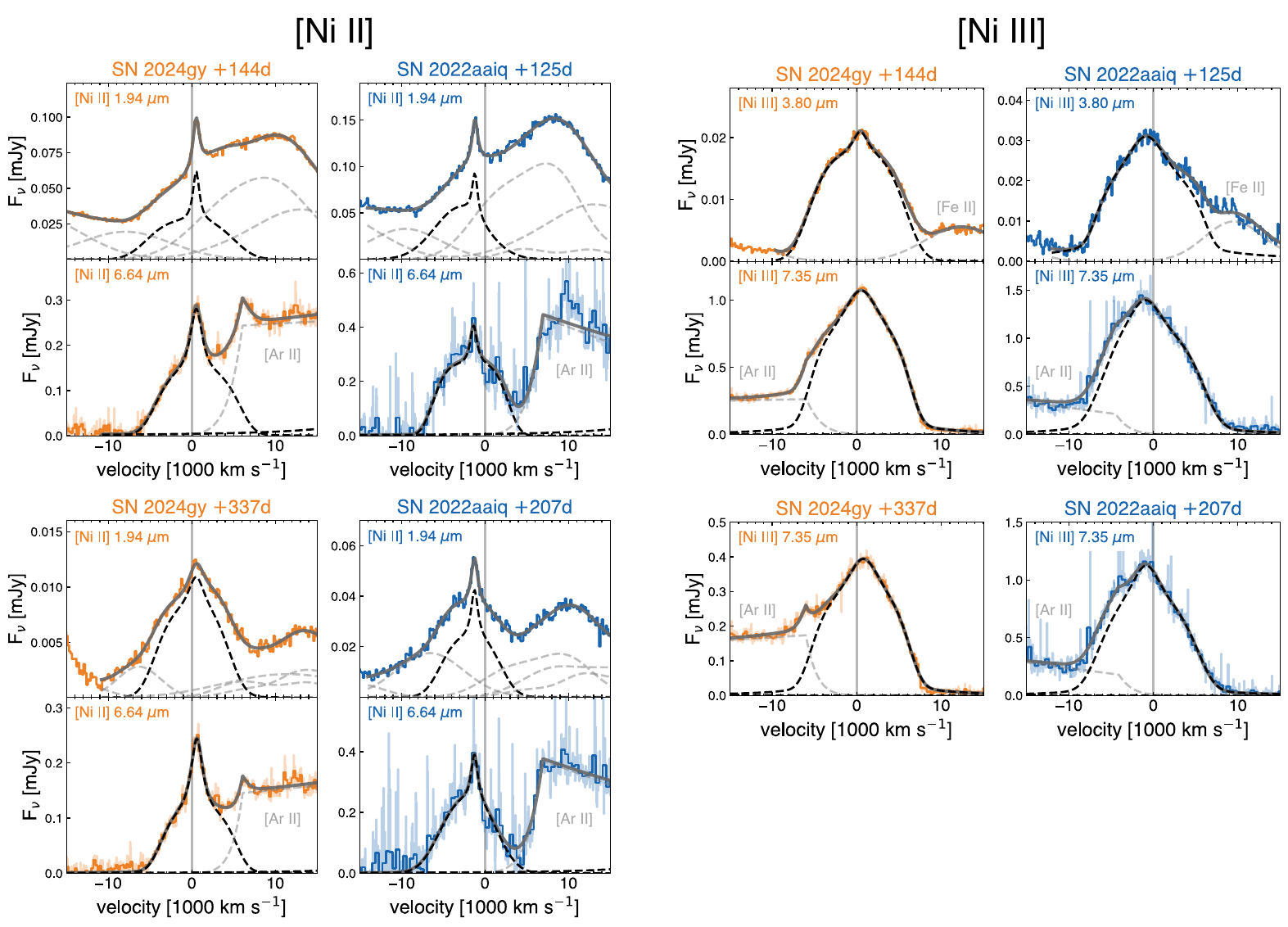}
    \caption{Line profile fits for [\ion{Ni}{2}]~1.94 and 6.64\um\ (\textit{left}) and [\ion{Ni}{3}]~3.80 and 7.35\um\ (\textit{right}) in SN\,2024gy (orange) at $+$144 and $+$337~days, and SN\,2022aaiq (blue) at $+$125 and $+$207~days. The contribution from Ni is shown in dashed black, and contributions from other nearby lines are shown in dashed gray. The composite fit is shown in solid gray.}
    \label{fig:Ni2_fits}
\end{figure*}

Our rereduction further shows that [\ion{Ar}{3}]~8.98\um\ lacks the strong blue peak previously attributed to [\ion{Ni}{4}]~8.95\um, [\ion{Ti}{2}]~8.92\um, or [\ion{Ti}{2}]~9.19\um\ \citep{DerKacy2024}. We do not find clear evidence for these lines. Given the weakness of [\ion{Ni}{4}]~8.40\um\ (expected to be the strongest [\ion{Ni}{4}] transition) and the absence of other Ti features, we attribute the 9\um\ feature almost entirely to [\ion{Ar}{3}]. Fluctuations atop the Ar lines had been suggested to arise from weak lines, including neutral ions \citep{DerKacy2023}; however, we do not detect other expected neutral transitions (e.g., [\ion{Ni}{1}]~7.51\um, which would be isolated at the velocities of SN\,2022xkq), and the phase ($+$114~days) is likely too early for such low ionization (e.g., [\ion{Ni}{1}] appears in SN\,2024gy at $+$337~days but not in SN\,2022aaiq at $+$207~days). We therefore favor ejecta inhomogeneities and asymmetries as the source of these variations \citep[e.g.,][]{Pollin2025, Simotas2025}.

Finally, the rereduction reduces the noise in the continuum between prominent lines. \citet{DerKacy2024} discuss a pseudocontinuum from a lingering photosphere; however, we find that the continuum in SN\,2022xkq is not more pronounced than in the later-epoch observations of the other SN.

\begin{figure*}
    \centering
    \includegraphics[width=\textwidth]{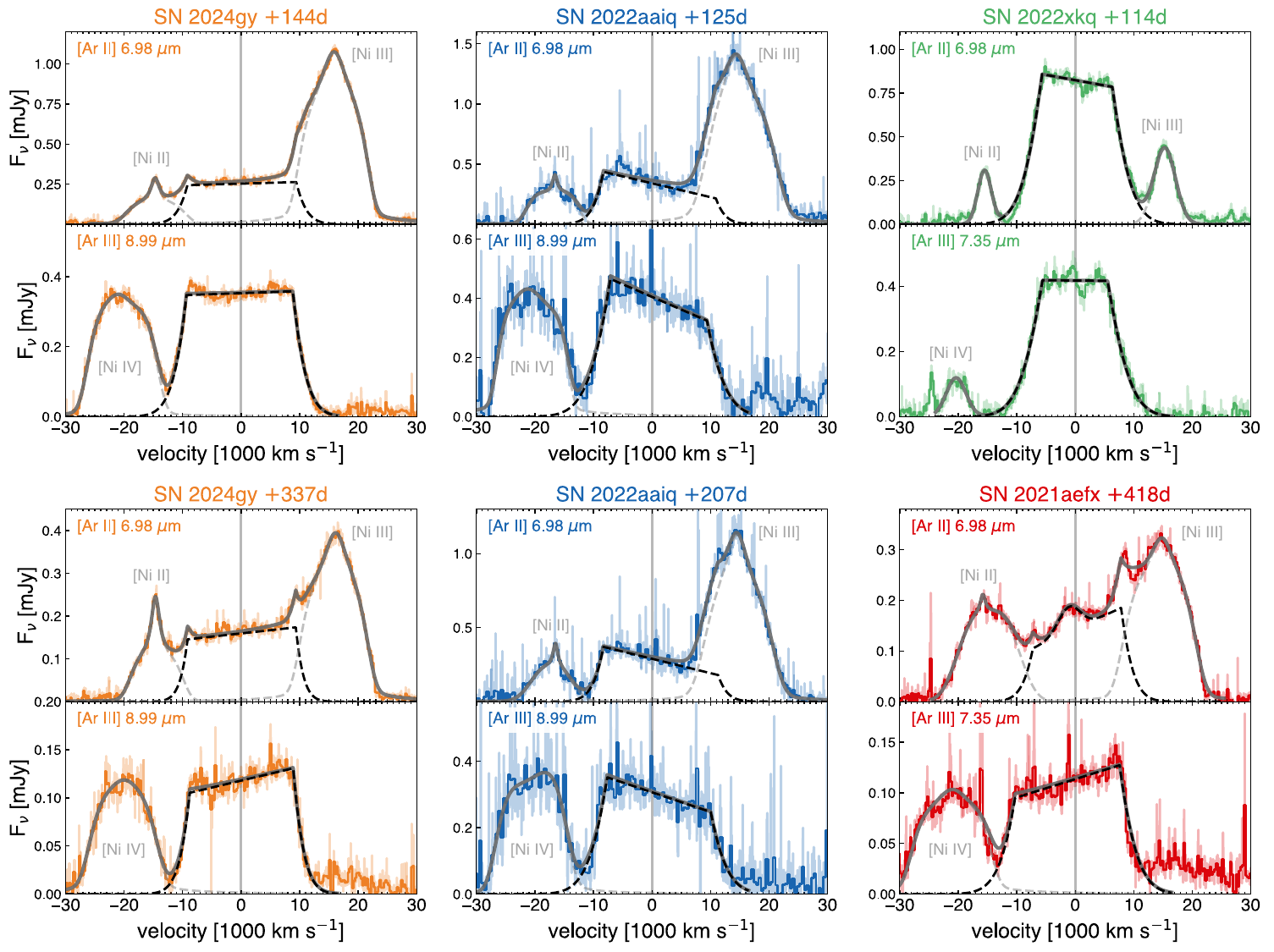}
    \caption{Line profile fits for [\ion{Ni}{2}]\,6.98\um, [\ion{Ar}{2}]\,6.98\um, [\ion{Ni}{3}]\,7.35\um\ and [\ion{Ni}{4}]\,8.40\um, [\ion{Ar}{3}]~8.99\um\ in SN\,2024gy (orange) at $+$144 and $+$337~days, SN\,2022aaiq (blue) at $+$125 and $+$207~days, SN\,2022xkq (green) at $+$114~days, and SN\,2021aefx (red) at $+$418~days. The contribution from Ar is shown in dashed black, and contributions from Ni lines are shown in dashed gray. The composite fit is shown in solid gray. The [\ion{Ar}{2}] and [\ion{Ar}{3}] lines are well-fit by slanted flat-topped profiles indicating a mildly asymmetric shell. The [\ion{Ni}{2}] lines in SN\,2022aaiq and SN\,2024gy are well-fit by a narrow core atop a broad base, indicating enhanced central emission. The [\ion{Ni}{3}] lines in SN\,2022aaiq, SN\,2024gy, and SN\,2021aefx are well-fit by a broken-slope profile with a broad base that narrows toward peak.}
    \label{fig:Ar_fits}
\end{figure*}

\subsubsection{SN\,2021aefx \label{sec:obs_21aefx}}

SN\,2021aefx was observed at a later phase than the other SN. It shows strong [\ion{Fe}{2}]~17.93\um\ and [\ion{Fe}{3}]~22.92\um, consistent with the time evolution seen in SN\,2022aaiq and SN\,2024gy. Our rereduction more clearly reveals flat-topped [\ion{S}{3}]~18.71\um\ beneath [\ion{Fe}{2}]~17.93\um, as well as the 14--17\um\ Co complex. Near 7\um\ there is an excess above expectations for a flat-topped [\ion{Ar}{2}]~6.98\um, given the [\ion{Ar}{3}]~8.99\um\ line profile (\autoref{fig:Ar_fits}). This bump has been present since $+$255~days \citep{Kwok2023, DerKacy2023, Ashall2024}, but earlier epochs had lower resolution. The excess is not consistent with [\ion{Ni}{2}]~6.92\um, a weak nearby line. We suggest this excess may be additional central [\ion{Ar}{2}] emission, perhaps due to mixing.\\

\subsection{Comparisons}

The spectra of SN\,2022aaiq, SN\,2024gy, and SN\,2021aefx are broadly similar despite their phase differences. They share the same dominant lines with comparable morphologies: IME features are flat-topped, while IGE features are centrally peaked. Yet within this similarity, variations emerge that are physically meaningful -- for example, tilted IME profiles and skewed [\ion{Co}{3}]~11.88\um\ indicate mild asymmetries, consistent with off-center explosions viewed at different angles \citep{DerKacy2023}. A detailed multiobject comparison will be presented in future work; here we predominantly focus on the Ni lines.

By contrast, the subluminous, spectroscopically SN\,1991bg-like \citep{filippenko_1991bg_1992,leibundgut_1991bg_1993} SN\,2022xkq \citep{Pearson2024} shows marked differences. Its Ni lines are single-component, narrower than the broad bases of normal SN\,Ia yet broader than their narrow Ni cores. Relative to [\ion{Ni}{3}]~7.35\um\ and [\ion{Co}{3}]~11.88\um, the [\ion{Ar}{2}]~6.98\um\ and [\ion{Ar}{2}]~8.99\um\ lines are much stronger than in normal SN\,Ia. Velocity measurements and interpretation are given in \autoref{sec:vel_measurements} and \autoref{sec:discussion}.

\section{Emission-Line Profiles \label{sec:line_profiles}}

\begin{figure}
    \centering
    \includegraphics[width=0.5\textwidth]{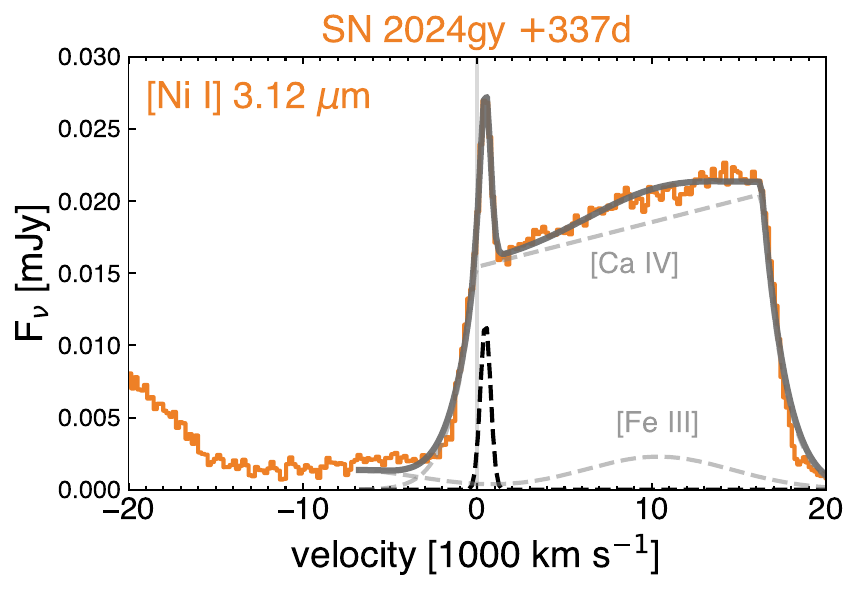}
    \caption{Line profile fit for [\ion{Ni}{1}]~3.12\um\ in SN\,2024gy (orange) at $+$337~days. The contribution from Ni is shown in dashed black, and contributions from other nearby lines are  in dashed gray. The composite fit is displayed in solid gray. The [\ion{Ni}{1}] line is well-fit by only a narrow component (without a broad base), suggesting that it is only present in the center-most regions.}
    \label{fig:Ni1}
\end{figure}

\subsection{Line-fitting Procedure}

We model the nebular spectra using the fitting approach from \citet{Kwok2023, Kwok2024, Kwok2025} in which blended spectral lines are fit simultaneously. We choose wavelength intervals for our fits that encompass all relevant overlapping transitions (e.g., 6.4--7.8\um\ for [\ion{Ni}{2}]\,6.64\um, [\ion{Ar}{2}]\,6.98\um, and [\ion{Ni}{3}]\,7.35\um\ in \autoref{fig:Ar_fits}). If multiple transitions of the same ion fall within the specified spectral window, they are required to share the same kinematic parameters (velocity offset, width, and profile shape), while their relative line strengths are fixed according to a model line list computed from the N100 DDT model \citep{Seitenzahl2013} at 165~days post-explosion (see \autoref{sec:model}). The overall normalization of each ion is free. Thus, the fits determine the velocity centroid, characteristic width, and relative flux contribution of individual lines.

Different ions are described by functional forms chosen to reproduce their observed profile shapes. Specifically, [\ion{Ar}{2}] and [\ion{Ar}{3}] are modeled with a slanted flat-top plus Gaussian wings, [\ion{Co}{2}] and [\ion{Co}{3}] with a Gaussian, and [\ion{Ni}{2}], [\ion{Ni}{3}], and [\ion{Ni}{4}] with a Lorentzian plus super-Gaussian (described below). All velocities are measured relative to the adopted heliocentric systemic velocity of the host galaxy, and \textit{JWST} transitions are fit using vacuum rest wavelengths.

\subsection{Ni and Ar in SN\,2024gy and SN\,2022aaiq}

The 6--9.5\um\ spectral window contains the dominant [\ion{Ni}{2}]~6.64\um, [\ion{Ar}{2}]~6.98\um, [\ion{Ni}{3}]~7.35\um, [\ion{Ni}{4}]~8.40\um, and [\ion{Ar}{3}]~8.99\um\ transitions. While [\ion{Ar}{2}] is blended on both sides, the blue wing of [\ion{Ni}{2}] and red wing of [\ion{Ni}{3}] in SN\,2022aaiq and SN\,2024gy are relatively clean and reveal a two-component structure: a narrow central peak superimposed on a broader base.

We therefore fit the Ni profiles with a two-component function consisting of a Lorentzian (narrow core) and a super-Gaussian (broad base), allowing independent velocity centroids and widths for each component (see \autoref{sec:appendix_fitting_22aaiq_24gy} for functional details). Alternative representations, such as narrow $+$ broad Gaussians, do not simultaneously reproduce the observed sharp narrow component and steep outer wings. Our Lorentzian $+$ super-Gaussian provides a more flexible empirical description for the line shape, without affecting centroid measurements.

We constrain [\ion{Ar}{2}]~6.98\um\ using the nearly isolated [\ion{Ar}{3}]~8.98\um\ line, adopting its best-fit kinematic parameters as priors ($\pm500$\kms). Applying the two-component Ni model described above, we then fit the full 6--9.5\um\ complex in both supernovae, shown in \autoref{fig:Ar_fits} (and see \autoref{sec:appendix_fitting_22aaiq_24gy}). We find that [\ion{Ni}{2}] is systematically narrower than [\ion{Ni}{3}] at a given phase, consistent with ionization stratification in which recombination occurs more efficiently in the denser inner ejecta. This trend is also observed in the narrow components of the NIR transitions [\ion{Ni}{2}]~1.94\um\ and [\ion{Ni}{3}]~3.80\um.

In \autoref{fig:Ni2_fits}, we use MIR-derived parameters (with $\pm500$\kms\ priors on [\ion{Ni}{2}] narrow-component parameters, and $\pm1000$\kms\ priors for others) and find that the NIR lines can be fit well with these kinematic constraints. Without MIR constraints, the narrow component in [\ion{Ni}{2}]\,1.94\um\ is still well constrained, but the broad base becomes highly sensitive to bounds and blending, showing that the NIR alone cannot robustly constrain it. Additional details and discussion of NIR line fitting is given in \autoref{sec:appendix_fitting_22aaiq_24gy}.

\begin{figure*}
    \centering
    \includegraphics[width=0.8\linewidth]{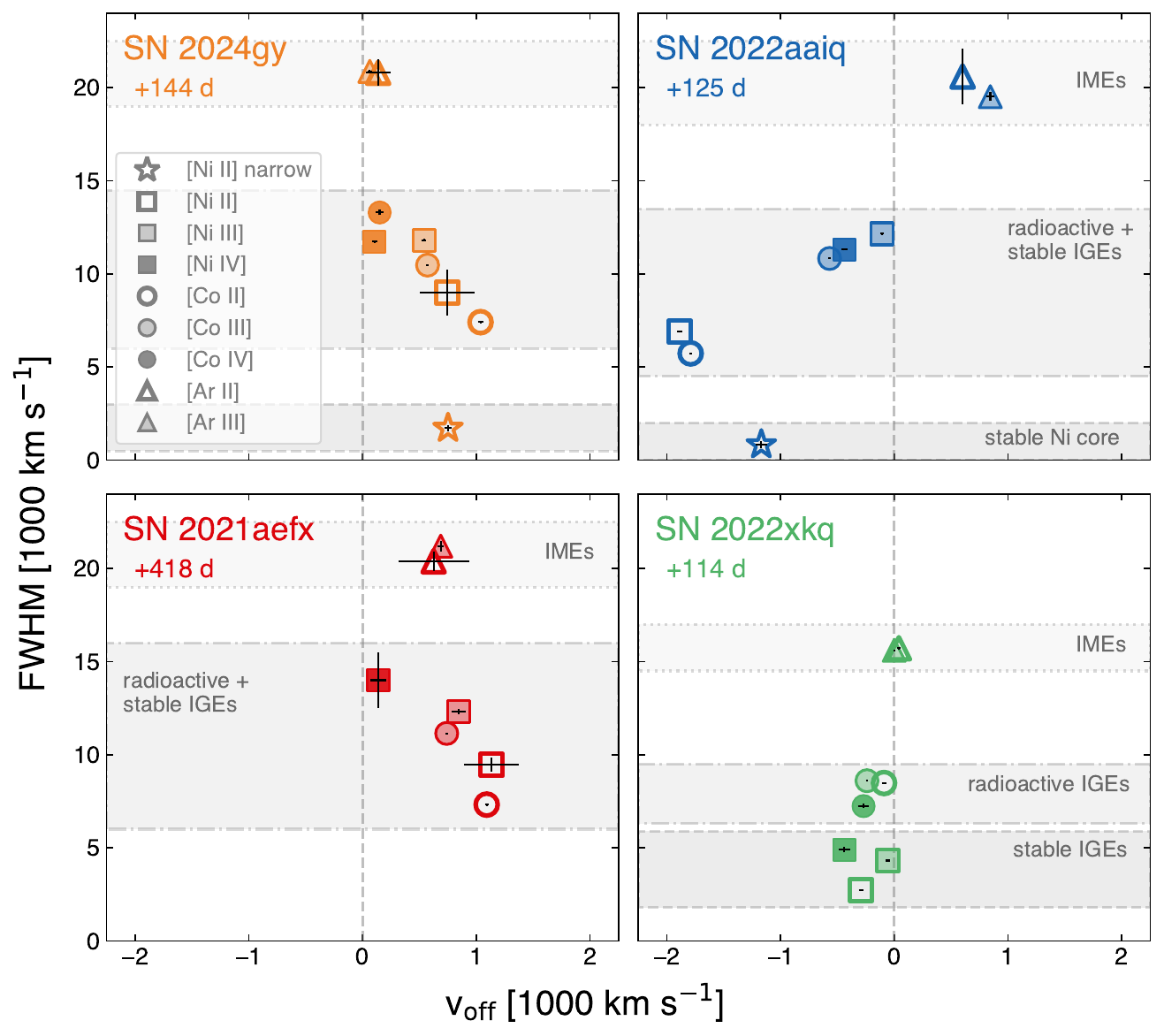}
    \caption{Kinematic offsets ($v_{\rm off}$) and FWHM measurements from fits to the emission-line profiles of SN\,2024gy (orange), SN\,2022aaiq (blue), SN\,2021aefx (red), and SN\,2022xkq (green). For multicomponent fits, the plotted FWHM corresponds to the broader component, except for the narrow [\ion{Ni}{2}] component which is marked by a star. Highly blended or poorly constrained lines (e.g., [\ion{S}{4}]) are omitted. Elements are distinguished by marker shape, while ionization stage is indicated by fill. SN\,2024gy, SN\,2022aaiq, and SN\,2021aefx exhibit clear stratification between IMEs and IGEs, while SN\,2022xkq shows additional separation between radioactive and stable IGEs. In all objects, IGEs exhibit larger velocity offsets than IMEs, consistent with off-center ignition. Values for the fit measurements can be found in \autoref{tab:fit_params}.}
    \label{fig:voff_fwhm}
\end{figure*}

\subsection{[\ion{Ni}{1}] in SN\,2024gy}

\autoref{fig:Ni1} shows our fit to the narrow [\ion{Ni}{1}]~3.12\um\ line in SN\,2024gy at $+$337~days. The line sits on the blue edge of [\ion{Ca}{4}]~3.21\um, which itself is blended on the red side with [\ion{Fe}{3}]~3.23\um. We find no evidence for a broad [\ion{Ni}{1}] component (see \autoref{sec:narrow_Ni_ids}), and therefore fit only a Lorentzian. The line is nearly identical in width and offset to the narrow [\ion{Ni}{2}] components, supporting ionization stratification: at this epoch, only the physical conditions in the core (temperature, density, nonthermal processes, etc.) permit recombination to [\ion{Ni}{1}]. Taken together, the line profiles show an ionization-dependent trend: the narrow core contribution increases toward lower ionization states.

\subsection{SN\,2021aefx and SN\,2022xkq}

At $+$418\,days, SN\,2021aefx exhibits Ni profiles broadly similar to those of SN\,2022aaiq and SN\,2024gy, though the [\ion{Ni}{2}] narrow core is only marginally detected and the two components in [\ion{Ni}{3}] are less distinct. The [\ion{Ar}{2}]~6.98\um\ line in SN\,2021aefx exhibits an additional central component not present in the other objects (\autoref{fig:Ar_fits}). The nearby [\ion{Ni}{2}]~6.92\um\ transition is too weak and velocity-inconsistent to account for this excess, leaving its origin uncertain. This feature may reflect some inward mixing of Ar into lower-velocity regions.

In contrast, SN\,2022xkq (a transitional SN\,1991bg-like event) shows [\ion{Ni}{2}]~6.64\um\ and [\ion{Ni}{3}]~7.35\um\ lines that are well described by single Gaussians with no narrow cores, and are substantially narrower than in the normal events (\autoref{fig:Ar_fits}). The Ar lines are likewise narrower but relatively stronger compared to Ni. These widths indicate that stable Ni in SN\,2022xkq is confined to the low-velocity inner ejecta and that the overall ejecta velocities are reduced relative to normal SN\,Ia.

Additional details and discussion of line fitting for SN\,2021aefx and SN\,2022xkq is given in \autoref{sec:appendix_fitting_21aefx_22xkq}.

\subsection{Kinematic Trends} \label{sec:vel_measurements}

\autoref{fig:voff_fwhm} presents our measurements of the velocity offset, $v_{\rm off}$, and full width at half-maximum intensity (FWHM) from the fitted profiles of the dominant emission lines for SN\,2024gy (orange; \textit{top left}), SN\,2022aaiq (blue; \textit{top right}), SN\,2021aefx (red; \textit{bottom left}), and SN\,2022xkq (green; \textit{bottom right}). These quantities are derived from the fitted line profiles obtained by simultaneously fitting the 6.5--17\um\ spectral range, with uncertainties estimated via a bootstrap method (\citealt{Kwok2023, Kwok2025}). Several physically meaningful trends emerge.

In each SN, the FWHM measurements reveal a clear ion-dependent velocity stratification, with the IMEs (Ar and S) exhibiting significantly broader profiles than the IGEs. SN\,2024gy, SN\,2022aaiq, and SN\,2021aefx share a similar structure in which the IMEs extend to $\sim$20,000\kms, while the IGEs (Ni and Co) cluster around 8000--13,000\kms. In contrast, SN\,2022xkq shows a more pronounced separation between nucleosynthetic layers: the IMEs extend to $\sim$16,000\kms, the radioactive IGEs (Co) to $\sim$8000\kms, and the stable IGEs (Ni) to only $\sim$4000\kms. This suggests that in the more luminous explosions the radioactive and stable IGEs are broadly cospatial, whereas in the subluminous SN\,2022xkq the nucleosynthetic layers remain more distinctly separated.

Notably, in SN\,2024gy and SN\,2022aaiq the narrow core component of [\ion{Ni}{2}] is even narrower than the Ni emission observed in SN\,2022xkq. This indicates that although stable Ni and radioactive Co extend to similar velocities overall, a portion of the stable Ni remains highly concentrated in the innermost ejecta while Co is more broadly distributed.

The $v_{\rm off}$ measurements show that the IGEs tend to exhibit larger velocity offsets than the IMEs. In SN\,2024gy and SN\,2022aaiq, the lowest ionization states ([\ion{Ni}{2}] and [\ion{Co}{2}]) have the largest offsets, with the narrow [\ion{Ni}{2}] core displaying a similar shift to their broader counterparts. This behavior is consistent with expectations for a modestly off-center ignition ($\sim500$--1500\kms). The magnitude and sign of the offsets (redshift versus blueshift) vary between the SN, consistent with viewing-angle effects. Among the objects studied here, SN\,2022aaiq exhibits the strongest trend in $v_{\rm off}$; its Ar profiles are also the most strongly tilted and are slanted opposite to those of SN\,2024gy and SN\,2021aefx, with the velocity offsets showing a similar reversal.

These results support the findings of \citet{Maeda2010a} and \citet{Maguire2018}, and provide independent MIR evidence for spatially distinct emitting regions of nucleosynthetic products, ionization-dependent stratification, and velocity offsets driven by off-center ignition and viewing-angle effects.

\begin{figure*}
    \centering
    \includegraphics[width=0.8\textwidth]{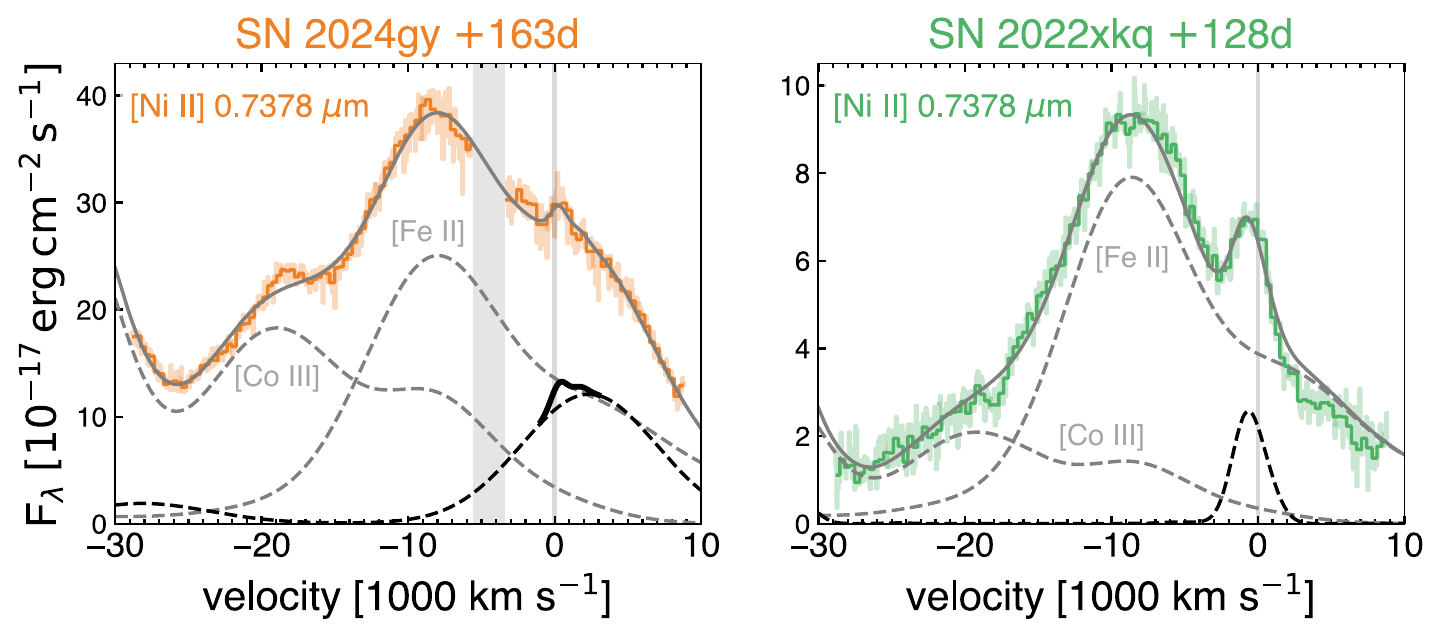}
    \caption{Line profile fit for [\ion{Ni}{2}]~0.7378\um\ in SN\,2024gy (orange) at $+$163~days (\textit{left}) and SN\,2022xkq at $+$128~days (\textit{right}). The contribution from Ni is shown in dashed black, and contributions from other nearby lines are  in dashed gray. The composite fit is displayed in solid gray. \textit{Left}: The solid thick black line shows the narrow component of the [\ion{Ni}{2}] emission that we detect in our Keck/LRIS spectrum of SN\,2024gy.}
    \label{fig:optical_Ni}
\end{figure*}

\subsection{Optical [\ion{Ni}{2}] Fits}
\label{sec:opt_ni}

The most prominent Ni feature in the optical is [\ion{Ni}{2}]~$\lambda$7378, which is heavily blended with lines from [\ion{Fe}{2}] and [\ion{Co}{3}]. Estimating its flux therefore requires joint modeling of the full 7300\,\AA\ complex, accounting for contributions from multiple species. In our fitting, we do not find any need for contributing emission from [\ion{Ca}{2}]. The optical data were taken from the ground, so we analyze this region with respect to the air wavelengths of the transitions. We model this region using the Bayesian fitting code \texttt{sn\_line\_velocities}\footnote{\url{https://github.com/slowdivePTG/sn_line_velocities}} \citep{Liu_20jgb_2023}, incorporating priors on line velocities and widths informed by the isolated NIR and MIR lines, but not fixing them. Specifically, the [\ion{Ni}{2}]~6.64~\um\ and [\ion{Co}{3}]~11.88~\um\ lines provide central velocity ($v_\mu$) and FWHM constraints for those ions, while [\ion{Fe}{2}] parameters are averaged over multiple NIR transitions. We fit the 5750--7600\,\AA\ rest-frame region with single-Gaussian components for [\ion{Fe}{2}], [\ion{Ni}{2}], and [\ion{Co}{3}], fixing each ion to share the same $v_\mu$ and dispersion across its transitions. The line list and relative strengths for each species are based on late-time models of a normal SN\,Ia, SN~2015F \citep{Flors2020}. Gaussian priors of $\mathcal N(v_{\mu,\mathrm{IR}},500)$~km~s$^{-1}$ and $\mathcal N[\ln(\mathrm{FWHM}_\mathrm{IR}/2.355),0.05]$ are adopted for $v_\mu$ and $\ln v_\sigma$, respectively. We fix the continuum level to zero in the fit, as host-galaxy contamination was removed prior to fitting using \texttt{hostsub\_gp} (see \autoref{sec:opt_spec}; \citealt{Liu_hostsub_2025}).

\autoref{fig:optical_Ni} demonstrates that both SN\,2022xkq and SN\,2024gy are well described by this three-ion model. For SN\,2022xkq, small residuals near 7500\,\AA\ ($\sim$5000\kms; \autoref{fig:optical_Ni}, right panel) may reflect deviations in [\ion{Fe}{2}] ratios due to its lower ionization. The distinct secondary peak at $\sim$7400\,\AA, centered near zero velocity, is reproduced by the [\ion{Ni}{2}] multiplet with a velocity dispersion of FWHM~$=2320\pm80$\kms---significantly narrower than the [\ion{Fe}{2}] and [\ion{Co}{3}] components ($\sim$$10^4$\kms). This is consistent with the MIR spectra, where [\ion{Ni}{2}] is also much narrower (FWHM $\approx$ 3000\kms) than [\ion{Co}{3}] (FWHM $\approx$ 8000\kms). 

In SN\,2024gy, the overall 7300\,\AA\ complex is similarly well fit with [\ion{Fe}{2}], [\ion{Ni}{2}], and [\ion{Co}{3}]. A narrow [\ion{Ni}{2}] component (thick black line, \autoref{fig:optical_Ni}, left panel) is detected at a velocity offset of $\sim$$+500$\kms, consistent with that measured in the NIR and MIR [\ion{Ni}{2}] lines. We detect optical [\ion{Ni}{2}]~$\lambda$7378 emission in both SN\,2022xkq and SN\,2024gy, and find --- as expected --- that the optical and NIR [\ion{Ni}{2}] lines share all velocity characteristics in our fits. The close correspondence across wavelengths establishes that the narrow-core [\ion{Ni}{2}] emission discovered with \textit{JWST} is also present in the optical.

\begin{figure*}
    \centering
    \gridline{
    \includegraphics[width=0.48\textwidth]{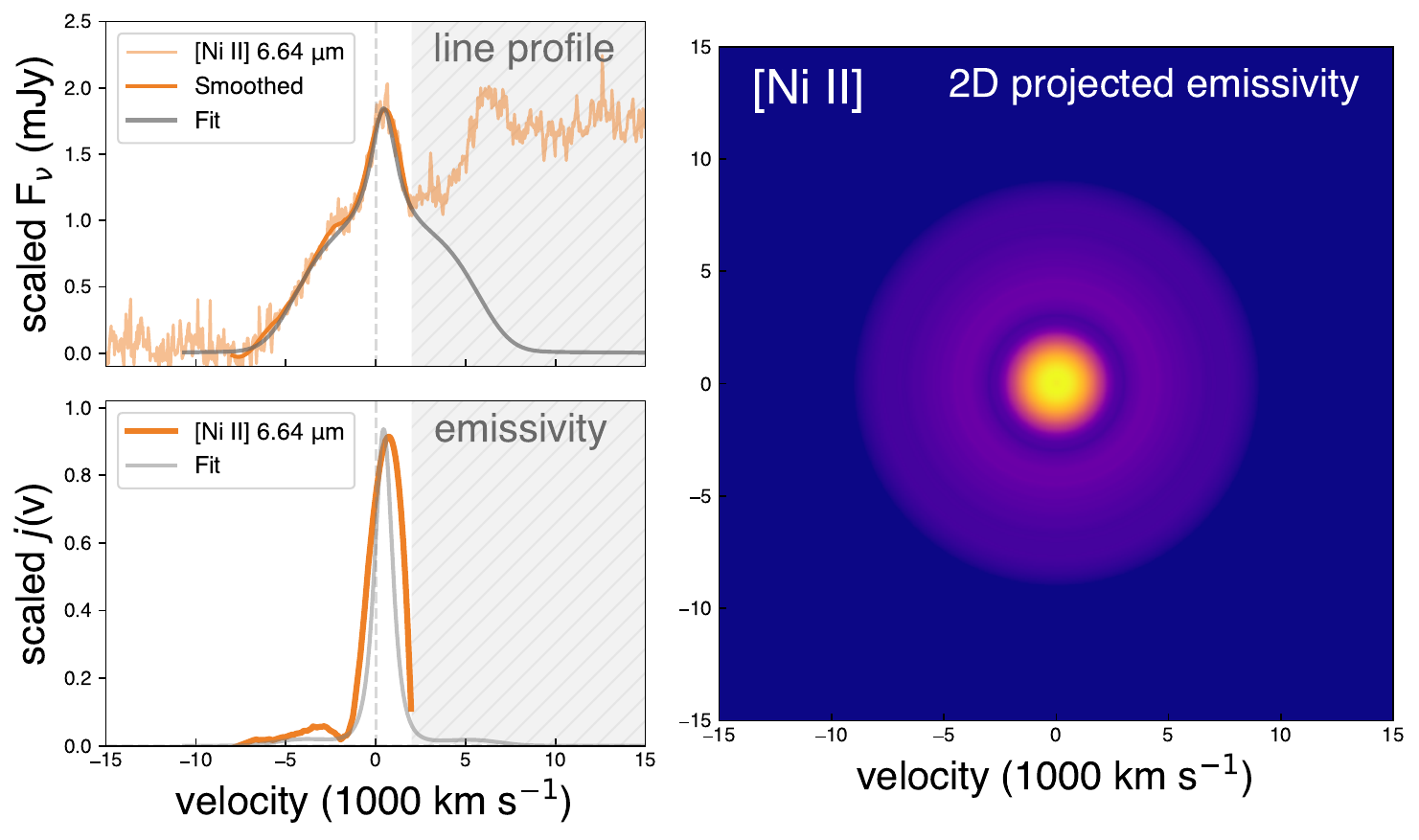}
    \includegraphics[width=0.48\textwidth]{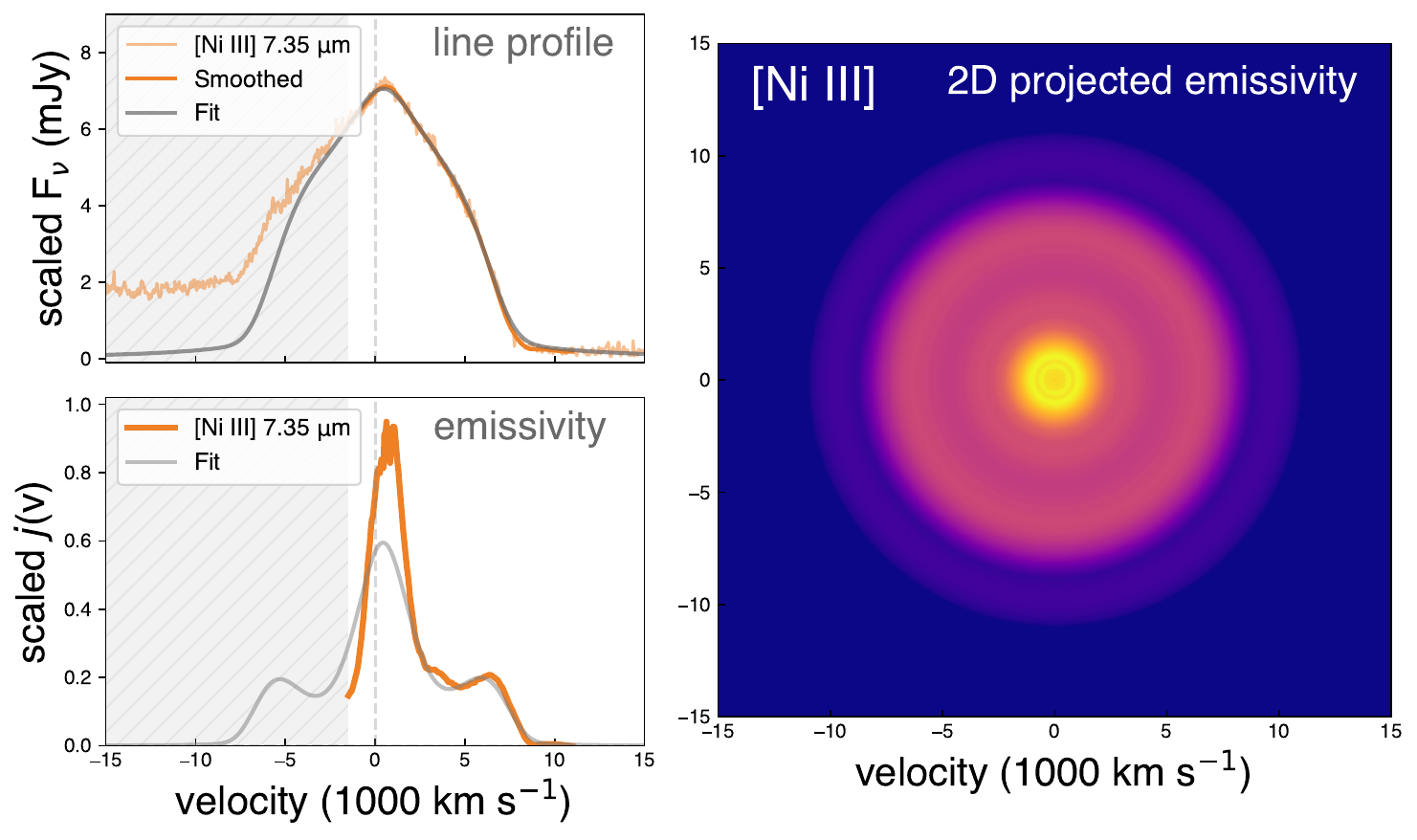}
    }
    \gridline{
    \includegraphics[width=0.48\textwidth]{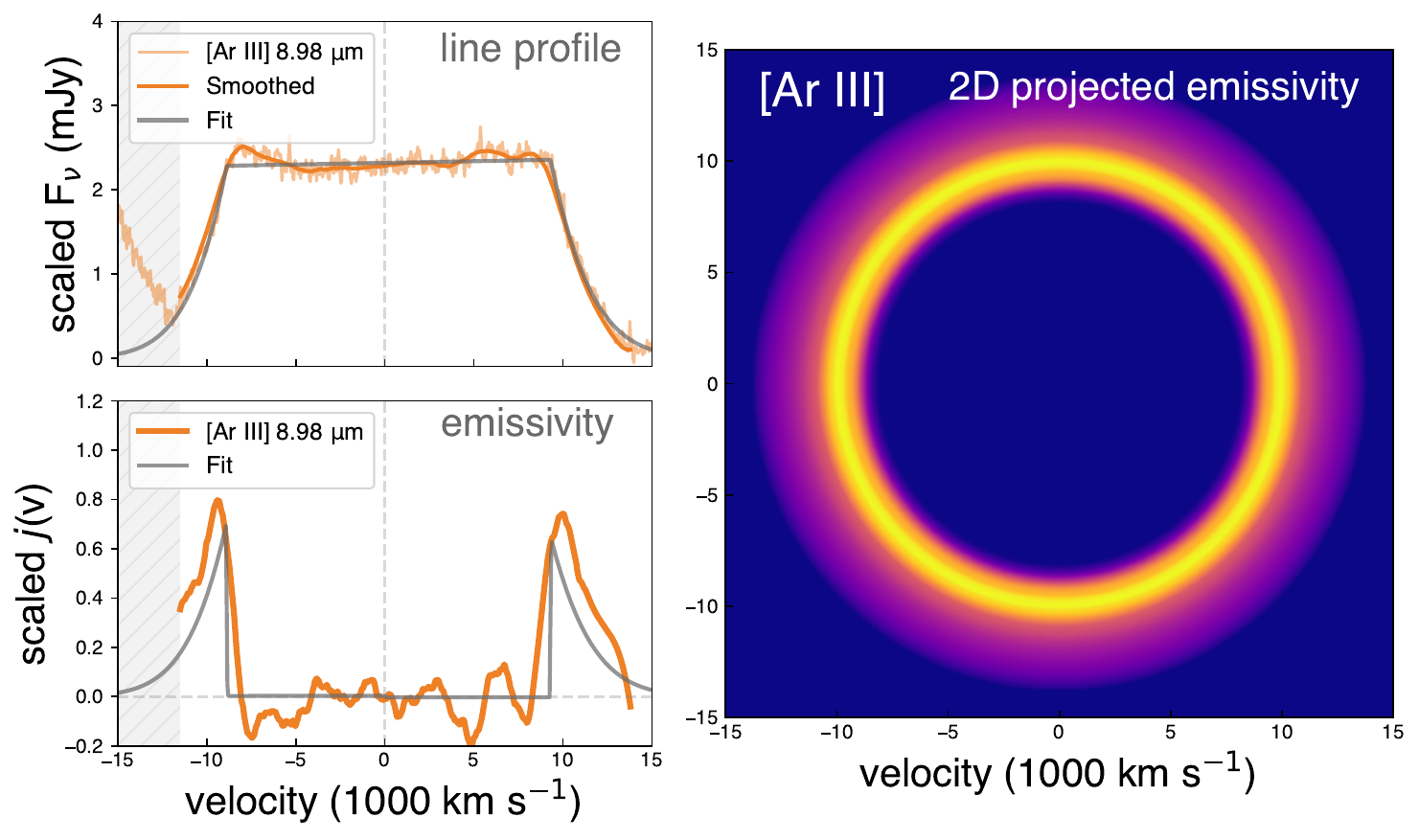}
    \includegraphics[width=0.48\textwidth]{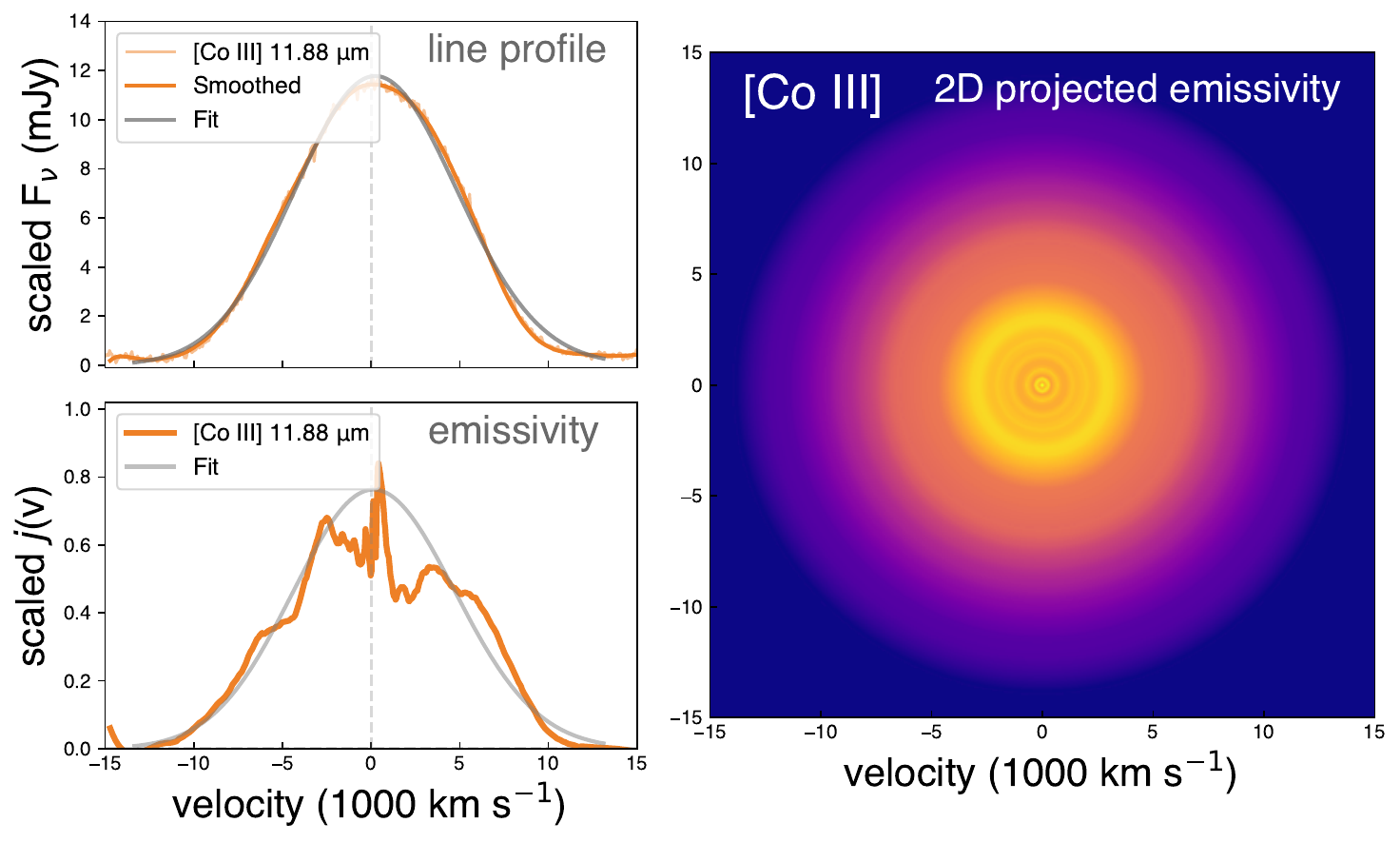}
    }
    \caption{Line profile, emissivity, and two-dimensional (2D) projected emissivity of [\ion{Ni}{2}]~6.64\um\ (\textit{upper left}), [\ion{Ni}{3}]~7.35\um\ (\textit{upper right}), [\ion{Ar}{3}]~8.99\um\ (\textit{lower left}), and [\ion{Co}{3}]~11.88\um\ (\textit{lower right}) for SN\,2024gy at $+$144 days. For each ion, the upper-left panel displays the line profile, the lower-left panel shows the emissivity $j(v)$, and the right panel indicates the emissivity projected into 2D. The data are shown in light orange, the smoothed data in dark orange, our line fits in gray, and regions contaminated by line overlap in hatched light gray. Radial symmetry is imposed on the 2D projected emissivities, which are visual aids rather than physically accurate.}
    \label{fig:emissivity_24gy}
\end{figure*}

\section{Line Profile Inversions \label{sec:line_inversions}}

Nebular line profiles directly trace the emissivity of the ejecta along the line of sight. In principle, they can be inverted to recover the emissivity distribution $j(v)$, and ultimately the density structure $\rho(v)$. In practice, this is difficult because (i) lines may be blended, (ii) deprojecting the line-of-sight requires symmetry assumptions, and (iii) connecting $j(v)$ to $\rho(v)$ requires knowledge of temperature $T(v)$ and electron density $n_e(v)$ \citep{jerkstrand_spectra_2017}.

\citet{Fransson1989} showed that assuming symmetry about the line-of-sight axis and homologous expansion, the emissivity as a function of velocity can be estimated from the derivative of the line profile. Rearranging Equation~8 of \citet{jerkstrand_spectra_2017} gives
\begin{equation}\label{eq:1}
    j(v) \propto \frac{1}{v}\,\frac{{\rm d}F_\nu(v)}{{\rm d}v}.
\end{equation}
We apply this method to the \textit{JWST}/MIRI data of SN\,2024gy. The high resolution and S/N of our MRS observations, together with several minimally blended MIR lines, make this approach particularly effective for probing ejecta structure. We focus on four representative ions: [\ion{Ni}{2}]~6.64\um\ (clean blue side), [\ion{Ni}{3}]~7.35\um\ (clean red side), [\ion{Ar}{3}]~8.99\um\ (minor blue overlap only), and [\ion{Co}{3}]~11.88\um\ (essentially uncontaminated). These trace stable IGEs (Ni), radioactive IGEs (Co), and IMEs (Ar). The nearly flat [\ion{Ar}{3}]~8.99\um\ profile supports spherical symmetry as a reasonable first-order assumption for this object. However, this approach relies on symmetry assumptions. In particular, Equation~\ref{eq:1} depends on the 3D velocity, whereas in practice we use the line-of-sight velocity as a proxy, introducing additional uncertainty. Fully capturing the true ejecta structure will therefore require future multidimensional modeling.

\subsection{Emissivity Profiles: SN 2024gy}

\autoref{fig:emissivity_24gy} shows the observed line profiles (upper left), derived emissivity profiles (lower left), and 2D projections (right; for visualization purposes). To mitigate sensitivity to noise, the data were gently smoothed before differentiation; results are insensitive to smoothing parameter choices. Velocities were offset by the value where the derivative crosses zero, consistent with a bulk offset of the explosion center, and the $v=0$\kms\ bin was interpolated to avoid division by zero. Regions affected by line overlap (hatched gray) are excluded from the derivative, though overplotted fits (gray) from \autoref{sec:line_profiles} illustrate approximate contributions. We show the same calculation for emissivity based on the fits and find that it reveals subtle shortcomings in our fitted shapes. The two-dimensional (2D) projected emissivity visualization panels in \autoref{fig:emissivity_24gy} are generated assuming radial symmetry and we stress that they are a visual guide more than physically accurate. 

In this work we do not attempt to derive ion abundance or density profiles from our emissivity profiles owing to the large number of model-dependent assumptions that are required. Instead, our analysis stays closer to the data and we suggest that modeling comparisons be made between the output emissivity profiles or nebular lines themselves.

\subsubsection{[\ion{Ni}{2}]: enhanced central emissivity}

The [\ion{Ni}{2}]~6.64\um\ emissivity profile (\autoref{fig:emissivity_24gy}, top left) shows strong enhancement in the inner few thousand\kms, corresponding to the narrow core. The broader base reflects a shell-like distribution at larger velocities, with a thin gap in emission between the core and outer material. We construct the radially symmetric 2D projected emissivity using only the clean blue side of the line. This morphology indicates a centrally concentrated reservoir of stable Ni.

\subsubsection{[\ion{Ni}{3}]: broken-slope structure}

The [\ion{Ni}{3}]~7.35\um\ profile (\autoref{fig:emissivity_24gy}, top right) displays a broken-slope morphology: a shallower inner slope that steepens beyond $\sim$4000\kms. The emissivity dips only slightly between the two slopes, creating a broad, double-peaked structure. Compared to [\ion{Ni}{2}], the [\ion{Ni}{3}] emission is more extended, consistent with expected ionization stratification. We construct the radially symmetric 2D projected emissivity using only the clean red side of the line.

\subsubsection{[\ion{Co}{3}]: extended radioactive IGEs}

The [\ion{Co}{3}]~11.88\um\ line (\autoref{fig:emissivity_24gy}, bottom right) is effectively isolated and shows weaker central emission than either Ni line, with slight asymmetry between blue and red wings. The profile suggests that radioactive IGEs (80--90\% of Co is radioactive at this phase; \citealt{Blondin2023}) are more evenly distributed and less centrally concentrated than stable Ni. Emission in the innermost regions is weaker than predicted by a Gaussian fit, consistent with high-density material being converted to stable Ni at the expense of radioactive Ni which rapidly decays to radioactive Co \citep{Gerardy2007}.

\subsubsection{[\ion{Ar}{3}]: shell-like IMEs}

The [\ion{Ar}{3}]~8.99\um\ line (\autoref{fig:emissivity_24gy}, bottom left) exhibits a nonuniform shell morphology: emissivity peaks near the inner-shell edge at $\sim$8000\,km\,s$^{-1}$, shows a clear shoulder, and falls steeply at the outer boundary. The nearly colocated [\ion{Ar}{3}] inner-shell edge and outer extent of Ni emission highlights the layered structure of IMEs surrounding stable IGEs.

For centrally filled emission, the inversion in \autoref{eq:1} relies on directly mapping the line-of-sight velocity to the 3D velocity. However, for flat-topped profiles indicating shell-like emission, material with $v_{\rm los}\approx0$ arises mostly from ejecta at large 3D velocities nearly perpendicular to the line of sight, rather than from the center. Applying the standard $1/v$ factor at line-of-sight velocities smaller than the inner-shell boundary therefore overestimates the emissivity at small radii. To account for this projection effect, we impose a floor on $v$ and compute
\[
j(v) \propto 
\begin{cases}
    \dfrac{1}{v_{\rm inner}} \dfrac{{\rm d} F_\nu}{{\rm d}v} & \text{if } v < v_{\rm inner} \\[12pt]
    \dfrac{1}{v} \dfrac{{\rm d} F_\nu}{{\rm d}v} & \text{if } v \geq v_{\rm inner}\, ,
\end{cases}
\] 
where $v_{\rm inner}$ is the inner velocity boundary of the shell. This approach preserves the form of the inversion while incorporating the expected shell geometry.

\begin{figure*}
    \centering
    \includegraphics[width=\linewidth]{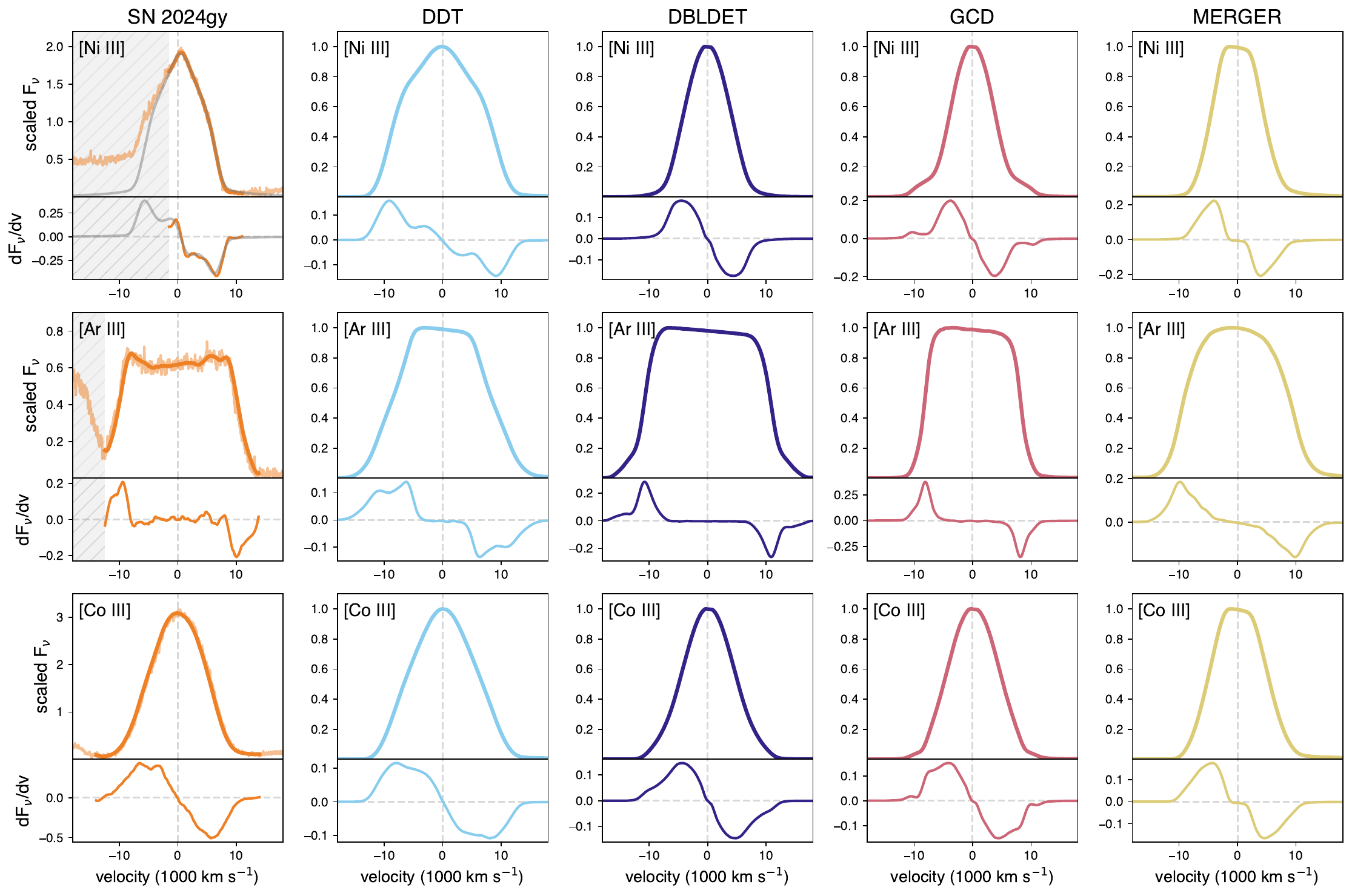}
    \caption{Line profiles (upper panels) and their derivatives (lower panels) for [\ion{Ni}{3}]~7.35\um, [\ion{Ar}{3}]~8.99\um, and [\ion{Co}{3}]~11.88\um\ for SN\,2024gy at $+$144~days (orange), and the DDT (cyan), DBLDET (green), GCD (pink), and MERGER (yellow) models from \cite{Blondin2023}. For SN\,2024gy, the data are shown in light orange, the smoothed data in dark orange, our fit to the [\ion{Ni}{3}] profile in gray, and regions contaminated by line overlap in hatched light gray. The derivative morphologies of SN\,2024gy are most similar to those of the DDT model. Owing to the 1D nature of CMFGEN, the model line profiles are symmetric about zero velocity, with a slight blueward tilt because of relativistic effects \citep{Blondin2023}.}
    \label{fig:derivatives}
\end{figure*}

\subsection{Line-Profile Derivatives}

While emissivity profiles are useful for visualization, line-profile derivatives provide an assumption-independent way to highlight subtle changes in slope and morphology. In \autoref{fig:derivatives}, we compare the line profiles and derivatives of [\ion{Ni}{3}]~7.35\um, [\ion{Ar}{3}]~8.99\um, and [\ion{Co}{3}]~11.88\um\ in SN\,2024gy to four representative explosion models from \citet{Blondin2023}: 

\begin{itemize}
    \item DDT (delayed detonation): a Chandrasekhar-mass ($1.4\,{\rm M}_\odot$) WD ignited at 100 points, based on the N100 model of \citet{Seitenzahl2013}.
    \item DBLDET (double detonation): a $1.0\,{\rm M}_\odot$ C/O WD with a $0.2\,{\rm M}_\odot$ He shell, following \citet{Gronow2021}.
    \item GCD (gravitationally confined detonation): ignition offset at 10~km in a $1.4\,{\rm M}_\odot$ WD with a central density of $10^9$ g cm$^{-3}$ \citep{Lach2022}.
    \item MERGER (violent merger): dynamic disruption of two WDs with masses $1.1\,{\rm M}_\odot$ and $0.9\,{\rm M}_\odot$ \citep{Pakmor2012}.
\end{itemize}

Each model was rerun at 165~days post-explosion to match our first epoch of SN\,2024gy, and \autoref{fig:derivatives} shows the isolated single-ion nebular line profiles. Distinct profile morphologies arise from the different ejecta compositions and structures produced by each explosion mechanism. Slight tilts in the model [\ion{Ar}{3}] profiles reflect relativistic effects \citep{Blondin2023}.

None of the models reproduces all details of the observed line widths, ionization balance, or relative line strengths. These discrepancies likely reflect limitations in current nebular radiative-transfer calculations and model assumptions (discussed further in \autoref{sec:model}). By contrast, the overall morphology of the nebular lines---set by how different elements are distributed in velocity space---is uniquely imprinted by the explosion physics. The line morphologies are one of the most discriminating signatures of the explosion mechanism.

\subsubsection{SN\,2024gy: signatures of a delayed detonation}

Against this model backdrop, we examine the line morphologies of SN\,2024gy in detail. The high S/N and resolution of our MRS data of SN\,2024gy allow us to recover subtle features in the line morphologies. Both the line profiles and their derivatives carry the same diagnostic information, but the derivatives highlight slope changes more clearly.

The [\ion{Ni}{3}]~7.35\um\ profile derivative shows a shallower slope at inner velocities that steepens by roughly a factor of two at higher velocities. This broken-slope behavior, also visible directly in the profile, is a characteristic feature of the DDT model (\autoref{fig:derivatives}, top row). The velocity of the slope break and the overall line width are somewhat smaller in SN\,2024gy than in the DDT model, but the qualitative shape, with distinct inner and outer stable Ni components, is strikingly similar.

The [\ion{Ar}{3}]~8.99\um\ profile also resembles the DDT model in overall morphology, though with notable differences: the observed Ar shell begins significantly farther out and is confined to a smaller velocity range than in the model (\autoref{fig:derivatives}, middle row). In both data and model, the profile shows a peak at the inner shell edge, followed by a weaker shoulder that drops off. The extent of the observed [\ion{Ar}{3}] shell actually more closely matches the DBLDET model, which places Ar at higher velocities, although the derivative shape of the wings differs. This suggests that some aspects of the Ar distribution may be captured better by certain double-detonation realizations.

The [\ion{Co}{3}]~11.88\um\ profile in SN\,2024gy is faintly double-peaked, with a notch on the red side and a slight inner shoulder on the blue side (\autoref{fig:derivatives}, bottom row). The DDT model predicts an intermediate derivative shape between these two observed sides. The DBLDET and GCD models also show some similarity here, indicating that the distribution of radioactive material between models may be more similar and this line less discriminatory.

Taken together, the line morphologies of SN\,2024gy---particularly the broken-slope [\ion{Ni}{3}] profile---most closely resemble those of the DDT model. However, the observed [\ion{Ar}{3}] extent and, to some degree, the [\ion{Co}{3}] profile bear similarities to the DBLDET model. Since we consider only one realization of each scenario, further exploration of a broader landscape of models may reproduce these features more fully. We therefore conclude that the overall evidence favors a DDT origin for SN\,2024gy, but recommend more detailed modeling for confirmation. In \autoref{sec:discussion} we further discuss implications for explosion mechanisms.

\begin{table*}
\centering
\caption{Measured stable Ni fluxes and derived quantities.
Luminosities are computed using the adopted distances listed below.}
\label{tab:ni_luminosities}
\begin{tabular}{lcccccc}
\hline
Supernova & Distance & Phase & $F_{\rm Ni}$ & $L_{\rm Ni}$ & $L_{\rm Ni}(+130\,{\rm d})$ \\
   & (Mpc)    & (d)   & (erg s$^{-1}$ cm$^{-2}$) & (erg s$^{-1}$) & (erg s$^{-1}$) \\
\hline
SN\,2024gy & $17.2 \pm 0.9$ 
& +144 & $(9.4 \pm 0.5)\times10^{-14}$ & $(3.3 \pm 0.4) \times10^{39}$ 
& $\sim4.1\times10^{39}$ \\

SN\,2022xkq & $31 \pm 2$ 
& +114 & $(5.8 \pm 0.9)\times10^{-15}$ & $(6.7 \pm 1.3) \times10^{38}$ 
& $\sim5.4\times10^{38}$ \\

\hline
\end{tabular}
\end{table*}

\subsubsection{SN\,2022xkq, SN\,2022aaiq, and SN\,2021aefx}
\label{sec:derivatives_others}

SN\,2022xkq has sufficiently high S/N for derivative analysis (see \autoref{sec:other_lineinversions}). Its MIR [\ion{Ni}{2}] and [\ion{Ni}{3}] profiles are single-component and significantly narrower than those of normal SNe\,Ia, though broader than the narrow [\ion{Ni}{2}] core seen in SN\,2024gy and SN\,2022aaiq. The [\ion{Ni}{3}] derivative most closely resembles the DBLDET morphology (albeit at smaller velocity width), while [\ion{Ar}{3}] shows features more consistent with the DDT, including hints of an inner peak and outer shoulder. In contrast, the [\ion{Co}{3}] profile exhibits sharper, more triangular derivatives reminiscent of MERGER models. Taken together, SN\,2022xkq shows mixed morphological signatures across Ni, Ar, and Co lines and does not appear consistent with any of the models we investigated, and is markedly different from SN\,2024gy. We therefore regard its origin as inconclusive, consistent with the conclusion of \citet{Pearson2024} that no existing explosion model fully explains its optical photometric and spectroscopic dataset.

SN\,2022aaiq and SN\,2021aefx provide additional context (see \autoref{sec:other_lineinversions} for full derivative comparisons). The lower S/N of SN\,2022aaiq limits robust conclusions, though the presence of a narrow [\ion{Ni}{2}] component similar to that of SN\,2024gy may indicate a related origin, potentially viewed from a different angle. SN\,2021aefx, observed at a later phase ($+$418 days), exhibits [\ion{Ni}{3}] and [\ion{Co}{3}] derivative structures broadly consistent with the DDT model, though with some differences in the [\ion{Ar}{3}] morphology that may reflect phase-dependent visibility of outer ejecta. This interpretation is consistent with \citet{DerKacy2023} and \citet{Ashall2024}, who also suggest a DDT origin for SN\,2021aefx.

\subsection{Stable Ni Luminosities and Mass Estimates}
\label{sec:ni_masses}

\citet{Blondin2023} showed that the total Ni luminosity (integrated over the range 0.3--30\um\ at $\sim$270~days post-explosion) correlates tightly with the stable Ni mass across a wide range of explosion models (see their Fig.~13). We use this relation to estimate the relative stable Ni production for SN\,2024gy and SN\,2022xkq.

Before proceeding, we emphasize several limitations. The DDT points shown in Figure~13 of \cite{Blondin2023} do not represent seven independent explosion models, but rather one 3D model reconstructed in 1D along multiple viewing directions. These nonspherically averaged realizations were designed to probe asymmetry effects \citep{Blondin2023} and may not represent fully self-consistent explosion models. As a result, the Ni luminosity--mass relation is model dependent, and we therefore use the roughly linear relationship only in a comparative sense. Future work should investigate this relationship and its application to observations in more systematic detail.

Using the optical, NIR, and MIR line fits from \autoref{sec:line_profiles}, we integrate the observed [\ion{Ni}{1}] (negligible contribution), [\ion{Ni}{2}], [\ion{Ni}{3}], and [\ion{Ni}{4}] emission to obtain total stable Ni fluxes. The measured fluxes and derived luminosities are given in \autoref{tab:ni_luminosities}.

For SN\,2024gy, we estimate the total Ni flux at $+$144\,days to be $(9.4\pm0.5)\times10^{-14}$, with [\ion{Ni}{2}]~7378~\AA\ and [\ion{Ni}{3}]~7.35~\um\ being the dominant contributors. For SN\,2022xkq, we estimate the total Ni flux at $+$114~days to be $(5.7\pm0.9)\times10^{-15}$~erg~s$^{-1}$~cm$^{-2}$, using the optically thin ratio between [\ion{Ni}{2}]~$\lambda$7378 and 1.94~\um\ of 5.67 to account for missing NIR coverage (other NIR Ni lines contribute $<3\%$ in SN\,2024gy, so we neglect them in SN\,2022xkq).

Using the adopted distances (\autoref{tab:ni_luminosities}, from our BayeSN fits and \citealt{Pearson2024}), we compute total Ni luminosities at the observed phases. To adjust for the phase difference, we extrapolate the luminosities to a common phase of $+$130\,days, using a late-time decline rate of 1.5~mag per 100\,days for both objects. We consider a range of plausible late-time decline rates ($\sim1.2$--1.8~mag per 100~days), informed by the objects themselves and by similar events in the literature \citep[e.g.,][]{Cappellaro1997, Lair2006, Leloudas2009, Graur2020} and find that reasonable choices within this range do not significantly alter the qualitative comparison between the two objects. The resulting Ni luminosities at $+$130\,days are $\sim4.1\times10^{39}$~erg~s$^{-1}$ for SN\,2024gy and $\sim5.4\times10^{38}$~erg~s$^{-1}$ for SN\,2022xkq.

Comparing SN\,2024gy and SN\,2022xkq at the extrapolated common phase of $+$130\,days, we find that the stable Ni emission in SN\,2024gy is 5--10 times more luminous than in SN\,2022xkq. Despite significant systematic uncertainties in the absolute values, the tight luminosity-mass correlation for stable Ni ensures a robust relative comparison: SN\,2024gy produced substantially more stable Ni than SN\,2022xkq. Combined with the low $^{56}$Ni mass ($0.22\pm0.03$~M$_\odot$) inferred by \citet{Pearson2024}, the small stable Ni yield of SN\,2022xkq disfavors a suggested high-density DDT origin \citep{DerKacy2024} and instead may favor a sub-\mch\ explosion scenario. The origin of SN\,2022xkq is further discussed in \autoref{sec:22xkq_origin}.

\begin{figure*}
    \centering
    \includegraphics[width=0.8
    \linewidth]{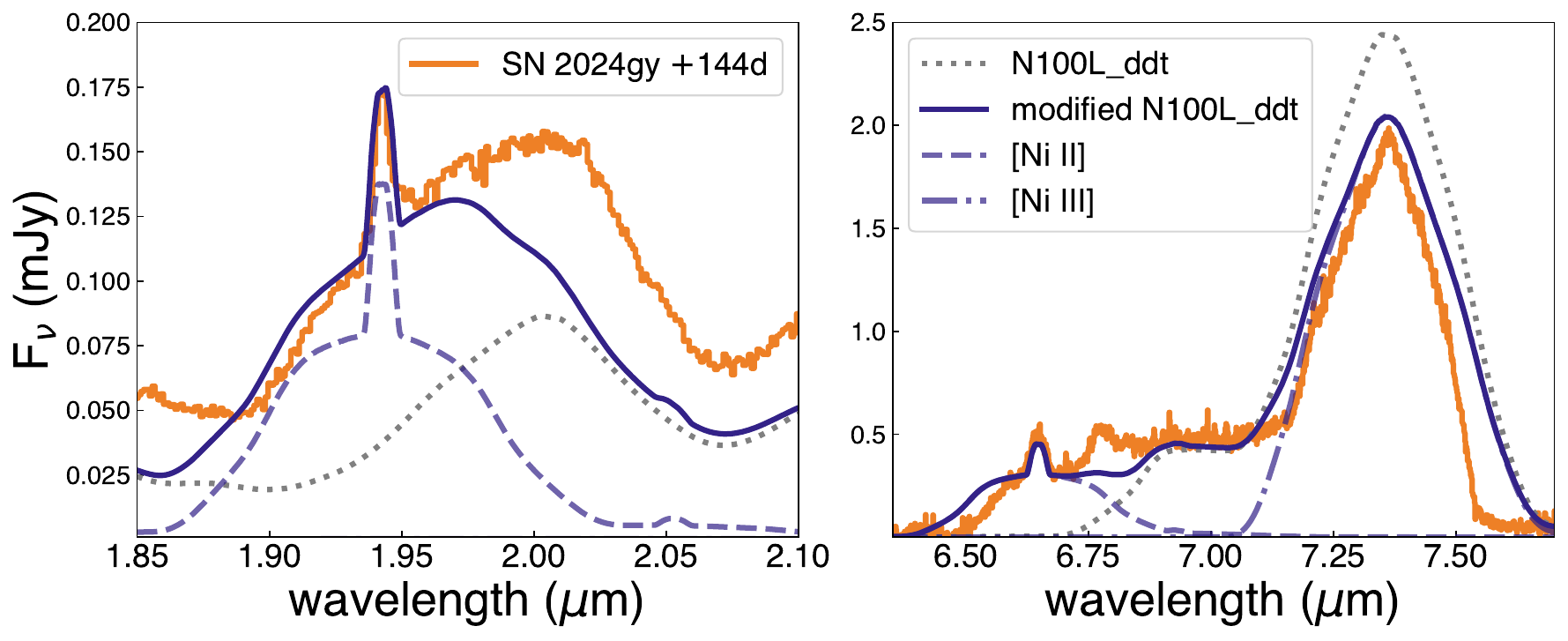}
    \caption{SN\,2024gy at $+$144~days ($\sim$165~days post-explosion; orange) compared to our radiative-transfer N100L\_ddt model (dotted gray) and our modified N100L\_ddt model (solid indigo) at 165~days post explosion. In our modified N100L\_ddt model we artificially adjust the [\ion{Ni}{2}] and [\ion{Ni}{3}] ion populations to match the observed [\ion{Ni}{2}]/[\ion{Ni}{3}] line strength ratio, and increase the abundance of $^{58}$Ni in the innermost $v < 1000$\kms\ by $10^{-3}$\,M$\odot$  to match the strength of the narrow feature in the [\ion{Ni}{2}]~1.94 and 6.64\um\ lines. The individual contributions of [\ion{Ni}{2}] and [\ion{Ni}{3}] to the composite features are shown in dashed indigo and dotted-dashed indigo, respectively. The model lines have been redshifted by 500\kms\ to match the observed offset in SN\,2024gy.}
    \label{fig:ni_models}
\end{figure*}

\section{Radiative-transfer experiments \label{sec:model}}

To explore the origin of the narrow [\ion{Ni}{2}] cores, we perform controlled radiative-transfer experiments using 1D non-LTE calculations with CMFGEN \citep{Hillier2012} applied to spherically averaged DDT models from \citet{Seitenzahl2013}, following \citet{Blondin2023} and \citet{Kwok2025}. These calculations are not intended to be fully self-consistent. Rather, they are designed to test how specific modifications to the inner ejecta structure affect the emergent nebular line profile.

Several limitations must be kept in mind. The underlying explosion models are 3D simulations that have been spherically averaged prior to 1D radiative-transfer calculations. As a result, the calculations do not capture intrinsic asymmetries present in the original 3D models. In addition, 1D nebular radiative-transfer calculations are known to exhibit systematic ionization discrepancies, often producing spectra that are more highly ionized than typically observed (e.g., \citealt{Shingles2022}; \citealt{Blondin2023}). These discrepancies primarily affect relative line strengths but have a smaller impact on velocity-space morphology, which is the primary diagnostic considered here.

Based on results from \autoref{sec:line_inversions} suggesting a DDT origin for SN\,2024gy, we perform our radiative-transfer experiments on the \texttt{N100\_ddt} model and its lower and higher central density variants (\texttt{N100L\_ddt} and \texttt{N100H\_ddt}; \citealt{Seitenzahl2013}), available on the Heidelberg Supernova Model Archive (HESMA\footnote{\url{https://hesma.h-its.org}}; \citealt{Kromer2017}). We note, however, that the qualitative result of the experiments described below---that enhancing the stable Ni abundance at the lowest velocities produces a narrow [\ion{Ni}{2}] core---is not specific to these particular DDT models. Any explosion model with a centrally concentrated stable Ni distribution, with energy deposited from radioactive decay, would be anticipated to exhibit similar qualitative morphological behavior, although the quantitative details would depend on the underlying density and abundance structure.

In the DDT models from \citet{Seitenzahl2013}, the number of ignition kernels sets the strength of the deflagration phase: fewer kernels yield less pre-expansion, higher central densities, and a higher abundance of $^{56}$Ni while stable Ni yields remain similar. These differences in burning strength produce distinct element and isotope distributions (see Fig.~8 of \citealt{Seitenzahl2013}) that imprint directly on nebular line profiles. Among them, the N100 model shows the most prominent broken-slope morphology in [\ion{Ni}{3}]~7.35\um\ (\autoref{fig:derivatives}), while the lower and higher central density variants (\texttt{N100L\_ddt} and \texttt{N100H\_ddt}) display subtler versions. The \texttt{N100L\_ddt} case produces narrower lines, closer to the [\ion{Ni}{3}] widths we observe, and retains a slope break. We therefore adopt this model for further modification, calculated at 165~days post explosion to match the first epoch of \textit{JWST} observations of SN\,2024gy.

We first test whether raising the density of the innermost ejecta ($v \lesssim 1000$\,km\,s$^{-1}$) can reproduce the narrow [\ion{Ni}{2}] cores. Scaling the central density with a Gaussian of FWHM $\approx 1000$\,km\,s$^{-1}$ to add 0.01--0.1\,M$_\odot$ enhances central emission, but in \emph{all} IGE lines. This contradicts the data: in SN\,2024gy, no comparable narrow peak is seen in [\ion{Co}{2}]~10.52\um, despite sufficient sensitivity to detect one. Pure density enhancement therefore fails to explain the observations, unless the core is composed almost entirely of stable Ni, and not radioactive Co.

We next enhance the abundance of $^{58}$Ni in the innermost layers ($v<1000$\,km\,s$^{-1}$), conserving total mass by reducing the Fe fraction. Adding $10^{-3}\,{\rm M}_\odot$ of stable Ni---a factor of $\sim$10 above the original $10^{-4}\,{\rm M}_\odot$ present---to a region of total mass 0.006\,$M_\odot$ significantly boosts the local Ni mass fraction. However, this corresponds to only a percent-level perturbation of the total stable Ni mass in the model. Variations in ignition configuration alter the central ejecta structure and stable IGE distribution in the model suite \citep{Seitenzahl2013}, indicating that central stable Ni concentration is sensitive to ignition configuration within the DDT framework. This modification produces the desired morphology: a narrow [\ion{Ni}{2}] core atop a broad base, while leaving other lines unaffected, in agreement with the observed line profiles.

As shown in \autoref{fig:ni_models}, the unmodified \texttt{N100L\_ddt} model (dotted gray) underpredicts the [\ion{Ni}{2}] flux, consistent with the ionization discrepancy noted above \citep[e.g.,][]{Shingles2022}. To isolate the morphological effect of enhanced central stable Ni, we follow \citet{Blondin2022} and adjust the Ni ionization state for comparison purposes. With this adjustment, the enhanced-Ni model reproduces both the absolute strength and the relative narrow-to-broad structure of the [\ion{Ni}{2}]~1.94 and 6.64\um\ lines.

These experiments demonstrate that introducing additional stable Ni at the lowest velocities naturally produces a narrow [\ion{Ni}{2}] core atop a broader base, while leaving other lines unaffected. The narrow [\ion{Ni}{2}] cores in SN\,2024gy and SN\,2022aaiq therefore point to an enhanced concentration of stable Ni in the innermost ejecta. Possible physical origins for such central Ni enhancement are discussed in \autoref{sec:discussion}.

\section{Discussion \label{sec:discussion}}

Our \textit{JWST} MRS observations provide new constraints on the distribution of stable IGEs, radioactive IGEs, and IMEs in SN\,Ia. The line morphologies---particularly the presence of narrow [\ion{Ni}{2}] cores and broken-slope [\ion{Ni}{3}] profiles---carry diagnostic information about the explosion physics. Below we interpret these results in the context of different explosion scenarios.

\subsection{Interpretation of Narrow Cores}
\label{sec:narrow_core_interpretation}

Narrow emission cores are common in nebular spectra of stripped-envelope SNe (SN\,Ib/c), most notably in [\ion{O}{1}] $\lambda\lambda$6300,~6364. \citet{Taubenberger2009} attribute them to several possible geometries: (a) enhanced central density in nearly spherical ejecta, (b) an equatorial torus or disk viewed face on, or (c) clumps ejected perpendicular to the line of sight. The latter two are strongly viewing-angle dependent and would statistically produce narrow peaks substantially offset from line center. By contrast, all three SN\,Ia in our sample show narrow cores with $|v_{\rm off}|<1300$\kms, more consistent with enhanced central density in a mildly off-center ignition. The likelihood of this occurring purely from orientation effects is low, though larger samples are needed to confirm this. By analogy with the dense O cores inferred in SN\,Ib/c \citep{Iwamoto2000, Mazzali2000, Maeda2003, Mazzali2007a, Mazzali2007b}, we find an enhanced central concentration of stable Ni to be the most compelling explanation for our SN\,Ia.

The observed ionization dependence of Ni line widths supports this interpretation. In SN\,2024gy at $+$337~days, [\ion{Ni}{3}] lines are broader than [\ion{Ni}{2}], which in turn are broader than the purely narrow [\ion{Ni}{1}], consistent with recombination proceeding most efficiently in the densest central regions at a given epoch. The line morphologies also depend on energy deposition from radioactive material, predominantly $^{56}$Co at these epochs. Since Co emission is strong and broadly distributed, most layers of the ejecta remain illuminated, implying that the Ni cores are not artifacts of uneven energy input.

We next consider whether the narrow Ni cores could instead arise solely from ionization effects. In SN\,2024gy at $+$144~days, the [\ion{Ni}{3}]~3.80\um\ line shows a weaker, flatter narrow component than [\ion{Ni}{2}]~1.94 and 6.64\um, hinting at stratification within the core. However, some degree of ionization stratification is expected even if the narrow feature results from an abundance enhancement. The widths of the [\ion{Ni}{2}] narrow components remain constant in both SN\,2022aaiq and SN\,2024gy---an evolution spanning nearly 200~days---and in SN\,2024gy the [\ion{Ni}{1}] width matches that of [\ion{Ni}{2}]. Ionization-driven changes would instead produce continuous temporal evolution and systematic width differences between ions as the density and temperature change. Moreover, no comparable narrow feature is seen in [\ion{Co}{2}] despite sufficient sensitivity. 

Taken together, these results indicate that ionization effects modulate line strengths and relative contributions from broad and narrow components but cannot by themselves account for the narrow cores. Their persistence across multiple ions and epochs, combined with their absence in Co, point to an intrinsic enhancement of stable Ni abundance in the innermost ejecta.

\subsection{Delayed-Detonation Models}
\label{sec:ddt_discussion}

In 2D and 3D DDT simulations, buoyant deflagration ashes rise through Rayleigh-Taylor (RT) instabilities, mixing outward while unburned C/O fuel flows inward. When the detonation follows, it burns the remaining fuel almost instantaneously, converting the core entirely into detonation ashes \citep[e.g.,][]{Maeda2010b, Seitenzahl2013, Pakmor2024}.

In 3D simulations, the central ejecta are therefore dominated by detonation ashes, while deflagration ashes occupy higher velocities, overlapping at intermediate regions in velocity space through rising plumes. The broken-slope morphology of [\ion{Ni}{3}] in the DDT model (\autoref{fig:derivatives}) directly reflects this two-zone structure. 1D models also produce two distinct zones, but the geometry is inverted: deflagration ashes are artificially confined to the center and surrounded by detonation ashes \citep{Blondin2013, Pakmor2010, Maeda2010b}. \citet{Pakmor2024} argue that such confinement is inherently unphysical. \citet{Hoeflich2021} calculate the deflagration phase assuming spherical symmetry, suggesting that strong magnetic fields might suppress RT instabilities and retain deflagration ashes in the core. Whether such fields can arise self-consistently in 3D remains uncertain.

In 3D DDT models, the deflagration phase produces $\sim$15--20\% stable Ni, while the detonation phase---occurring at lower densities after expansion---produces only $\sim$5\% \citep{Pakmor2024}. The [\ion{Ni}{2}] and [\ion{Ni}{3}] profiles of SN\,2024gy indicate that stable Ni is distributed across three regions: outer ($\sim$4000--8000\kms), inner ($\sim$1500--4000\kms), and a compact central core ($\lesssim$1500\kms). The broken-slope [\ion{Ni}{3}] morphology, with distinct inner and outer components, is qualitatively consistent with detonation ashes in the center and deflagration ashes farther out. However, the presence of an additional narrow core implies that the innermost layers burned at densities high enough to synthesize stable IGEs. This presents a challenge for current 3D DDT models, as it would require either (i) confinement of some deflagration ashes in the core---via an as-yet unidentified mechanism---or (ii) recompression of the core to high density prior to detonation, perhaps through pulsations or other dynamic processes.

The two-component Ni distribution inferred from the [\ion{Ni}{3}] broken slope, together with the high inferred stable Ni mass of SN\,2024gy (\autoref{sec:ni_masses}), points to a DDT origin as the most plausible scenario. The narrow cores require enhanced stable Ni in the innermost ejecta, but whether this originates from confined deflagration ashes or high-density detonation ashes remains uncertain. Current 3D simulations do not reproduce this structure, and further work is needed to test mechanisms such as suppression of RT instabilities or pulsational compression prior to detonation.

\subsection{Double-Detonation Models}
\label{sec:dbldet_discussion}

Sub-\mch\ double detonations predict a distinct distribution of stable IGEs compared to near-\mch\ DDTs, with stable IGEs concentrated in the C/O core detonation ashes and colocated with radioactive IGEs. The resulting nebular line profiles are more nearly single-component Gaussians (\autoref{fig:derivatives}). A small amount of stable Ni may also form in the outer He-burning ashes, but these layers are at low density and unlikely to be detectable in nebular spectra.

An important distinction from \mch\ explosions is that sub-\mch\ models generally produce less total stable Ni, owing to their lower progenitor masses and consequently lower burning densities \citep[e.g.,][]{Hoflich2004, Blondin2018, Shen2018, Gronow2021}. However, in double detonations, the initial He-shell detonation can exert additional pressure that compresses the underlying C/O core, allowing the inner regions to reach densities high enough for stable Ni synthesis \citep[e.g.,][]{Moran-Fraile2024}. As a result, the C/O core detonation in some double-detonation models can produce a comparable amount of stable Ni to the detonation phase of DDT models \citep{Pakmor2024}. The key difference is that DDTs also form stable Ni during the preceding deflagration phase, which sub-\mch\ explosions lack. Progenitor metallicity can further modify this outcome: higher metallicity increases the initial neutron excess, enhancing the production of stable Ni even in lower-mass WDs \citep{Timmes2003, Blondin2022}.

Double-degenerate scenarios in which both WDs undergo double detonation \citep[e.g.,][]{Shen2024} could, in principle, produce two distinct Ni-rich regions, as the secondary WD detonates within the ashes of the primary. In practice, however, the secondary WD burns at much lower densities and is not expected to synthesize significant stable Ni in the scenario from \citet{Shen2024}. This scenario would also introduce an intervening layer of IMEs between the two Ni zones---contrary to our observed [\ion{Ar}{3}] profiles in SN\,2024gy, which do not show such a structure. More equal binary mass ratios would lead to more stable Ni produced in the secondary WD, but also likely lead to bulk asymmetries \citep[e.g.,][]{Kwok2024}.

While the broken-slope [\ion{Ni}{3}] morphology is difficult to reproduce in a double-detonation framework, a potential mechanism for producing a narrow Ni core in sub-\mch\ progenitors is gravitational settling of $^{22}$Ne. In near-\mch\ WDs, a convective simmering phase homogenizes isotopes prior to explosion \citep{Woosley2004, Piro2008}, but in lower-mass WDs (which lack vigorous convection), neutron-rich $^{22}$Ne can gravitationally settle toward the center \citep{Bildsten2001, Garcia-Berro2008}. This enrichment could seed enhanced neutronization during burning and increase the yield of stable Ni in the innermost ejecta.

We cannot rule out sub-\mch\ double detonations, particularly if $^{22}$Ne settling enhances central neutronization. However, existing sub-\mch\ models predict smaller stable Ni masses and lack the observed broken-slope morphology. Future 3D simulations, coupled with late-time \textit{JWST} MIR spectroscopy, will be essential to distinguishing between these possibilities.

\subsection{Origin of SN\,2022xkq}
\label{sec:22xkq_origin}

SN\,2022xkq is spectroscopically distinct from the other SN\,Ia in our sample. Its [\ion{Ni}{2}] and [\ion{Ni}{3}] profiles are well described by single Gaussian components with FWHM~$\approx$ 3000\kms---narrower than the broad bases typical of normal SN\,Ia, yet broader than the narrow [\ion{Ni}{2}] cores seen in SN\,2022aaiq and SN\,2024gy. The line-profile derivatives do not closely resemble any of the explosion models in \autoref{fig:derivatives}. Consistent with this, \citet{Pearson2024} found that no existing model reproduces the full early-time optical photometric and spectroscopic dataset of SN\,2022xkq. These single-component, moderately narrow Ni features, together with comparatively strong Ar emission, indicate a meaningful physical difference in the explosion mechanism relative to the brighter events.

\citet{DerKacy2024} proposed a high-central-density DDT model to explain the narrow Ni features, arguing that they require little to no central mixing. However, the deflagration phase in that model was computed in 1D (imposing spherical symmetry) and therefore cannot capture the mixing behavior expected in 3D deflagrations. This raises the possibility that the limited central mixing instead reflects detonation-dominated burning. Moreover, the model fit from \citet{DerKacy2024} is degraded when compared with the rereduced spectra: Ti lines are not clearly present, and the model predicts [\ion{Ni}{1}]~7.51\um\ emission that is not observed. The absence of this line, even at late times in SN\,2024gy when [\ion{Ni}{1}]~3.12\um\ is detected, may indicate that the modeled central densities are too high.

From our analysis, combined with previous measurements by \citet{Li2026} and \citet{Pearson2024}, SN\,2024gy produces both more $^{56}$Ni and more $^{58}$Ni than SN\,2022xkq. A sub-\mch\ origin could naturally explain the relatively narrow single-component Ni lines. However, \citet{Pearson2024} reported persistent carbon absorption in SN\,2022xkq, which is challenging to reconcile with standard sub-\mch\ models \citep{Blondin2017, polin_observational_2019}. It should also be noted that we do not explore the full diversity of DDT models in this work, and we cannot rule out that a DDT scenario with differing conditions could explain the differences between SN\,2022xkq and the brighter events.

The NIR [\ion{Ni}{2}]~1.94\um\ line provides additional context. \citet{Kumar2025} found strong, narrow [\ion{Ni}{2}] emission in subluminous SN\,Ia and suggested it might be evidence for DDTs of near-\mch\ WDs. Their detection method, however, favors narrower features that rise above blended neighboring lines. Our \textit{JWST}/MIRI data reveal that [\ion{Ni}{2}] emission can include both broad and narrow components, suggesting their method may overlook objects with substantial amounts of stable Ni, but no strong narrow component, such as SN\,2021aefx. In subluminous events such as SN\,2022xkq, the lower ejecta velocities and ionization state likely enhance the visibility of [\ion{Ni}{2}] by reducing blending and allowing it to dominate cooling over [\ion{Ni}{3}]. The faster transition to optically thin conditions may further isolate the line.

Taken together, these observations suggest that SN\,2022xkq occupies a physically distinct regime from the DDT-like SN\,2022aaiq and 2024gy. We suggest that double-detonation models remain potentially viable---especially those incorporating $^{22}$Ne settling, which could enhance central neutronization and produce the observed narrower, stable Ni lines without a broader base component. Further multidimensional modeling and time-dependent radiative-transfer calculations will be essential to determine whether such sub-\mch\ explosions can fully reproduce the Ni morphology and luminosity of SN\,2022xkq.

\subsection{Asymmetries}
\label{sec:asymmetries}

The small sample of \textit{JWST}/MRS observations of SN\,Ia analyzed in this work reveals that, while the main MIR lines are broadly similar, even normal SN\,Ia exhibit notable variations. One key difference is the presence of asymmetries in several line profiles, especially [\ion{Ar}{3}]~8.99\um\ and [\ion{Co}{3}]~11.88\um, reflecting intrinsic asymmetries in the ejecta. Our line-inversion analysis in \autoref{sec:line_inversions} assumes spherical symmetry---an inherently unphysical approximation for a 3D explosion, but also a good place to start. In addition to the lower S/N of the SN\,2022aaiq MRS data, such asymmetry may explain why the analysis performs well for SN\,2024gy but yields a less coherent result for SN\,2022aaiq.

The line profiles in SN\,2022aaiq are less symmetric than those in SN\,2024gy or SN\,2021aefx, most clearly in [\ion{Ar}{3}]~8.99\um, which appears distinctly slanted in SN\,2022aaiq but comparatively flat in SN\,2024gy. Even in SN\,2024gy, the [\ion{Co}{3}]~11.88\um\ emissivity and derivative profiles deviate from perfect symmetry. \citet{DerKacy2023} demonstrated that similar asymmetries in SN\,2021aefx can arise from viewing-angle effects in an off-center ignition. The [\ion{Ar}{3}]~8.99\um\ line in SN\,2021aefx slopes upward toward the red, whereas it slopes to the blue in SN\,2022aaiq, and is flatter in SN\,2024gy. Such variation is consistent with a population of off-center explosions viewed from different orientations. SN\,2022aaiq may therefore represent a more intrinsically asymmetric event, or we may simply view SN\,2024gy from a direction where it appears more symmetric. Asymmetries may also become more pronounced at later times, as seen in SN\,2024gy where the [\ion{Ar}{3}]~8.99\um\ line is more slanted in the $+$337~day observation.

Recently, \citet{Pollin2025} computed multidimensional models and nebular-phase spectra for the dynamically driven double-degenerate double-detonation (D$^{6}$) model. They found that multidimensional effects significantly influence the ionization, velocity, and width of emission features, producing distinct line-profile morphologies depending on viewing angle. This underscores the need for further 3D modeling and radiative-transfer post-processing across multiple explosion channels. Our comparison of SN\,2024gy with angle-averaged 3D model outputs demonstrates that assuming spherical symmetry can still provide valuable insights, but improved approximations that better reflect fully 3D geometries will be essential for interpreting the asymmetries observed in \textit{JWST} spectra.

While asymmetries may arise from the explosions themselves (i.e., off-center ignition), \citet{Simotas2025} showed that certain asymmetric signatures detectable with \textit{JWST}/MRS could instead result from wakes formed as the ejecta encounter and flow around a companion star. The predicted line-profile distortions from such wakes depend strongly on viewing angle and should appear consistently across lines from multiple ions. We exclude this scenario as the origin of the narrow [\ion{Ni}{2}] cores in SN\,2022aaiq and SN\,2024gy: no corresponding features are seen in the Ar and Co lines, and the narrow Ni peaks are centered in both objects despite different inferred viewing angles from the [\ion{Ar}{3}] lines. Nonetheless, companion-induced wakes remain an intriguing observational signature to test in future \textit{JWST}/MRS studies of SN\,Ia.

\section{Summary and Conclusions \label{sec:summary}}

We have presented medium-resolution \textit{JWST} NIR$+$MIR spectroscopy of the normal SN\,Ia\,2022aaiq ($+$125 and $+$207~days) and 2024gy ($+$144 and $+$337~days), which reveal novel narrow-core ($v_{\textrm{FWHM}}<1500$\kms) emission in both the NIR and MIR [\ion{Ni}{2}] lines. The relative isolation of the MIR [\ion{Ni}{2}]~6.64\um\ line shows that this narrow feature sits atop a broader base of emission. A later epoch of SN\,2024gy further reveals [\ion{Ni}{1}]~3.12\um\ emission composed solely of this narrow component. The high S/N and spectral resolution of the SN\,2024gy data also reveal a broken-slope line profile in [\ion{Ni}{3}], indicating distinct inner and outer regions of stable IGEs.

The emissivity and derivative profiles we derive for SN\,2024gy show close agreement with predictions from DDT models, though the narrow-core Ni component remains challenging to explain with current models. We summarize several new observational signatures from SN\,2022aaiq and SN\,2024gy.
\begin{itemize}
    \item The narrow-core $+$ broad-base [\ion{Ni}{2}] structure demonstrates that stable Ni extends to intermediate velocities, while the innermost $v<1500$\kms\ exhibits a concentrated enhancement of stable Ni mass.
    \item The narrow-core Ni peak is blueshifted by $\sim$1000\kms\ in SN\,2022aaiq and redshifted by $\sim$500\kms\ in SN\,2024gy. These small offsets suggest mildly off-center ignition, as expected for DDT explosions \citep[e.g.,][]{Maeda2010a, Maeda2010b, DerKacy2023, DerKacy2024}. The direction of the offset likely depends on viewing angle.
    \item The derived emissivity and derivative profiles for SN\,2024gy reveal three distinct Ni-emitting zones: an innermost core ($\lesssim1500$\kms), an inner region ($\sim$1500--3000\kms) and an outer region ($\sim$3000--8000\kms). In the [\ion{Ni}{3}]~7.35\um\ line, these inner and outer regions appear as a ``broken-slope'' morphology, matching the separation between detonation and deflagration ashes predicted by DDT models. The [\ion{Ar}{3}]~8.99\um\ and [\ion{Co}{3}]~11.88\um\ lines, albeit more tentatively, show signatures consistent with this interpretation.
    \item The [\ion{Co}{3}]~11.88\um\ emissivity profile, tracing radioactive $^{56}$Co decay, differs from the Ni profiles: it shows less emission in the core and more at high velocities.
    \item Only the innermost, stable Ni-dominated regions---traced primarily by [\ion{Ni}{2}] and, at later epochs, [\ion{Ni}{1}]---show a narrow enhancement. The [\ion{Ni}{3}] and [\ion{Ni}{4}] lines, which trace less-dense outer regions, lack strong narrow components. No narrow cores are detected in [\ion{Co}{2}] or [\ion{Co}{3}], which trace radioactive material. Radiative-transfer models reproduce the observed narrow [\ion{Ni}{2}] component by artificially increasing the inner Ni abundance by $10^{-3}$\msun\ within $v<1000$\kms, roughly a tenfold enhancement.
    \item The [\ion{Ar}{3}]~8.99\um\ and [\ion{Co}{3}]~11.88\um\ lines exhibit asymmetries, differing in slope and skew between SN\,2022aaiq, SN\,2024gy, and SN\,2021aefx. These variations are consistent with off-center ignition and viewing-angle effects predicted for DDT explosions, suggesting that even normal SN\,Ia can display moderate large-scale asymmetries in their ejecta.
\end{itemize}

Overall, we favor a near-\mch\ progenitor undergoing a DDT with mild off-center ignition for both SN\,2022aaiq and SN\,2024gy. The narrow [\ion{Ni}{2}] cores trace enhanced stable Ni in the innermost ejecta, while the broken-slope [\ion{Ni}{3}] profiles record the separation of deflagration and detonation ashes. Ionization effects alone likely cannot reproduce the persistent narrow component, implying a compositional concentration of stable Ni at low velocities. The narrow-core Ni remains a key challenge for current 3D DDT models, requiring either confinement of deflagration ashes or recompression of the core before detonation. Although we cannot entirely rule out sub-\mch\ origins, our findings most strongly support near-\mch\ DDTs for SN\,2024gy, and more tentatively for SN\,2022aaiq and SN\,2021aefx.

The X-ray spectra of some SN remnants (SNRs) thought to be of SN\,Ia origin have also provided interesting constraints on the mass of the exploding WDs. Notably, \citet{Yamaguchi2015} found that the Ni/Fe and Mn/Fe mass ratios inferred from the X-ray spectrum of SNR 3C397 effectively rule out a sub-\mch\ origin (see also \citealt{Mehta2024}). Using a more indirect method based on the Ca/S mass ratio, which is sensitive to the level of neutronization in the SN ejecta, \citet{Martinez-Rodriguez2017} found independent evidence for high neutronization in three SNRs: 3C397 and G337.2-0.7 in the Milky Way, and N103 in the Large Magellanic Cloud. This evidence for high neutronization in the relatively small sample of X-ray bright SNRs in the Milky Way and the Magellanic Clouds suggests that there is a near-\mch\ channel to SNe\,Ia, and that this channel makes a significant contribution to the SN\,Ia rate in star-forming galaxies.

We encourage high-S/N observations between $\sim$100--200\,days post-peak in the optical and NIR to search for narrow [\ion{Ni}{2}] components. A weak narrow feature in [\ion{Ni}{2}]~7378\,\AA\ is detected in our highest-S/N Keck/LRIS spectrum of SN\,2024gy, and \citet{Kumar2025} reported a similar feature in SN\,2011iy in the NIR. Because \textit{JWST} samples will be comparatively limited, establishing the frequency of narrow-core [\ion{Ni}{2}] emission should also be pursued by ground-based facilities.

The origin of the subluminous SN\,2022xkq remains uncertain. Additional modeling---including sub-\mch\ double-detonation scenarios---should be explored to understand its narrow, single-component Ni lines and the implications for the SN\,1991bg-like subclass.

This analysis demonstrates that detailed mapping of stable Ni emission is possible only with the medium-resolution modes of \textit{JWST}/MIRI (\autoref{sec:resolution_comp}). Ni lines in the optical and NIR are too blended, while low-resolution \textit{JWST} modes, though efficient for large samples, wash out diagnostic structure. MRS observations are essential for detailed case studies of the nearest SN\,Ia. Expanding this sample will provide critical tests of explosion models and progenitor masses in shaping SN\,Ia diversity.

\clearpage

\input{acknowledgements}

\facilities{\textit{JWST} (NIRSpec/MIRI), Keck:I (LRIS), Keck:II (NIRES), Keck:II (DEIMOS, ESI), LCO/GSP}

\software{Astropy \citep{astropycollaboration_astropy:_2013, astropycollaboration_astropy_2018, AstropyCollaboration2022}, 
Matplotlib \citep{hunter_matplotlib:_2007}, 
NumPy \citep{oliphant_guide_2006}, PyRAF \citep{Pyraf}, PySALT \citep{PySALT}, dust extinction \citep{karl_gordon_2022_6397654}, jwst \citep{Bushouse_JWST_Calibration_Pipeline_2022}, {\tt YSE-PZ} \citep{CoulterZenodo, CoulterYSEPZ}, CMFGEN \citep{Hillier2012}, ChatGPT \citep{openai_chatgpt5_1}
}

\begin{appendix}

\section{Additional line comparisons and fits \label{sec:appendix_more_lines}}

\subsection{All optical, NIR, and MIR stable Ni lines \label{sec:appendix_all_Ni}}

\begin{figure*}
    \centering
    \includegraphics[width=\linewidth]{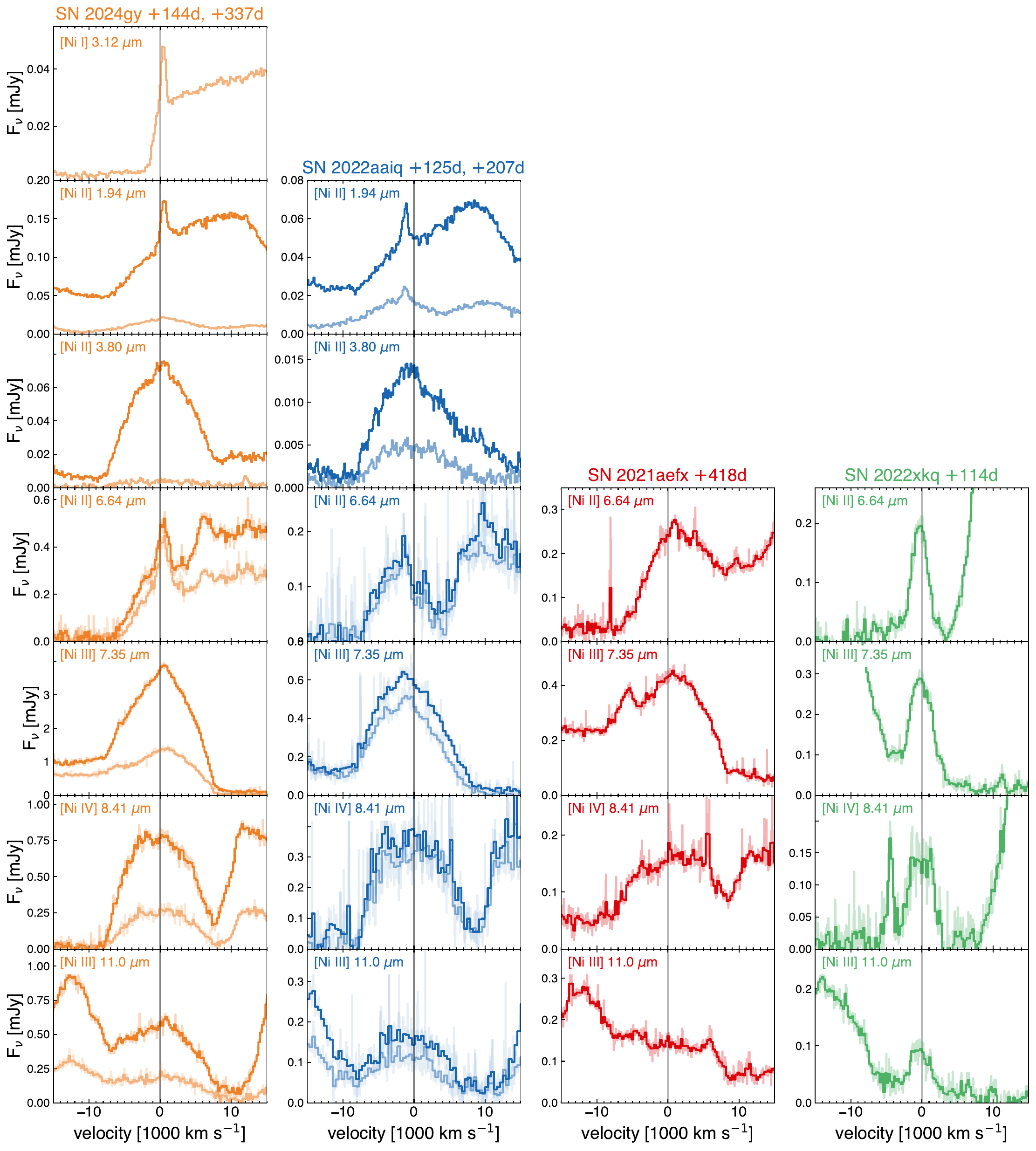}
    \caption{All prominent Ni lines in the \textit{JWST} data of the normal SN\,2024gy at $+$144~days (orange) and $+$337~days (light orange), SN\,2022aaiq at $+$125~days (blue) and $+$207~days (light blue), and SN\,2021aefx at $+$418~days (red), and the subluminous SN\,2022xkq at $+$114~days (green). For each SN, the velocity of the peak  agrees for all lines.}
    \label{fig:all_Ni_lines}
\end{figure*}

\autoref{fig:all_Ni_lines} shows all NIR and MIR Ni lines in SN\,2024gy, SN\,2022aaiq, SN\,2021aefx, and SN\,2022xkq. All Ni lines peak at a consistent velocity across a large wavelength range. This agreement in velocity confirms the narrow spike in the NIR spectra of SN\,2022aaiq and SN\,2024gy is [\ion{Ni}{2}]~1.94\um.

\subsection{Fitting [\ion{Ni}{2}] and [\ion{Ni}{3}] in SN\,2022aaiq and SN\,2024gy \label{sec:appendix_fitting_22aaiq_24gy}}

In \autoref{fig:Ni2_fits}, the unblended [\ion{Ni}{2}] wing reveals a narrow core profile: a sharp central peak atop a broader base. The sharp narrow peak is better modeled by a Lorentzian than a Gaussian, while the base shows a bend near $3000$\kms\ and steeper wings than a Gaussian, empirically motivating the use of a super-Gaussian. Thus, [\ion{Ni}{2}] is fit with a Lorentzian $+$ super-Gaussian function,
\begin{equation}
f(v) = A \big[\text{Lorentzian}(v) + R~\text{Super-Gaussian}(v) \big]
\label{eq:L+SG}
\end{equation}
\noindent where
\begin{align}
\text{Lorentzian}(v) &= \frac{1}{1 + \left( \dfrac{2(v - v_{\rm L}}{\mathrm{FWHM}_{\rm L}} \right)^2}\, , \\
\text{Super-Gaussian}(v) &=
\exp\left[
-\left(
\dfrac{2(\ln 2)^{1/(2n)}(v - v_{\rm SG})}{\mathrm{FWHM}_{\rm SG}}
\right)^{2n}
\right].
\end{align}
\noindent
Here \( A \) is the line amplitude, \( R \) is the Lorentzian-to-super-Gaussian ratio, \( v_L \) and \( v_{SG} \) are velocity offsets, \( \text{FWHM}_L \) and \( \text{FWHM}_{SG} \) are the full width at half-maximum intensity (FWHM), and \( n \) is the order of the super-Gaussian, which controls flatness near the peak and steepness of the falloff. This six-parameter fit captures the sharp core and broad wings, and is flexible enough for both [\ion{Ni}{2}] and [\ion{Ni}{3}].

The [\ion{Ni}{3}] lines show a less pronounced core and a more pronounced bend around $4000$\kms\ (\autoref{fig:Ni2_fits}, right panel). We refer to this bend in the line profile (seen clearly in the derivative in \autoref{sec:line_inversions}) as a ``broken-slope” profile. At $+$144~days, the red side of the [\ion{Ni}{4}]~8.40\um\ line in SN\,2024gy also shows a broken slope (\autoref{fig:Ar_fits}).

In our fits, we constrain [\ion{Ar}{2}]~6.98\um\ using the nearly isolated [\ion{Ar}{3}]~8.98\um\ line, which is well-fit by a slanted flat-top with Gaussian wings. Allowing $\pm500$\kms\ bounds around the parameters of this fit for [\ion{Ar}{2}], and adopting Lorentzian$+$super-Gaussian profiles for Ni, we fit the full 6--8\um\ complex in both SN. Fit results are shown in \autoref{fig:Ar_fits} and \autoref{fig:Ni2_fits}.

The sharp uptick on the blue edge of [\ion{Ar}{2}] is likely overfit; a similar feature appears in [\ion{Ar}{3}]~8.99\um\ and [\ion{Ca}{4}]~3.21\um, suggesting ejecta inhomogeneity or asymmetry. Offsets between Lorentzian and super-Gaussian centers are allowed but remain small (\autoref{fig:Ni2_fits}, left panel). Allowing such offsets and tilted flat-tops implicitly assumes axisymmetry along the line of sight.

We extend these fits to the NIR. Using MIR-derived parameters as bounds ($\pm500$\kms\ for [\ion{Ni}{2}] narrow components, $\pm1000$\kms\ for others), we fit [\ion{Ni}{2}]~1.94\um\ and [\ion{Ni}{3}]~3.80\um\ with the same two-component profiles (\autoref{fig:Ni2_fits}). Emission from the same ion should arise from the same ejecta regions, making these constraints physically well motivated. This is especially important for [\ion{Ni}{2}]~1.94\um, where blending with [\ion{Co}{3}] and [\ion{Fe}{2}] lines complicates the fit. Without MIR constraints, the narrow [\ion{Ni}{2}] component is still well constrained, but the broad base becomes highly sensitive to bounds and blending, showing that the NIR alone cannot robustly constrain it. At much later phases ($>$400~days), the [\ion{Ni}{2}]~1.94\um\ line becomes more isolated \citep{Dhawan2018}, but it is also much fainter and therefore challenging to observe.

The [\ion{Ni}{3}]~3.80\um\ line, newly accessible with NIRSpec, is well isolated (\autoref{fig:Ni2_fits}, right panel). It shows a distinct narrow peak like [\ion{Ni}{2}], though with smaller relative strength. Unlike many of the strong MIR lines, this transition does not terminate at the ground state; instead, it connects an excited upper level to a lower excited state. This may explain why its narrow component appears more prominent than in [\ion{Ni}{3}]~7.35\um. By the second epoch in both SN\,2024gy and SN\,2022aaiq, the line weakens and flattens (see \autoref{fig:all_Ni_lines}, lower opacity lines), so we do not fit the [\ion{Ni}{3}]~3.80\um\ line at these epochs in \autoref{fig:Ni2_fits} (right panel).

\subsection{Fitting [\ion{Ni}{2}] and [\ion{Ni}{3}] in SN\,2021aefx and SN\,2022xkq \label{sec:appendix_fitting_21aefx_22xkq}}

\begin{figure}
    \centering
    \includegraphics[width=0.5\textwidth]{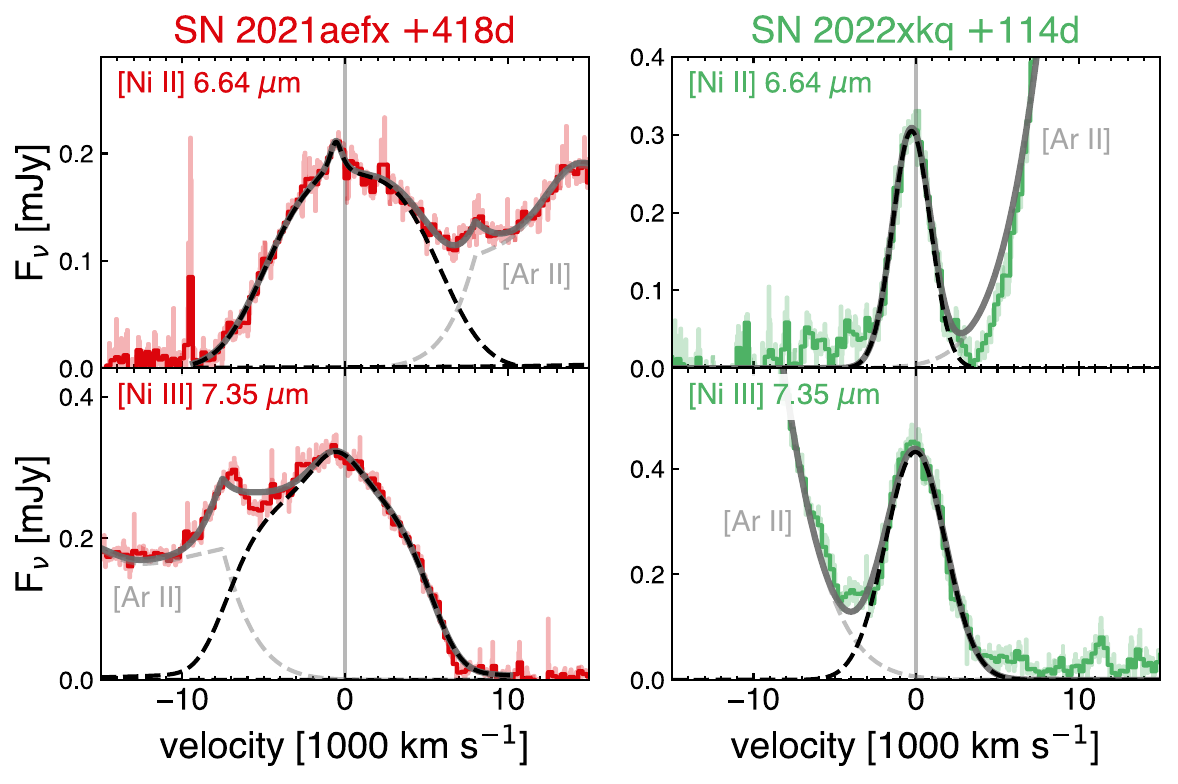}
    \caption{Line profile fits for [\ion{Ni}{2}]~6.64\um\ and [\ion{Ni}{3}]~7.35\um\ in SN\,2021aefx (red) at $+$418~days and SN\,2022xkq (green) at $+$114~days. The contribution from Ni is shown in dashed black, and contributions from other nearby lines are  in dashed gray. The composite fit is displayed in solid gray. The SN\,2021aefx Ni lines display similar shape to SN\,2024gy and SN\,2022aaiq, but have less contribution from the narrower components. The SN\,2022xkq Ni lines are well-fit by a simple Gaussian.}
    \label{fig:Ni_21aefx_22xkq}
\end{figure}

For SN\,2021aefx, our fits reproduce the overall [\ion{Ni}{2}] and [\ion{Ni}{3}] line profiles. We marginally detect a weak [\ion{Ni}{2}] narrow core, and the two-component structure in [\ion{Ni}{3}] is only weakly developed, with a bend near $\sim5000$\kms\ (\autoref{fig:Ni_21aefx_22xkq}). The blue wing of [\ion{Ar}{3}]~8.99\um\ appears to decline more steeply than permitted by the adopted profile shape, so the blue wing of [\ion{Ar}{2}]~6.98\um\ likely does as well. To reproduce the central excess in [\ion{Ar}{2}]~6.98\um, we include an additional component modeled as a slanted flat-top profile with parabolic wings plus a central Gaussian (\autoref{fig:Ar_fits}).

For SN\,2022xkq, the [\ion{Ni}{2}]~6.64\um\ and [\ion{Ni}{3}]~7.35\um\ lines are well described by single Gaussians, with [\ion{Ni}{2}] slightly narrower than [\ion{Ni}{3}] (\autoref{fig:Ni_21aefx_22xkq}). The narrow width of the [\ion{Ar}{2}] feature allows clean separation of the Ni and Ar contributions.

\input{velocity_fitting_table}

\section{Impact of Spectral Resolution}
\label{sec:resolution_comp}

Medium-resolution \textit{JWST} modes (NIRSpec gratings and MIRI/MRS; $R \approx 1000$--3000) provide an order-of-magnitude increase in resolving power compared to low-resolution modes (NIRSpec/PRISM and MIRI/LRS; $R \approx 40$--300). This increased resolution is essential for resolving low-velocity features, such as the narrow [\ion{Ni}{2}] emission and detailed line-profile structure. The tradeoff is that medium-resolution observations are significantly more time-intensive, requiring high-S/N targets that are typically nearby.

This is illustrated by SN\,2024gy ($\sim$17.2~Mpc) and SN\,2022aaiq ($\sim$36~Mpc): for the same exposure times, SN\,2024gy achieves substantially higher S/N. Its MIRI/MRS spectrum clearly resolves the broad-base, narrow-core structure of [\ion{Ni}{2}]~6.64\um, whereas the narrow core is much less clearly distinguishable in SN\,2022aaiq.

\begin{figure*}
    \centering
    \includegraphics[width=\linewidth]{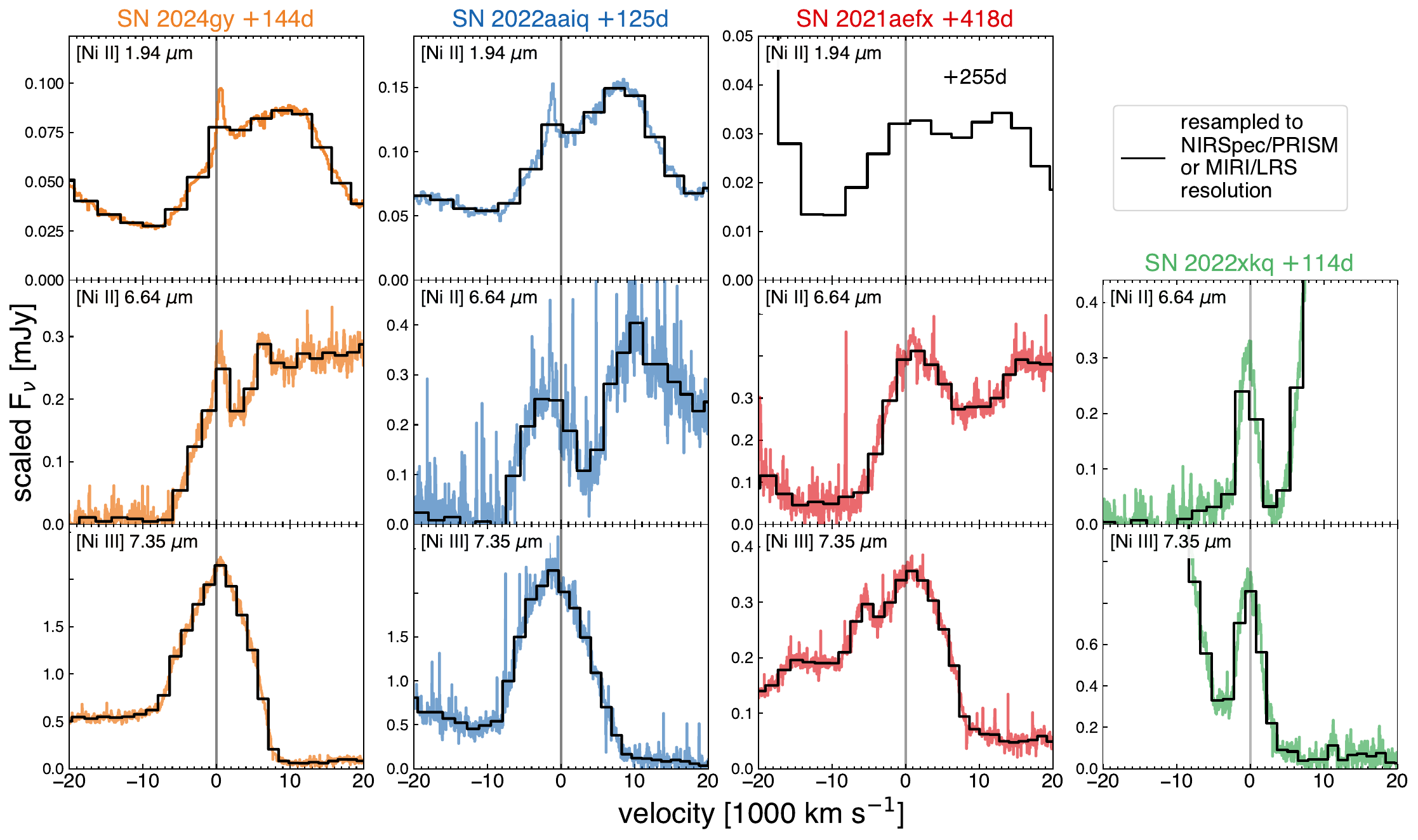}
    \caption{\textit{JWST} NIRSpec/G235M grating and MIRI/MRS data of SN\,2024gy (orange), SN\,2022aaiq (blue), SN\,2021aefx (red), SN\,2022xkq (green) resampled onto the low-resolution wavelength grids of NIRSpec/PRISM and MIRI/LRS (black). The low-resolution modes cannot capture fine details such as narrow lines.}
    \label{fig:res_comp}
\end{figure*}

As shown in \autoref{fig:res_comp}, degrading the spectra to the resolving power of MIRI/LRS and NIRSpec/PRISM significantly blurs both the narrow [\ion{Ni}{2}] core and the broken-slope structure of [\ion{Ni}{3}]~7.35\um, preventing separation of narrow and broad components. The [\ion{Ni}{2}]~1.94\um\ line is detected with NIRSpec/G235M, but blending with nearby features complicates isolation of the broad component.

Although the [\ion{Ni}{2}]~1.94\um\ line is accessible from the ground, strong telluric absorption in this region makes such observations challenging \citep[e.g.,][]{Kumar2025}. We therefore recommend \textit{JWST} NIRSpec/G235M as the optimal mode for future observations of this feature.

\section{Line-profile derivatives for SN\,2022aaiq, SN\,2021aefx, and SN\,2022xkq}
\label{sec:other_lineinversions}

\begin{sidewaysfigure}
    \centering
    \includegraphics[width=\linewidth]{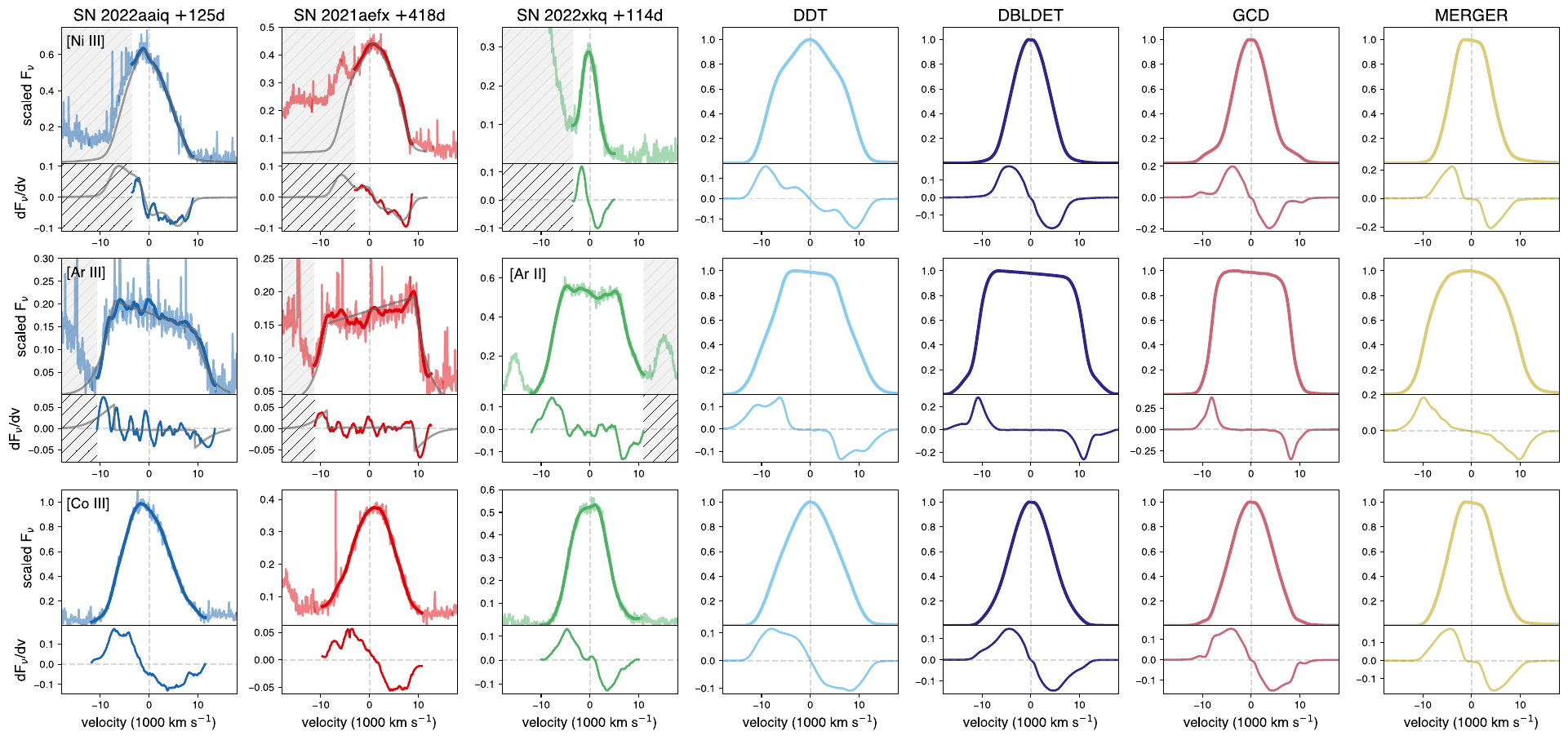}
    \caption{Line profiles (upper panels) and their derivatives (lower panels) for [\ion{Ni}{3}]~7.35\um, [\ion{Ar}{3}]~8.99\um, and [\ion{Co}{3}]~11.88\um\ for SN\,2022aaiq at $+$125~days (blue), SN\,2021aefx at $+$418~days (red), SN\,2022xkq at $+$114~days (green) and the DDT (cyan), DBLDET (green), GCD (pink), and MERGER (yellow) models from \cite{Blondin2023}. The data are shown in lower opacity and the smoothed data in full opacity. Our fits to blended lines are shown in gray, and regions contaminated by line overlap in hatched light gray.}
    \label{fig:derivatives_others}
\end{sidewaysfigure}

\autoref{fig:derivatives_others} presents the line-profile derivatives for SN\,2022aaiq, SN\,2021aefx, and SN\,2022xkq, with comparisons to DDT, DBLDET, GCD, and MERGER models (see \autoref{fig:derivatives_others}).

For SN\,2022xkq, the [\ion{Co}{3}] profile exhibits a flat-topped structure at $v \lesssim 2000$\,km\,s$^{-1}$ and sharper, more triangular derivatives than seen in normal SNe\,Ia. While some features resemble DDT or DBLDET morphologies, the combination of Ni, Ar, and Co structures differs from both SN\,2024gy and the N100 DDT model \citep{Seitenzahl2013}, underscoring the distinct nature of SN\,1991bg-like events.

For SN\,2022aaiq, the [\ion{Ni}{3}], [\ion{Ar}{3}], and [\ion{Co}{3}] profiles appear more asymmetric than in SN\,2024gy. The [\ion{Co}{3}] line shows a possible notch feature reminiscent of DDT or GCD models, though the limited S/N precludes a firm interpretation.

SN\,2021aefx, observed at $+418$~days, shows a double-peaked [\ion{Ni}{3}] derivative with a shallow-to-steep slope break, similar to SN\,2024gy and DDT models. The [\ion{Co}{3}] profile also exhibits a notch-like feature, while the [\ion{Ar}{3}] line has a steeper red wing without a clear outer shoulder. At this late phase, reduced density in the outer ejecta may suppress higher-velocity emission, and we therefore regard the derivative-based interpretation as tentative.

\section{BayeSN Fits to SN\,2022aaiq and SN\,2024gy Photometry \label{sec:bayesn}}

\begin{figure}
    \centering
    \includegraphics[width=0.45\linewidth]{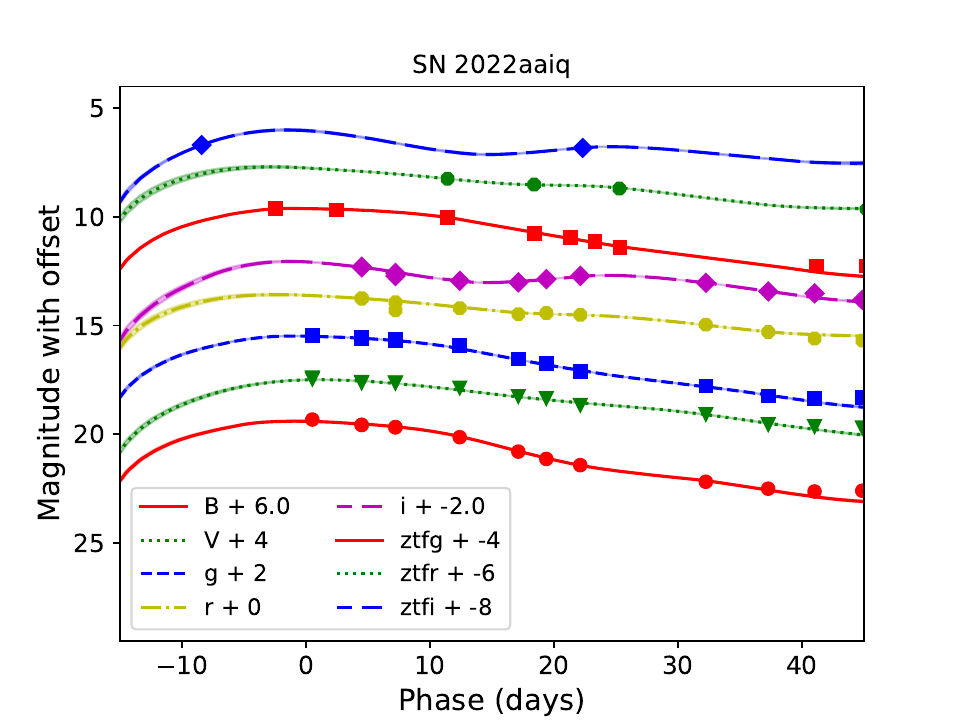}
    \includegraphics[width=0.45\linewidth]{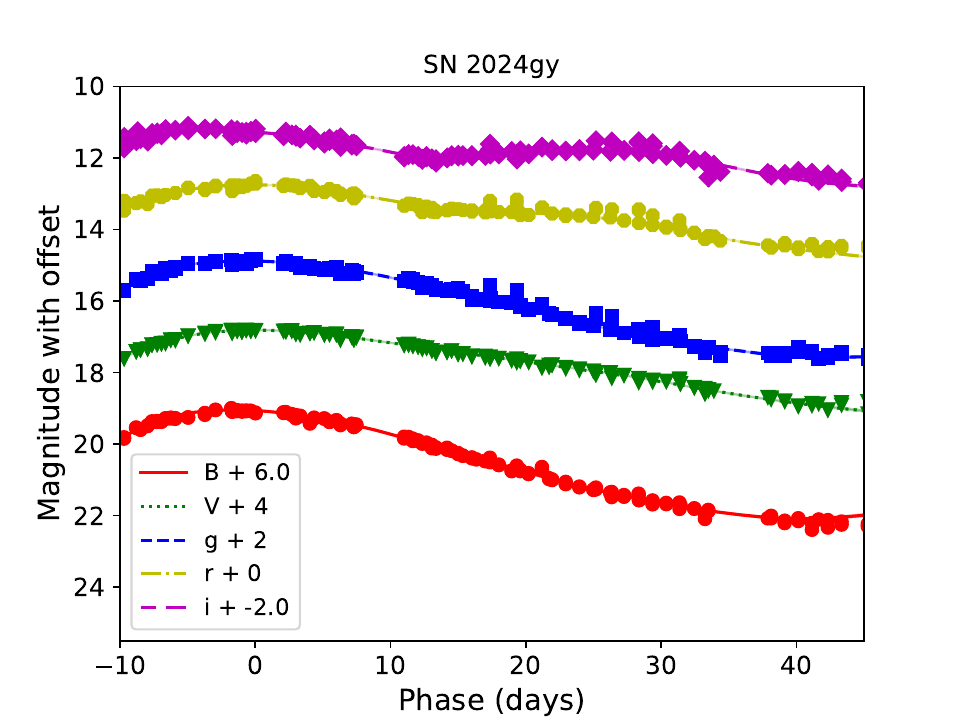}
    \caption{\textit{Left:} LCO (from the GSP collaboration) and ZTF photometry for SN~2022aaiq. The BayeSN light-curve fit is overplotted and the filters are offset for clarity. \textit{Right:} Light-curve fit for SN~2024gy (with only LCO data). The data are in the rest-frame phase for each SN.}
    \label{fig:bayesn_24gy}
\end{figure}

\autoref{fig:bayesn_24gy} shows our BayeSN fits to the light curves of SN\,2022aaiq and SN\,2024gy from \autoref{sec:obs}. The photometry is given in \autoref{tab:2022aaiq_photometry} and \autoref{tab:2024gy_photometry}. Photometry in the $BV$ bands is reported in Vega magnitudes using standards from \citet{2000PASP..112..925S}, while $gri$ band data are presented in AB magnitudes \citep{1983ApJ...266..713O}, calibrated against Sloan Digital Sky Survey (SDSS) sources \citep{Smith2002}.

\input{2022aaiq_photometry_table}
\input{2024gy_photometry_table}

\end{appendix}

\normalsize
\bibliography{zotero_abbrev}

\end{CJK*}
\end{document}

%% file: affiliations.tex
\newcommand{\CIERA}{\affiliation{Center for Interdisciplinary Exploration and Research in Astrophysics (CIERA), Northwestern University, Evanston, IL 60201, USA}}
\newcommand{\Carnegie}{\affiliation{The Observatories of the Carnegie Institution for Science, Pasadena, CA 91101, USA}}
\newcommand{\ICE}{\affiliation{Institute of Space Sciences (ICE\textendash CSIC), Campus UAB, 08193 Barcelona, Spain}}
\newcommand{\IEEC}{\affiliation{Institut d\'Estudis Espacials de Catalunya (IEEC), 08034 Barcelona, Spain}}
\newcommand{\Konkoly}{\affiliation{Konkoly Observatory, HUN\textendash REN Research Center for Astronomy and Earth Sciences, Budapest, 1121, Hungary}}

\newcommand{\LCO}{\affiliation{Las Cumbres Observatory, Goleta, CA 93117, USA}}
\newcommand{\Northwestern}{\affiliation{Department of Physics and Astronomy, Northwestern University, Evanston, IL 60208, USA}}
\newcommand{\Princeton}{\affiliation{Department of Astrophysical Sciences, Princeton University, Princeton, NJ 08544, USA}}
\newcommand{\Purdue}{\affiliation{Department of Physics and Astronomy, Purdue University, West Lafayette, IN 47907, USA}}
\newcommand{\Rutgers}{\affiliation{Department of Physics and Astronomy, Rutgers, The State University of New Jersey, 136 Frelinghuysen Road, Piscataway, NJ 08854, USA}}
\newcommand{\SkAI}{\affiliation{NSF\textendash Simons AI Institute for the Sky (SkAI), Chicago, IL 60611, USA}}
\newcommand{\Thailand}{\affiliation{National Astronomical Research Institute of Thailand (NARIT), Chiang Mai 50180, Thailand}}
\newcommand{\UA}{\affiliation{Steward Observatory, University of Arizona, Tucson, AZ 85721, USA}}
\newcommand{\UCD}{\affiliation{Department of Physics and Astronomy, University of California, Davis, CA 95616, USA}}
\newcommand{\UCSC}{\affiliation{Department of Astronomy and Astrophysics, University of California, Santa Cruz, CA 95064, USA}}
\newcommand{\UPitt}{\affiliation{Department of Physics and Astronomy, University of Pittsburgh, Pittsburgh, PA 15260, USA}}
\newcommand{\USzeged}{\affiliation{Department of Experimental Physics, Institute of Physics, University of Szeged, D\'om t\'er 9, Szeged, 6720 Hungary}}
\newcommand{\UTexas}{\affiliation{Department of Astronomy, The University of Texas at Austin, Austin, TX 78712, USA}}
\newcommand{\ESOgermany}{\affiliation{European Southern Observatory, Karl-Schwarzschild-Stra\ss{}e 2, D-85748, Garching bei M\"unchen, Germany}}
\newcommand{\AixMarseille}{\affiliation{Aix Marseille Univ, CNRS, CNES, LAM, Marseille, France}}
\newcommand{\UCSB}{\affiliation{Department of Physics, University of California, Santa Barbara, CA 93106-9530, USA}}
\newcommand{\TUM}{\affiliation{Technical University of Munich, TUM School of Natural Sciences, Physics Department, D-85748, Garching bei M\"unchen, Germany}}

%% file: authors.tex
\author[0000-0003-3108-1328]{Lindsey A.\ Kwok}
\thanks{NASA Hubble Fellow}
\CIERA

\author[0000-0002-7866-4531]{Chang~Liu (刘畅)}
\CIERA
\Northwestern

\author[0000-0001-8738-6011]{Saurabh W.\ Jha}
\Rutgers

\author[0000-0002-9388-2932]{St\'{e}phane Blondin}
\ESOgermany,\AixMarseille

\author[0000-0003-2037-4619]{Conor~Larison}
\Rutgers

\author[0000-0001-9515-478X]{Adam~A.~Miller}
\Northwestern
\CIERA
\SkAI

\author[0000-0002-5995-9692]{Mi~Dai}
\UPitt

\author[0000-0002-2445-5275]{Ryan J.\ Foley}
\UCSC

\author[0000-0003-3460-0103]{Alexei~V.~Filippenko}
\affiliation{Department of Astronomy, University of California, Berkeley, CA 94720-3411, USA}

\author[0000-0003-0123-0062]{Jennifer~E.~Andrews}
\affiliation{Gemini Observatory/NSF's NOIRLab, 670 N. A'ohoku Place, Hilo, HI 96720, USA}

\author[0000-0002-1895-6639]{Moira~Andrews}
\LCO
\UCSB

\author[0000-0002-4449-9152]{Katie~Auchettl}
\UCSC

\author[0000-0003-3494-343X]{Carles~Badenes}
\UPitt

\author[0000-0002-4924-444X]{K.~Azalee~Bostroem}
\thanks{LSST-DA Catalyst Fellow}
\UA

\author[0000-0001-5955-2502]{Thomas G.\ Brink}
\affiliation{Department of Astronomy, University of California, Berkeley, CA 94720-3411, USA}

\author[0009-0000-9117-8995]{Cristine~Koelln}
\ESOgermany
\TUM

\author[0000-0002-5680-4660]{Kyle~W.~Davis}
\UCSC

\author[0000-0003-2024-2819]{Andreas~Fl\"ors}
\affiliation{GSI Helmholtzzentrum f{\"u}r Schwerionenforschung, Planckstra{\ss}e 1, D-64291 Darmstadt, Germany}

\author[0000-0002-1296-6887]{Llu\'is Galbany}
\ICE
\IEEC

\author[0000-0002-4391-6137]{Or~Graur}
\affiliation{Institute of Cosmology \& Gravitation, University of Portsmouth, Dennis Sciama Building, Portsmouth, PO1 3FX, UK}
\affiliation{Department of Astrophysics, American Museum of Natural History, New York, NY 10024, USA}

\author[0000-0003-4253-656X]{D.~Andrew~Howell}
\LCO
\UCSB

\author[0000-0001-8367-7591]{Sahana~Kumar}
\affiliation{Department of Astronomy, University of Virginia, 530 McCormick Rd, Charlottesville, VA 22904, USA}

\author[0000-0002-8770-6764]{R\'eka K\"onyves-T\'oth}
\Konkoly
\USzeged

\author[0000-0002-2249-0595]{Natalie~LeBaron}
\affiliation{Department of Astronomy, University of California, Berkeley, CA 94720-3411, USA}
\affiliation{Berkeley Center for Multi-messenger Research on Astrophysical Transients and Outreach (Multi-RAPTOR), University of California, Berkeley, CA 94720-3411, USA}

\author[0000-0002-9209-2787]{Colin~W.~Macrie}
\affiliation{Department of Physics and Astronomy, Purdue University, West Lafayette, IN 47907, USA}
\Rutgers

\author[0000-0003-2611-7269]{Keiichi~Maeda}
\affiliation{Department of Astronomy, Kyoto University, Kitashirakawa-Oiwake-cho, Sakyo-ku, Kyoto 606-8502, Japan}

\author[0000-0002-9770-3508]{Kate Maguire}
\affiliation{School of Physics, Trinity College Dublin, The University of Dublin, Dublin 2, Ireland.}

\author[0000-0001-5807-7893]{Curtis~McCully}
\LCO

\author[0000-0002-7015-3446]{Nicolas~E.~Meza-Retamal}
\UCD

\author[0000-0003-0209-9246]{Estefania~Padilla~Gonzalez}
\affiliation{Johns Hopkins University, Department of Physics and Astronomy, 3400 N. Charles Street Baltimore, MD 21218}

\author[0000-0003-3308-2420]{R\"udiger~Pakmor}
\affiliation{Max-Planck-Institut f\"{u}r Astrophysik, Karl-Schwarzschild-Str. 1, D-85748, Garching, Germany}

\author[0000-0002-0744-0047]{Jeniveve Pearson}
\UA

\author[0000-0001-6806-0673]{Anthony~L.~Piro}
\Carnegie

\author[0000-0002-1633-6495]{Abigail~Polin}
\Purdue

\author[0000-0002-5683-2389]{Nabeel~Rehemtulla}
\CIERA
\Northwestern
\SkAI

\author[0000-0002-7559-315X]{C\'esar~Rojas-Bravo}
\affiliation{School of Astronomy and Space Science, University of Chinese Academy of Sciences, Beijing 100049, People's Republic of China}
\affiliation{National Astronomical Observatories, Chinese Academy of Sciences, Beijing 100101, People's Republic of China}

\author[0000-0003-4102-380X]{David~J.~Sand}
\UA

\author[0009-0007-7934-1543]{Chita~Sangkachan}
\Thailand

\author[0009-0002-5096-1689]{Michaela Schwab}
\affiliation{Department of Astronomy, University of Virginia, 530 McCormick Rd, Charlottesville, VA 22904, USA}

\author[0000-0001-8023-4912]{Huei~Sears}
\Rutgers

\author[0000-0001-6706-2749]{Mridweeka Singh}
\affiliation{Indian Institute of Astrophysics, Koramangala 2nd Block, Bangalore 560034, India}

\author[0000-0001-8073-8731]{Bhagya~M.~Subrayan}
\UA

\author[0000-0002-5748-4558]{Kirsty~Taggart}
\UCSC

\author[0000-0001-7380-3144]{Tea Temim}
\Princeton

\author[0000-0001-9834-3439]{Jacco~H.~Terwel}
\affiliation{School of Physics, Trinity College Dublin, The University of Dublin, Dublin 2, Ireland}

\author[0000-0002-1481-4676]{Samaporn~Tinyanont}
\Thailand

\author[0000-0001-8764-7832]{J{\'o}zsef Vink{\'o}}
\Konkoly
\USzeged
\UTexas
\affiliation{ELTE E\"otv\"os Lor\'and University, Institute of Physics and Astronomy,
P\'azm\'any P\'eter s\'et\'any 1A, Budapest 1117, Hungary }

\author[0000-0002-7334-2357]{Xiaofeng~Wang}
\affiliation{Physics Department, Tsinghua University, Beijing, 100084, China}

\author[0000-0003-1349-6538]{J.~Craig~Wheeler}
\UTexas

\author[0000-0002-6535-8500]{Yi~Yang}
\affiliation{Physics Department, Tsinghua University, Beijing, 100084, China}

\author[0000-0002-2636-6508]{WeiKang~Zheng}
\affiliation{Department of Astronomy, University of California, Berkeley, CA 94720-3411, USA}

%% file: acknowledgements.tex
We thank the anonymous referee for a thorough review that substantially improved this manuscript. We also thank Fionntan Callan, Alexander Holas, Christine Collins, Ken Shen, Stefan Taubenberger, Priyam Das, and others at the One Hundred Years of Supernova Science conference in Stockholm, Sweden for inspiring discussions related to the results in this manuscript.

This work is based on observations made with the NASA/ESA/CSA \textit{JWST} as part of programs \#02072 and \#04516. We thank Shelly Meyett for her consistently excellent work scheduling the \textit{JWST} observations, Sarah Kendrew for assistance with the MIRI observations, and Glenn Wahlgren for help with the NIRSpec observations. 
The data were obtained from the Mikulski Archive for Space Telescopes at the Space Telescope Science Institute (STScI), which is operated by the Association of Universities for Research in Astronomy (AURA), Inc., under National Aeronautics and Space Administration (NASA) contract NAS 5-03127 for \textit{JWST}. Support for this program at Rutgers University was provided by NASA through grants JWST-GO-02072.001 and JWST-GO-04516.001. 

L.A.K. is supported by NASA through a Hubble Fellowship grant HF2-51579.001-A awarded by STScI, which is operated by the Association of Universities for Research in Astronomy, Inc., for NASA, under contract NAS5-26555. C. Liu, A.A.M., and N.R.~are supported by DoE award \#\,DE-SC0025599 to Northwestern University. A.A.M.~is also supported by Cottrell Scholar Award \#\,CS-CSA-2025-059 from Research Corporation for Science Advancement.
C. Larison acknowledges support from National Science Foundation (NSF) award \#\,AST-2407567 and DOE award \#\,DE-SC0010008 to Rutgers University.

M.A. is supported by NSF grant AST-2308113.
J.E.A.\ is supported by the international Gemini Observatory, a program of NSF's NOIRLab, which is managed by the Association of Universities for Research in Astronomy (AURA) under a cooperative agreement with the NSF, on behalf of the Gemini partnership of Argentina, Brazil, Canada, Chile, the Republic of Korea, and the United States of America.
C.B. acknowledges support from Chandra Theory grants TM0-21004X and TM1-22004X, XRISM Guest Scientist grant 80NSSC23K0634, and NSF-AST grant 2307865.
Support for M.D. was provided by Schmidt Sciences, LLC.
A.F. acknowledges support by the European Research Council (ERC) under the European Union's Horizon 2020 research and innovation programme (ERC Advanced Grant KILONOVA No.~885281), the Deutsche Forschungsgemeinschaft (DFG, German Research Foundation) -- Project-ID 279384907 -- SFB 1245, and MA 4248/3-1.

L.G. acknowledges financial support from AGAUR, CSIC, MCIN, and AEI 10.13039/501100011033 under projects PID2023-151307NB-I00, PIE 20215AT016, CEX2020-001058-M, ILINK23001, COOPB2304, and 2021-SGR-01270.
This work makes use of data from the Las Cumbres Observatory global network of telescopes.  The LCO group is supported by NSF grants AST-1911151 and AST-2308113. R.K.T. is supported by the NKFIH/OTKA FK-134432 of the National Research, Development and Innovation (NRDI) Office of Hungary.
K. Maeda acknowledges support from JSPS KAKENHI grant (JP24KK0070, JP24H01810).
K. Maguire acknowledges funding from Horizon Europe ERC grant 101125877.
This work includes data obtained with the Swope Telescope at Las Campanas Observatory, Chile, as part of the Swope Time Domain Key Project (PI: Piro, Co-Is: Drout, Phillips, Holoien, Burns, Madore, Foley, Coulter, Rojas-Bravo, Dimitriadis, Kilpatrick, Hsiao).
We thank the Swope observers Jorge Anais, Abdo Campillay, and Yilin Kong Riveros for their useful observations.

Time-domain research by the University of Arizona team and D.J.S. is supported by NSF grants 2108032, 2308181, 2407566, and 2432036 and the Heising-Simons Foundation under grant $\#$2020-1864. K.A.B. is supported by an LSST-DA Catalyst Fellowship; this publication was thus made possible through the support of grant 62192 from the John Templeton Foundation to LSST-DA.
M.S. acknowledges financial support provided under the National Post Doctoral Fellowship (N-PDF; File Number PDF/2023/002244) by the Science \& Engineering Research Board (SERB), Anusandhan National Research Foundation (ANRF), Government of India.
T.T. acknowledges support from NSF grant 2205314.
J.H.T. acknowledges support from EU H2020 ERC grant 758638.
J.V. is supported by Hungarian NKFIH OTKA Grant K142534. 
X.~F.~Wang is supported by the National Science Foundation of China (NSFC grants 12288102 and 12033003) and the Tencent Xplorer Prize.
A.V.F.’s research group at UC Berkeley acknowledges financial       assistance from the Christopher R. Redlich Fund, as well as donations from Gary and Cynthia Bengier, Clark and Sharon Winslow, Alan Eustace and Kathy Kwan, Timothy and Melissa Draper, Briggs and Kathleen Wood, Sanford Robertson, and Alan and Ellyn Seelenfreund (W.Z. is a Bengier-Winslow-Eustace Specialist in Astronomy, T.G.B. is a Draper-Wood-Seelenfreund Specialist in Astronomy, Y.Y. was a Bengier-Winslow-Robertson Fellow in Astronomy), and numerous other donors. Y.Y.'s research is partially supported by the Tsinghua University Dushi Program. 

Some of the data presented herein were obtained at the W.~M. Keck Observatory, which is operated as a scientific partnership among the California Institute of Technology, the University of California, and NASA. The Observatory was made possible by the generous financial support of the W.~M. Keck Foundation. The authors wish to recognize and acknowledge the very significant cultural role and reverence that the summit of Maunakea has always had within the indigenous Hawaiian community. We are most fortunate to have the opportunity to conduct observations from this mountain.
W. M. Keck and MMT Observatory access was supported by Northwestern University and the Center for Interdisciplinary Exploration and Research in Astrophysics (CIERA).

A major upgrade of the Kast spectrograph on the Shane 3~m telescope at Lick Observatory, led by Brad Holden, was made possible through generous gifts from the Heising-Simons Foundation, William and Marina Kast, and the University of California Observatories. Research at Lick Observatory is partially supported by a generous gift from Google.

The Hobby–Eberly Telescope (HET) is a joint project of the University of Texas at Austin, the Pennsylvania State University, Ludwig-Maximilians-Universit\"at M\"unchen, and Georg-August-Universit\"at G\"ottingen. The HET is named in honor of its principal benefactors, William P. Hobby and Robert E. Eberly. The Low Resolution Spectrograph 2 (LRS2) was developed and funded by the University of Texas at Austin McDonald Observatory and Department of Astronomy, and by Pennsylvania State University. We thank the Leibniz-Institut f\"ur Astrophysik Potsdam (AIP) and the Institut für Astrophysik Göttingen (IAG) for their contributions to the construction of the integral field units. The authors are grateful to the HET Resident Astronomers and staff members at McDonald Observatory for their excellent work.

We acknowledge the Texas Advanced Computing Center (TACC) at The University of Texas at Austin for providing high-performance computing, visualization, and storage resources that have contributed to the results reported within this paper.
This work was supported by the ``Action Th\'ematique de Physique Stellaire'' (ATPS) of CNRS/INSU PN Astro cofunded by CEA and CNES.
This work made use of the Heidelberg Supernova Model Archive (HESMA), \url{https://hesma.h-its.org}. Some grammar improvement and graphic design was aided by an LLM.

%% file: velocity_fitting_table.tex
\begin{table*}
\centering
\footnotesize
\setlength{\tabcolsep}{4pt}

\caption{Best-fit MIR line profile parameters for each ion from \autoref{fig:voff_fwhm}.
All velocity quantities are given in units of 1000\kms.
Columns that are not relevant to a given profile type are marked with a dash. The reported uncertainties may underestimate the true uncertainties, as they do not include potential systematic effects (e.g., peculiar velocities within the host galaxy) or additional sources of uncertainty in the fitting procedure. The $v_c$ parameter, which controls the slope of the flat-top, represents the velocity offset of the plateau with respect to $v_{\rm off}$ while $v_{\rm inner}$ represents the radius of the inner edge of the shell.}
\label{tab:fit_params}

\begin{tabular}{llcccccccc}

\toprule
Ion & Profile &
\multicolumn{2}{c}{Outer} &
\multicolumn{2}{c}{Inner} &
$v_{\rm inner}$ &
$v_c$ &
Ratio &
$n$ \\

\cmidrule(lr){3-4}
\cmidrule(lr){5-6}

& &
FWHM & $v_{\rm off}$ &
FWHM & $v_{\rm off}$ &
& & ($A_{\rm out}/A_{\rm in}$) & \\

\midrule

\multicolumn{10}{l}{\textbf{SN\,2024gy $+$144\,d}} \\
\midrule
{[Ni II]} & L+SG\tablenotemark{\scriptsize a} & $9.01\pm1.24$ & $0.75\pm0.24$ & $1.57\pm0.24$ & $0.77\pm0.03$ & --- & --- & $1.44\pm0.74$ & $3.27\pm0.92$ \\
{[Ni III]} & L+SG\tablenotemark{\scriptsize a} & $11.82\pm0.06$ & $0.54\pm0.02$ & $5.98\pm0.08$ & $0.74\pm0.03$ & --- & --- & $1.20\pm0.04$ & $5.56\pm0.13$ \\
{[Ni IV]} & L+SG\tablenotemark{\scriptsize a} & $11.75\pm0.07$ & $0.11\pm0.02$ & $7.68\pm0.33$ & $0.40\pm0.20$ & --- & --- & $3.00\pm0.01$ & $5.38\pm0.14$ \\
\addlinespace
{[Co II]} & Gaussian & $7.41\pm0.08$ & $1.04\pm0.02$ & --- & --- & --- & --- & --- & --- \\
{[Co III]} & Gaussian & $10.48\pm0.01$ & $0.57\pm0.01$ & --- & --- & --- & --- & --- & --- \\
{[Co IV]} & Gaussian & $13.30\pm0.14$ & $0.15\pm0.03$ & --- & --- & --- & --- & --- & --- \\
\addlinespace
{[Ar II]} & Flat-top & $20.80\pm0.72$ & $0.14\pm0.11$ & --- & --- & $9.30\pm0.06$ & $0.04\pm0.02$ & --- & --- \\
{[Ar III]} & Flat-top & $20.88\pm0.09$ & $0.06\pm0.02$ & --- & --- & $9.07\pm0.03$ & $0.05\pm0.01$ & --- & --- \\
\\
\multicolumn{10}{l}{\textbf{SN\,2022aaiq $+$125\,d}} \\
\midrule
{[Ni II]} & Gaussian & $6.90\pm0.06$ & $-1.89\pm0.02$ & $0.83\pm0.17$\tablenotemark{\scriptsize b}  & $-1.17\pm0.05$\tablenotemark{\scriptsize b}  & --- & --- & --- & --- \\
{[Ni III]} & L+SG\tablenotemark{\scriptsize a} & $12.16\pm0.05$ & $-0.11\pm0.01$ & $6.76\pm0.07$ & $-0.83\pm0.03$ & --- & --- & $1.16\pm0.04$ & $4.33\pm0.08$ \\
{[Ni IV]} & Gaussian & $11.32\pm0.06$ & $-0.44\pm0.03$ & --- & --- & --- & --- & --- & --- \\
\addlinespace
{[Co II]} & Gaussian & $5.73\pm0.04$ & $-1.79\pm0.01$ & --- & --- & --- & --- & --- & --- \\
{[Co III]} & Gaussian & $10.83\pm0.01$ & $-0.57\pm0.01$ & --- & --- & --- & --- & --- & --- \\
\addlinespace
{[Ar II]} & Flat-top & $20.61\pm1.50$ & $0.60\pm0.01$ & --- & --- & $9.51\pm0.06$ & $-0.65\pm0.20$ & --- & --- \\
{[Ar III]} & Flat-top & $19.52\pm0.24$ & $0.84\pm0.03$ & --- & --- & $8.56\pm0.09$ & $-0.50\pm0.03$ & --- & --- \\
\\
\multicolumn{10}{l}{\textbf{SN\,2021aefx $+$418\,d}} \\
\midrule
{[Ni II]} & Gaussian & $9.45\pm0.38$ & $1.14\pm0.24$ & --- & --- & --- & --- & --- & --- \\
{[Ni III]} & Gaussian & $12.34\pm0.14$ & $0.85\pm0.06$ & --- & --- & --- & --- & --- & --- \\
{[Ni IV]} & Gaussian & $14.00\pm1.50$ & $0.14\pm0.07$ & --- & --- & --- & --- & --- & --- \\
\addlinespace
{[Co II]} & Gaussian & $7.32\pm0.08$ & $1.09\pm0.01$ & --- & --- & --- & --- & --- & --- \\
{[Co III]} & Gaussian & $11.14\pm0.03$ & $0.74\pm0.01$ & --- & --- & --- & --- & --- & --- \\
\addlinespace
{[Ar II]} & FT+G\tablenotemark{\scriptsize c} & $20.39\pm0.51$ & $0.63\pm0.31$ & $8.00\pm1.50$ & $0.16\pm0.06$ & $9.54\pm0.31$ & $0.50\pm0.20$ & $1.33\pm0.27$ & --- \\
{[Ar III]} & Flat-top & $21.20\pm0.27$ & $0.69\pm0.03$ & --- & --- & $8.89\pm0.05$ & $0.45\pm0.02$ & --- & --- \\
\\
\multicolumn{10}{l}{\textbf{SN\,2022xkq $+$114\,d}} \\
\midrule
{[Ni II]} & Gaussian & $2.74\pm0.06$ & $-0.29\pm0.02$ & --- & --- & --- & --- & --- & --- \\
{[Ni III]} & Gaussian & $4.32\pm0.07$ & $-0.06\pm0.02$ & --- & --- & --- & --- & --- & --- \\
{[Ni IV]} & Gaussian & $4.90\pm0.15$ & $-0.44\pm0.05$ & --- & --- & --- & --- & --- & --- \\
\addlinespace
{[Co II]} & Gaussian & $8.48\pm0.05$ & $-0.09\pm0.02$ & --- & --- & --- & --- & --- & --- \\
{[Co III]} & Flat-top & $8.60\pm0.02$ & $-0.24\pm0.01$ & --- & --- & $2.12\pm0.03$ & $0.20\pm0.02$ & --- & --- \\
{[Co IV]} & Gaussian & $7.24\pm0.12$ & $-0.27\pm0.05$ & --- & --- & --- & --- & --- & --- \\
\addlinespace
{[Ar II]} & Flat-top & $15.72\pm0.06$ & $0.04\pm0.01$ & --- & --- & $6.00\pm0.03$ & $-0.14\pm0.01$ & --- & --- \\
{[Ar III]} & Flat-top & $15.61\pm0.05$ & $0.00\pm0.01$ & --- & --- & $5.56\pm0.03$ & $-0.01\pm0.01$ & --- & --- \\
\\

\bottomrule
\end{tabular}
\tablenotetext{\scriptsize a}{\raggedright Lorentzian$+$SuperGaussian (\autoref{eq:L+SG})}
\tablenotetext{\scriptsize b}{\raggedright Parameter is from the narrow core of the NIR [\ion{Ni}{2}]\,1.94\um\ line}
\tablenotetext{\scriptsize c}{\raggedright Flat-top$+$Gaussian}
\end{table*}

%% file: 2022aaiq_photometry_table.tex
\begin{longtable}{cccc}
\caption{Optical photometry for SN~2022aaiq. Phases are relative to B-band maximum MJD $59914.0$. Only a portion of this table is shown here; the full machine-readable version is provided as supplementary material.}\label{tab:2022aaiq_photometry}\\
\hline
MJD & Phase (d) & Magnitude (mag) & Filter \\
\hline
\endfirsthead
\multicolumn{4}{c}{\small\itshape (continued)}\\
\hline
MJD & Phase (d) & Magnitude (mag) & Filter \\
\hline
\endhead
\hline
\endfoot
\hline
\endlastfoot
59912.516 & $-1.5$ & $13.32 \pm 0.01$ & B \\
59912.519 & $-1.5$ & $13.40 \pm 0.02$ & V \\
59912.520 & $-1.5$ & $13.46 \pm 0.02$ & V \\
59912.521 & $-1.5$ & $13.48 \pm 0.01$ & g \\
59912.522 & $-1.5$ & $13.46 \pm 0.01$ & g \\
59916.514 & $2.5$ & $13.57 \pm 0.01$ & B \\
59916.515 & $2.5$ & $13.57 \pm 0.01$ & B \\
59916.517 & $2.5$ & $13.63 \pm 0.02$ & V \\
59916.518 & $2.5$ & $13.60 \pm 0.02$ & V \\
59916.519 & $2.5$ & $13.56 \pm 0.01$ & g \\
$\cdots$ & $\cdots$ & $\cdots$ & $\cdots$\\
\end{longtable}

%% file: 2024gy_photometry_table.tex
\begin{longtable}{cccc}
\caption{Optical photometry for SN~2024gy. Phases are relative to B-band maximum MJD $60329.5$. Only a portion of this table is shown here; the full machine-readable version is provided as supplementary material.}\label{tab:2024gy_photometry}\\
\hline
MJD & Phase (d) & Magnitude (mag) & Filter \\
\hline
\endfirsthead
\multicolumn{4}{c}{\small\itshape (continued)}\\
\hline
MJD & Phase (d) & Magnitude (mag) & Filter \\
\hline
\endhead
\hline
\endfoot
\hline
\endlastfoot
60314.310 & $-15.2$ & $16.74 \pm 0.02$ & B \\
60314.310 & $-15.2$ & $16.72 \pm 0.02$ & B \\
60314.310 & $-15.2$ & $15.89 \pm 0.03$ & V \\
60314.310 & $-15.2$ & $15.91 \pm 0.03$ & V \\
60314.320 & $-15.2$ & $16.27 \pm 0.01$ & g \\
60314.320 & $-15.2$ & $16.25 \pm 0.01$ & g \\
60314.320 & $-15.2$ & $15.77 \pm 0.02$ & r \\
60314.320 & $-15.2$ & $15.74 \pm 0.01$ & r \\
60314.320 & $-15.2$ & $15.71 \pm 0.03$ & r \\
60314.320 & $-15.2$ & $16.21 \pm 0.02$ & i \\
$\cdots$ & $\cdots$ & $\cdots$ & $\cdots$\\
\end{longtable}